\documentclass[a4paper,12pt]{report}
\usepackage{amssymb}
\usepackage{microtype}
\usepackage{url}
\usepackage{graphicx}
\usepackage{listings}
\usepackage{enumerate}
\usepackage{cite}
\usepackage{color}
\usepackage[english]{babel}
\usepackage[intlimits]{amsmath}

\newcommand{\epem}{e$^+$+e$^-$ }

\begin{document}

\begin{center}

DEPARTMENT OF PHYSICS \\
UNIVERSITY OF JYV\"ASKYL\"A \\
RESEARCH REPORT No. 6/2014

\bigskip
\bigskip
\bigskip

{\Large

{\bf SPATIALLY DEPENDENT PARTON DISTRIBUTION FUNCTIONS AND HARD PROCESSES IN NUCLEAR COLLISIONS}

\bigskip
\bigskip

{\bf BY \\ ILKKA HELENIUS}

\bigskip
\bigskip
\bigskip

}

Academic Dissertation \\
for the Degree of \\
Doctor of Philosophy 

\bigskip
\bigskip

\emph{
To be presented, by permission of the\\
Faculty of Mathematics and Natural Sciences\\
of the University of Jyv\"askyl\"a,\\
for public examination in Auditorium FYS 1 of the\\
University of Jyv\"askyl\"a on August 5, 2014\\
at 12 o'clock noon
}

\end{center}

\bigskip

\begin{figure}[b]
\begin{center}
\scalebox{.15}{\includegraphics{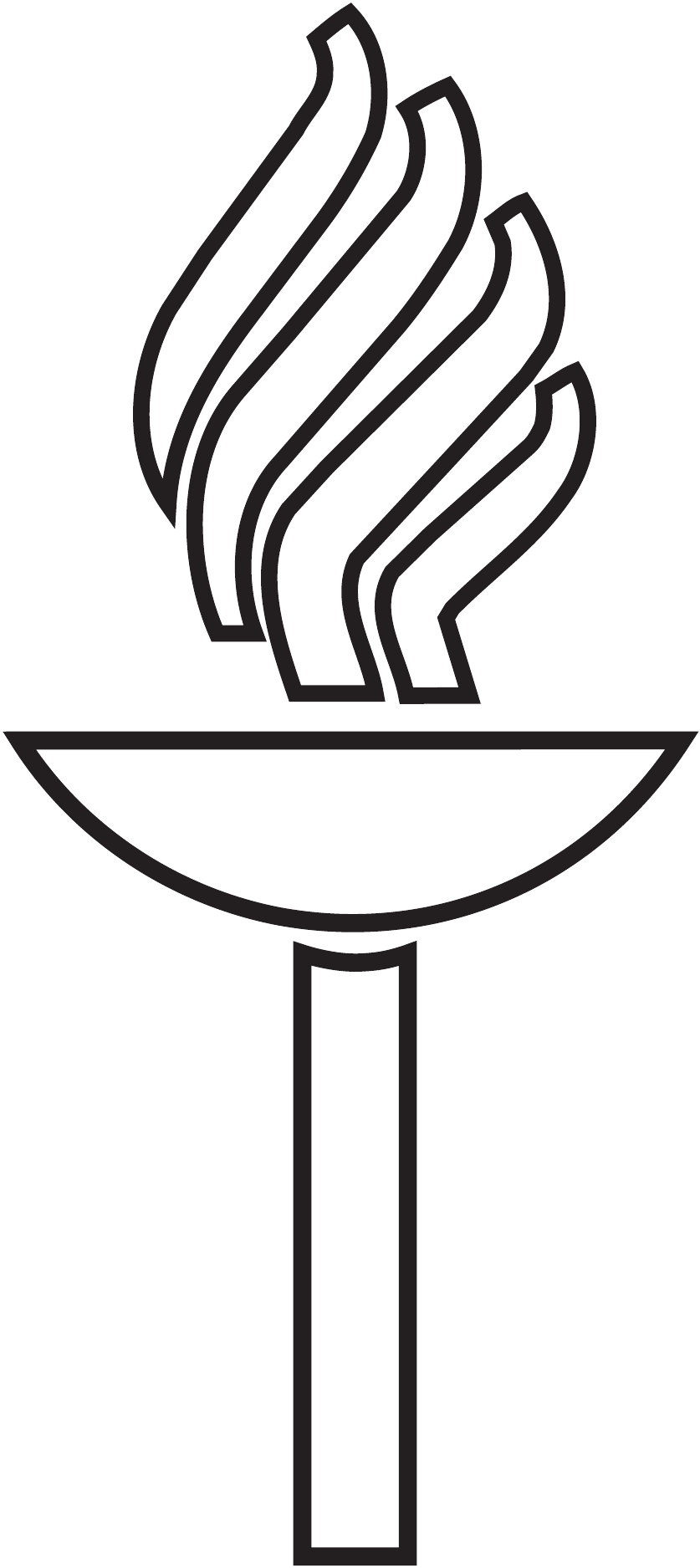}}

\medskip

\begin{minipage}[b]{3.5cm}
\begin{center}
Jyv\"askyl\"a, Finland \\
August 2014
\end{center}
\end{minipage}

\end{center}
\end{figure}

\thispagestyle{empty}

\chapter*{Preface}

\noindent The work presented in this thesis has been carried out during the years from 2010 to 2014 at the Department of Physics of the University of Jyv\"askyl\"a. The work has been supervised by Prof. Kari J. Eskola whom I would like to thank for guidance during my PhD studies. The supervision has been excellent and very friendly throughout the whole time.

I would also like to thank Dr. Thorsten Renk for his guidance at the beginning of my PhD studies and for the collaboration also later on. Large part of my research has been done together with Dr. Hannu Paukkunen whom I thank for the very smooth and effective collaboration. Prof. Carlos Salgado, Dr. David d'Enterria, Dr. Heli Honkanen, Dr. Rupa Chatterjee and Dr. Hannu Holopainen are also gratefully acknowledged for the collaboration. I am thankful also to the local ALICE people who have kept me connected also to the experimental side of particle physics. I am grateful to Prof. Paul Hoyer and Dr. Francois Arleo for reviewing the original manuscript and for providing useful comments regarding to it as well as to Dr. Marco Stratmann for promising to act as my opponent.

I want to thank the Department of Physics for the friendly atmosphere. The management and the office people have done great job to make things work smoothly. I thank the other PhD students in our group for numerous discussions and for keeping me company in several conference trips. Big thanks also to the ``Holvi'' community which has made our office a very pleasant place to work and has given a significant contribution also to my time outside the office. Especially the cruises on the Baltic sea are unforgettable.

Finally, I would like to thank my family for all the support you have provided during my studies and life in general. The special thanks goes to my girlfriend Annakaisa whose support has been priceless.

\vspace{10mm}
\noindent 
Financial support from the Magnus Ehrnrooth foundation, the Helsinki Institute of Physics, the Academy of Finland (Project No. 133005), the Graduate School for Particle and Nuclear Physics and the Department of Physics, University of Jyv\"askyl\"a, are gratefully acknowledged.

\vspace{10mm}
\noindent Jyv\"askyl\"a, July 2014\\
Ilkka Helenius

\thispagestyle{empty}


\chapter*{Abstract}

This work focuses on applications of perturbative QCD (pQCD) and collinear factorization theorem to hard particle production in nuclear and hadronic collisions at the BNL-RHIC and CERN-LHC colliders. The emphasis is on nuclear parton distribution functions (nPDFs) and their spatial dependence. Also parton-to-hadron fragmentation functions (FFs) are studied. A brief overview of the applied theoretical and numerical tools is given in the introductory part.

We have developed a framework for the spatial dependence of the nPDFs and published two new nPDF sets, EPS09s and EKS98s. We have applied these sets to study the centrality dependence of inclusive hadron and direct photon production in nuclear collisions and compared our results to existing data from different LHC and RHIC experiments. We have found a good agreement between our next-to-leading order (NLO) calculations and the published data, although the observed centrality dependence is rather mild and the experimental uncertainties are of the same order. According to our studies, the measurements at forward rapidities could provide more constraints for the nPDFs as the centrality dependence is more pronounced there. We have also shown that the addition of the NLO pQCD photon component on top of the thermal photons is necessary to explain the measured transverse momentum spectra of direct photons in nucleus-nucleus collisions, and that this significantly reduces the predicted photonic elliptic flow.

To study how the so far poorly known small-momentum fraction (x) gluon nPDFs could be constrained using the proton-lead collisions at the LHC, we have quantified which x regions are probed by inclusive hadron and direct photon production at different rapidities. We have found that the isolated photons at forward rapidities would be the best observable to study small-x effects. We have also shown that the NLO pQCD calculations with the present FF sets tend to overshoot the charged hadron data in proton-proton collisions at the LHC energies. The behaviour is identified to originate from too hard gluon-to-hadron FFs and a reanalysis is called for.

\thispagestyle{empty}

\newpage

\chapter*{}

\begin{tabular}{lll}
{\bf Author} & & Ilkka Helenius\\
	     & & Departments of Physics\\
	     & & University of Jyv\"askyl\"a\\
	     & & Finland\\
	     & & \\
{\bf Supervisor} & & Prof. Kari J. Eskola\\
	     & & Departments of Physics\\
	     & & University of Jyv\"askyl\"a\\
	     & & Finland\\
	     & & \\
{\bf Reviewers} & & Prof. Paul Hoyer\\
	     & & Division of elementary particle physics\\
	     & & Department of Physics\\
	     & & University of Helsinki\\
	     & & Finland\\
	     & & \\
             & & Dr. Francois Arleo\\
             & & Laboratoire Leprince-Ringuet\\
	     & & \'{E}cole polytechnique\\
	     & & Palaiseau, France\\
	     & & \\
{\bf Opponent} & & Dr. Marco Stratmann\\
             & & Institute for Theoretical Physics\\
             & & University of T\"ubingen\\
	     & & Germany\\
	     & & \\
\end{tabular}

\thispagestyle{empty}

\chapter*{List of publications}

\noindent This thesis consists of an introductory part and of the following publications:

\begin{enumerate}[\bf I]
\item {\bf Impact-parameter dependent nuclear parton distribution functions: EPS09s and EKS98s and their applications in nuclear hard processes}\\
  I.~Helenius, K.~J.~Eskola, H.~Honkanen and C.~A.~Salgado, \\
  JHEP {\bf 1207} (2012) 073,
  [arXiv:1205.5359 [hep-ph]]
\item {\bf Centrality dependence of inclusive prompt photon production in d+Au, Au+Au, p+Pb, and Pb+Pb collisions}\\
  I.~Helenius, K.~J.~Eskola and H.~Paukkunen, \\
  JHEP {\bf 1305} (2013) 030,
  [arXiv:1302.5580 [hep-ph]]
\item {\bf Elliptic flow of thermal photons from event-by-event hydrodynamic model}\\
  R.~Chatterjee, H.~Holopainen, I.~Helenius, T.~Renk and K.~J.~Eskola,\\
  Phys.~Rev.~C {\bf 88} (2013) 034901,
  [arXiv:1305.6443 [hep-ph]]
\item {\bf Confronting current NLO parton fragmentation functions with inclusive charged-particle spectra at hadron colliders} \\
  D.~d'Enterria, K.~J.~Eskola, I.~Helenius and H.~Paukkunen, \\
  Nucl.~Phys.~B {\bf 883} (2014) 615,
  [arXiv:1311.1415 [hep-ph]]
\item {\bf Probing the small-$x$ nuclear gluon distributions with isolated photons at forward rapidities in p+Pb collisions at the LHC}\\
  I.~Helenius, K.~J.~Eskola and H.~Paukkunen,\\
  arXiv:1406.1689 [hep-ph], to appear in JHEP
\end{enumerate}

\noindent The author performed all of the numerical work and wrote the original drafts for the publications [I, II, V]. Part of the results presented in the article [I] were obtained from a numerical code written from scratch by the author. The author prepared also the interface code and instructions for our www-release of EPS09s and EKS98s. For the article [IV] the author did all the numerical work and participated in the planning and writing of the publication. For the article [III] the author provided the new centrality dependent NLO pQCD calculations for the direct photon production and participated in the writing of the article.

\pagenumbering{roman} \setcounter{page}{1}	

\tableofcontents
\newpage

\pagenumbering{arabic} \setcounter{page}{1} 

\chapter{Collinear factorization and perturbative QCD}
\label{chap:pQCD}

In the standard model of particle physics the fundamental interactions between quarks and gluons are described by quantum chromodynamics (QCD). The peculiar feature of this quantum field theory is the running of the coupling constant $\alpha_s$: At small energy scales the strength of the coupling is large but weakens towards higher scales which leads to \emph{asymptotic freedom} at large energy scales. Thus, even though we do not observe free quarks in Nature but they are always confined to color neutral bound states, known as \emph{hadrons}, involving a quark-antiquark pair (mesons) or three (anti-)quarks (baryons), the quarks and gluons at large enough scales can be treated as free particles. The small coupling also allows us to use a perturbative expansion in the coupling constant to calculate the cross sections of QCD-induced reactions at high energies. There are other methods developed to study QCD at low energies and strong coupling, e.g.~the lattice QCD, effective field theories and gauge/gravity dualities. In this thesis, I will focus on the high-energy behavior of QCD and study the particle production in high-energy hadronic collisions using perturbative QCD (pQCD). 

The fundamental theorem for this thesis is the collinear factorization theorem \cite{Collins:1989gx, Brock:1993sz}. It states that the hard interactions between the quarks and gluons, which are often referred to as \emph{partons} due to historic reasons, taking place at large momentum scales can be factorized from the soft parts describing the exact partonic content of the hadrons. In this framework the cross section of inclusive hadron $(h_3)$ production in a collision of hadrons $h_1$ and $h_2$ can be written as
\begin{align}
&\mathrm{d} \sigma^{h_1 + h_2 \rightarrow h_3 + X}(\mu^2,Q^2,Q_F^2) \label{eq:coll_fact}\\ &= \sum\limits_{i,j,k,X'} f_{i}^{h_1}(x_1,Q^2) \otimes f_{j}^{h_2}(x_2,Q^2) \otimes \mathrm{d}\hat{\sigma}^{ij\rightarrow k + X'}(\mu^2,Q^2,Q_F^2) \otimes D_{k}^{h_3}(z,Q^2_F),\notag
\end{align}
where the $\otimes$ stands for a convolution between the distributions (the scales $\mu^2$, $Q^2$, and $Q_F^2$ will be defined later). There are three types of terms in equation (\ref{eq:coll_fact}):
\begin{itemize}
\item $f_{i}^{h_1}(x_1,Q^2)$ and $f_{j}^{h_2}(x_2,Q^2)$: The parton distribution functions (PDFs) describing the number densities of partons $i$ and $j$ in hadron $h_1$ and $h_2$, respectively. These will be discussed in detail in Chapter \ref{chap:PDF}.
\item $D_{k}^{h_3}(z,Q^2_F)$: The parton-to-hadron fragmentation functions (FFs) which describe the probability to produce the hadron $h_3$ from a parton $k$. These will be considered in Chapter \ref{chap:FF}. In the context of inclusive direct photon production, also parton-to-photon FFs will be discussed in Chapter \ref{chap:direct_photon}.
\item $\mathrm{d}\hat{\sigma}^{ij\rightarrow k + X'}(\mu^2,Q^2,Q_F^2)$: The partonic pieces which can be calculated using pQCD at a fixed order in $\alpha_s$. This part is discussed in more detail in Chapter \ref{chap:direct_photon} for the inclusive direct photon production.
\end{itemize}
It should be emphasized that the equation~(\ref{eq:coll_fact}) is not an exact result but receives corrections from the truncation of the perturbative series to a fixed order  and from higher twist effects. The order of the corrections depends on the given order of calculation but in each case the corrections are expected to be small at large enough energy scales.

In this thesis the collinear factorization framework is applied to study the particle production in hadronic and nuclear collisions, keeping the main emphasis on nuclear collisions performed with the Relativistic Heavy-Ion Collider (RHIC) at BNL and with the Large Hadron Collider (LHC) at CERN. Also the charged hadron production in proton-proton collisions at the LHC is discussed. The goal is to study how the PDFs and FFs could be improved with the existing and forthcoming data from these collisions and to present pQCD based predictions for so far unexplored kinematic regions to study the universality of the collinear factorization framework. A special emphasis is on the nuclear modifications of the PDFs.

As a completely original analysis, in the article [I] included in this thesis we developed also two sets of spatially dependent nuclear PDFs (nPDFs). These are reviewed in Chapter \ref{chap:spatial_nPDFs} after discussing the centrality class definitions in Chapter \ref{chap:centrality}. In Chapter \ref{chap:outlook} I discuss briefly the next possible steps in modeling the initial state nuclear modifications. The spirit of the thesis is to review the different components in the collinear factorization framework which are employed in the included articles, and present the main results of the articles in the related Chapters.


\chapter{Parton distribution functions}
\label{chap:PDF}

In this Chapter I will discuss the first two terms in the equation (\ref{eq:coll_fact}), i.e. the parton distribution functions. Rather than going through all lengthy derivations that are already well documented, e.g. in Refs.~\cite{Dokshitzer:1991wu, Paukkunen:2009ks}, the aim here is to get a grip on the relevant equations and to study the amount of uncertainty in the present fits.

\section{DGLAP equations}

\begin{figure}
\centering
\includegraphics[width=0.7\textwidth]{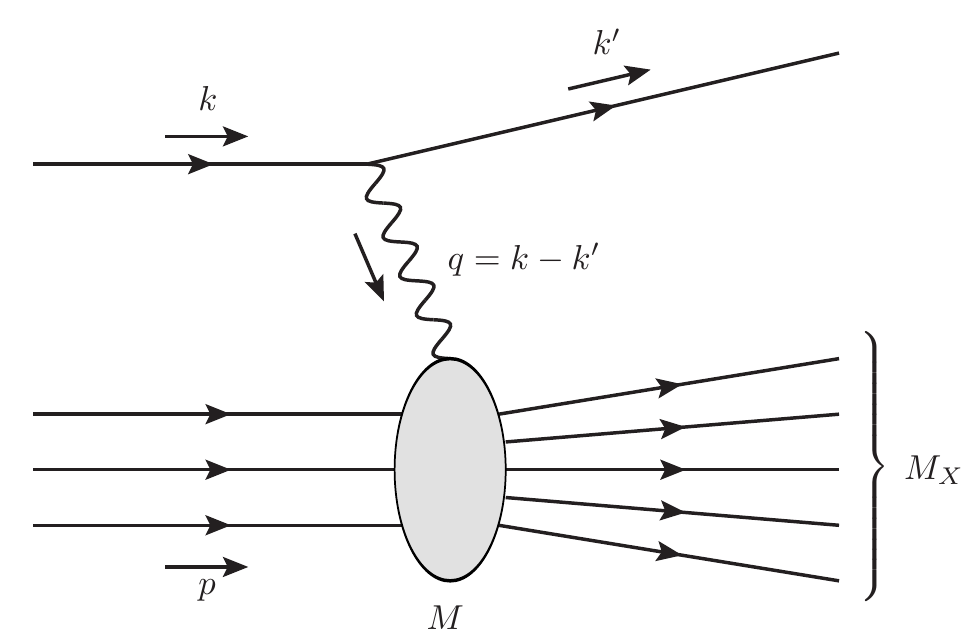}
\caption{The kinematics of a deep inelastic lepton-proton scattering.}
\label{fig:DISkin}
\end{figure}

Originally the inner structure of the proton was discovered in the 50's in elastic electron-proton scatterings measuring the proton form factors \cite{Mcallister:1956ng}. The deep inelastic scattering (DIS) experiments at SLAC-MIT then lead to the formulation of the parton model \cite{Feynman:1969ej, Bjorken:1969ja}. Later on these partons were identified as quarks and gluons which are described by QCD. Thus, it is natural to begin the PDF discussion with this process. In a DIS experiment a hadronic target is hit by a high-energy lepton and shattered to other hadrons, whose invariant mass $M_X$ is much larger than the original target-hadron mass $M$. As the collision kinematics, presented in figure \ref{fig:DISkin}, are determined entirely by the scattering angle and the momentum of the scattered lepton, the distribution of final state hadrons is not usually considered. However, for fragmentation studies one can use single-inclusive DIS (SIDIS) data, where the momentum of a hadron is measured.
The DIS kinematics can be defined in terms of the Lorentz-invariant quantities
\begin{eqnarray}
Q^2 & \equiv & -q^2\\
x & \equiv & \frac{Q^2}{2 p \cdot q},
\end{eqnarray}
where, in the leading order (LO) parton model, $x$ describes the momentum fraction of the struck parton w.r.t.~proton momentum. Using the parton model, the LO DIS cross section can be written as
\begin{equation}
\frac{\mathrm{d} \sigma}{\mathrm{d} Q^2 \mathrm{d} x} = \frac{2\pi \alpha^2}{Q^4}\sum_i e_{q_i}^2 f_i(x)\left[ 1 + \left( 1-\frac{Q^2}{xs} \right)^2 - \frac{Q^2M^2}{s^2} \right],
\end{equation}
where $e_{q_i}$ is the electric charge of the parton $i$ and $f_i(x)$ is the ``bare'', unrenormalized parton distribution function (PDF) of the proton. Noteworthy is that within the LO parton model the structure of the proton does not depend on the scale $Q^2$ at which the proton is probed. However, when one includes also the QCD corrections to the parton model, this so called Bjorken scaling does not hold anymore but the PDFs become scale dependent.

The scale evolution of the PDFs can be calculated using the \emph{Dokshitzer-Gribov-Lipatov-Altarelli-Parisi} (DGLAP) equations \cite{Lipatov:1974qm, Gribov:1972ri, Altarelli:1977zs, Dokshitzer:1977sg}. These can be derived by resumming the logarithmically divergent terms that arise from collinear emissions to all orders (for a useful review, see e.g. Ref.~\cite{Paukkunen:2009ks}). To see how these equations arise for quarks one can also study DIS at next-to-leading order (NLO) which corresponds to the order $\alpha \alpha_s$ in the electromagnetic and strong couplings. At this order there are ultraviolet, collinear and infrared divergences but after all contributions are taken into account, all but the collinear singularities for the emissions of initial partons cancel out. The remaining singularity is absorbed into the NLO definition of the PDFs. The definitions of the PDFs are, however, not unique but depend on the renormalization scheme. As shown e.g. in Refs.~\cite{Altarelli:1979ub, KJE_lectures} (and in my MSc-thesis \cite{Helenius:Thesis:2010}), the scale dependent NLO PDFs of quarks can be defined at a scale $Q^2$ as
\begin{align}
q^{scheme}(x,Q^2)\equiv& \int_x^1 \frac{\mathrm{d}\xi}{\xi} q_{0}(\xi)\left[ \delta(1-z) + \frac{\alpha_s}{2\pi}P_{qq}(z)\log\frac{Q^2}{\mu_{DR}^2}+\alpha_s f_q^{scheme}(z) \right] \notag\\
+& \int_x^1 \frac{\mathrm{d}\xi}{\xi} g_0(\xi)\left[ \frac{\alpha_s}{2\pi}P_{qg}(z)\log\frac{Q^2}{\mu_{DR}^2}+\alpha_s f_g^{scheme}(z) \right],
\end{align}
where $q_0(\xi)\equiv f_q(\xi)$ and $g_0(\xi)\equiv f_g(\xi)$ are the bare PDFs, $z=x/\xi$, and $\mu_{DR}$ is the (unphysical) scale arising from the dimensional regularization. The functions $f_g^{scheme}(z)$ and $f_q^{scheme}(z)$ depend on the renormalization scheme, e.g. in the modified minimal subtraction ($\rm \overline{MS}$) scheme these are defined as
\begin{equation}
f_q^{\rm \overline{MS}}(z) \equiv \frac{1}{2\pi}\left[ -\frac{1}{\hat{\epsilon}}P_{qq}(z) \right] \quad \text{and} \quad f_g^{\rm \overline{MS}}(z) \equiv \frac{1}{2\pi}\left[ -\frac{1}{\hat{\epsilon}}P_{qg}(z) \right],
\end{equation}
where $1/\hat{\epsilon} = 1/\epsilon-\gamma_E-\log(4\pi)$. The splitting functions $P_{qq}(z)$ and $P_{qg}(z)$ will be defined below. Taking now a derivative of $q(x,Q^2)$ with respect to $\log(Q^2)$ and replacing the bare PDFs with the scale dependent ones, gives 
\begin{equation}
\frac{\partial q(x,Q^2)}{\partial \log(Q^2)} = \frac{\alpha_s(Q^2)}{2\pi}\int_x^1 \frac{\mathrm{d}\xi}{\xi} \left[ q(\xi,Q^2)P_{qq}(x/\xi) + g(\xi,Q^2)P_{qg}(x/\xi) \right],
\end{equation}
which is now the DGLAP equation for quarks. It should be emphasized that also the strong coupling constant $\alpha_s$ depends on the scale through renormalization group equations, thus affecting also the scale evolution of the PDFs. At one loop this running can be written as \cite{Beringer:1900zz}
\begin{equation}
\alpha_s(Q^2)=\frac{12\pi}{\left(33-2n_f\right)\log (Q^2/\Lambda_{QCD}^2)},
\label{eq:alpha_s}
\end{equation}
where $n_f$ is the number of dynamical quark flavors and $\Lambda_{QCD}$ is the characteristic scale of QCD below which the coupling becomes strong. Measurements suggest $\Lambda_{QCD}\approx 200\,\mathrm{MeV}$, but the value used in different analyses varies.

Defining the convolution operator $\otimes$ as
\begin{equation}
P \otimes f \equiv \int_x^1 \frac{\mathrm{d}\xi}{\xi}P(x/\xi)f(\xi)
\end{equation}
we can write the full set of DGLAP equations as (leaving the scale dependence implicit)
\begin{align}
\frac{\partial q_i}{\partial \log(Q^2)} &= \frac{\alpha_s}{2\pi}\Big[\sum_j P_{q_i q_j}\otimes q_j + \sum_j P_{q_i \bar{q}_j}\otimes \bar{q}_j + P_{q_i g}\otimes g\Big] \label{eq:DGLAPq}\\
\frac{\partial \bar{q}_i}{\partial \log(Q^2)} &= \frac{\alpha_s}{2\pi}\Big[\sum_j P_{\bar{q}_i q_j}\otimes q_j + \sum_j P_{\bar{q}_i \bar{q}_j}\otimes \bar{q}_j + P_{\bar{q}_i g}\otimes g\Big] \label{eq:DGLAPqbar}\\
\frac{\partial g}{\partial \log(Q^2)} &= \frac{\alpha_s}{2\pi}\Big[\sum_j P_{g q_j}\otimes q_j + \sum_j P_{g \bar{q}_j}\otimes \bar{q}_j + P_{g g}\otimes g\Big]. \label{eq:DGLAPg}
\end{align}
This is a group of coupled integro-differential equations, which describe the scale evolution of the PDFs. The sum here runs over dynamical quark flavors. 

The splitting functions $P_{ij}(z)$ can be interpreted as a probability density to find a parton $i$ with a momentum fraction $z$ from a parton $j$. The LO splitting functions $P_{qq}(z)$ and $P_{qg}(z)$ can again be obtained from an NLO DIS calculation but for $P_{gq}(z)$ an $P_{gg}(z)$ one needs to consider some other process or DIS at NNLO. The outcome is
\begin{align}
P_{qq}(z) &= \frac{4}{3}\left[ \frac{1+z^2}{(1-z)_+} + \frac{3}{2}\delta(1-z) \right]\\
P_{qg}(z) &= \frac{1}{2}\left[ z^2 + (1-z)^2 \right]\\
P_{gq}(z) &= \frac{4}{3}\left[ \frac{1+(1-z)^2}{z}\right]\\
P_{gg}(z) &= 6\left[ \frac{z}{(1-z)_+} + \frac{1-z}{z}+z(1-z)+\frac{11-\frac{2}{3}n_f}{12} \delta(1-z) \right].
\end{align}
The definition of the standard ``plus'' distribution $\left(\frac{1}{1-z}\right)_+$, whose origin is at the correct treatment of the $z \rightarrow 1$ limit, can be found e.g. from Ref.~\cite{Helenius:Thesis:2010}. One can also calculate the splitting functions in higher order in $\alpha_s$. In principle this can be done to all orders, the current state of the art being NNLO \cite{Vogt:2004mw, Moch:2004pa}, but already at NLO the expressions for splitting functions become rather cumbersome \cite{Furmanski:1980cm,Curci:1980uw}.

\section{Free proton PDFs}

The DGLAP equations predict the scale dependence of the PDFs but do not provide the original $x$ dependence. For this, one needs to construct an ansatz at a chosen initial scale $Q_0$. The ansatz should be such that it leaves enough freedom to capture all the relevant features in the hard-process data while still keeping the number of parameters limited not to fit the fluctuations in the data but to catch the correct behaviour. The canonical form of the parametrization is
\begin{equation}
f_i(x,Q^2_0) = a_i x^{b_i}(1-x)^{c_i} F(x),
\end{equation}
where the modern PDF fits \cite{Pumplin:2002vw, Nadolsky:2008zw, Martin:2009iq, Lai:2010vv, Nocera:2014gqa} include more parameters in $F(x)$ to introduce more freedom to the $x$ behaviour. In principle the parameters are different for each flavor but some symmetries are often assumed to reduce the number of parameters in the fit, e.g. $q(x,Q^2) = \bar{q}(x,Q^2)$ for non-valence quarks. Some further constraints are obtained also from physical restrictions, e.g. baryon number and momentum sum rules. The values for the remaining parameters have to be obtained by a global analysis. The word ``global'' here means that one should take into account data from all possible different hard processes over a large kinematic reach. 

\begin{figure}[htb]
\centering
\includegraphics[width=0.9\textwidth]{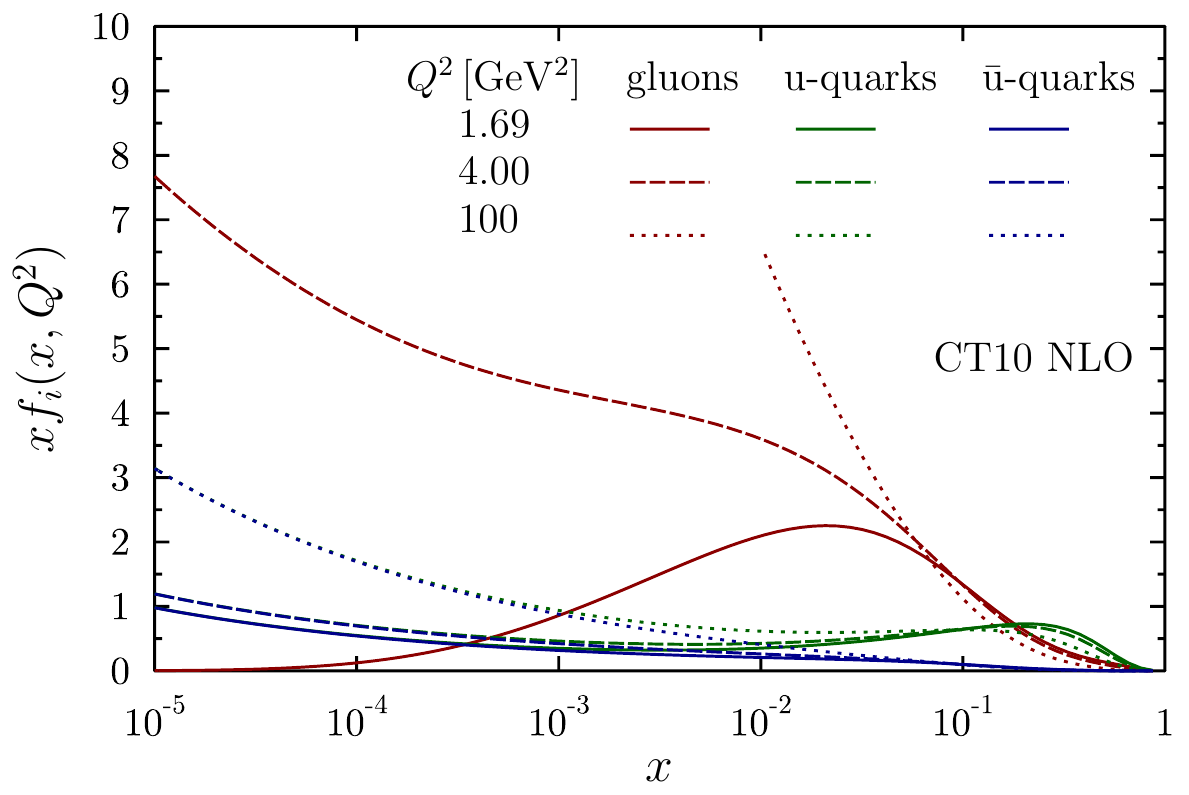}
\caption{The proton PDFs for gluons (red), u-quarks (green) and $\bar{\text{u}}$-quarks (blue) for scales $Q^2=1.69\,\mathrm{GeV^2}$ (solid), $4\,\mathrm{GeV^2}$ (dashed) and $100\,\mathrm{GeV^2}$ (dotted) from the CT10 NLO set.}
\label{fig:pdf_ct10}
\end{figure}

As an example of the free proton PDFs, I will consider those from a rather recent CT10 analysis \cite{Lai:2010vv}. Figure \ref{fig:pdf_ct10} shows the NLO gluon, u-quark and $\bar{\text{u}}$-quark PDFs at three different scales. Several common features of the PDF fits can be seen from the figure. At $x\gtrsim0.2$ the u-quark gives a larger contribution than the $\rm \bar{u}$ but towards lower values of $x$ the quark and antiquark distributions become equally important as the valence quark contribution vanishes. This is reflected by the very similar u- and $\bar{\text{u}}$ -distributions in the figure at $x<0.01$. Above the initial scale ($Q_0^2=1.69\,\mathrm{GeV^2}$ in CT10), the gluon PDFs become very large at $x<0.1$ due to the very rapid scale evolution at small $x$. The NLO scale evolution predicted by the DGLAP equations for quarks is significantly slower than for gluons in this kinematic region. 

\begin{figure}[htb]
\centering
\includegraphics[width=0.49\textwidth]{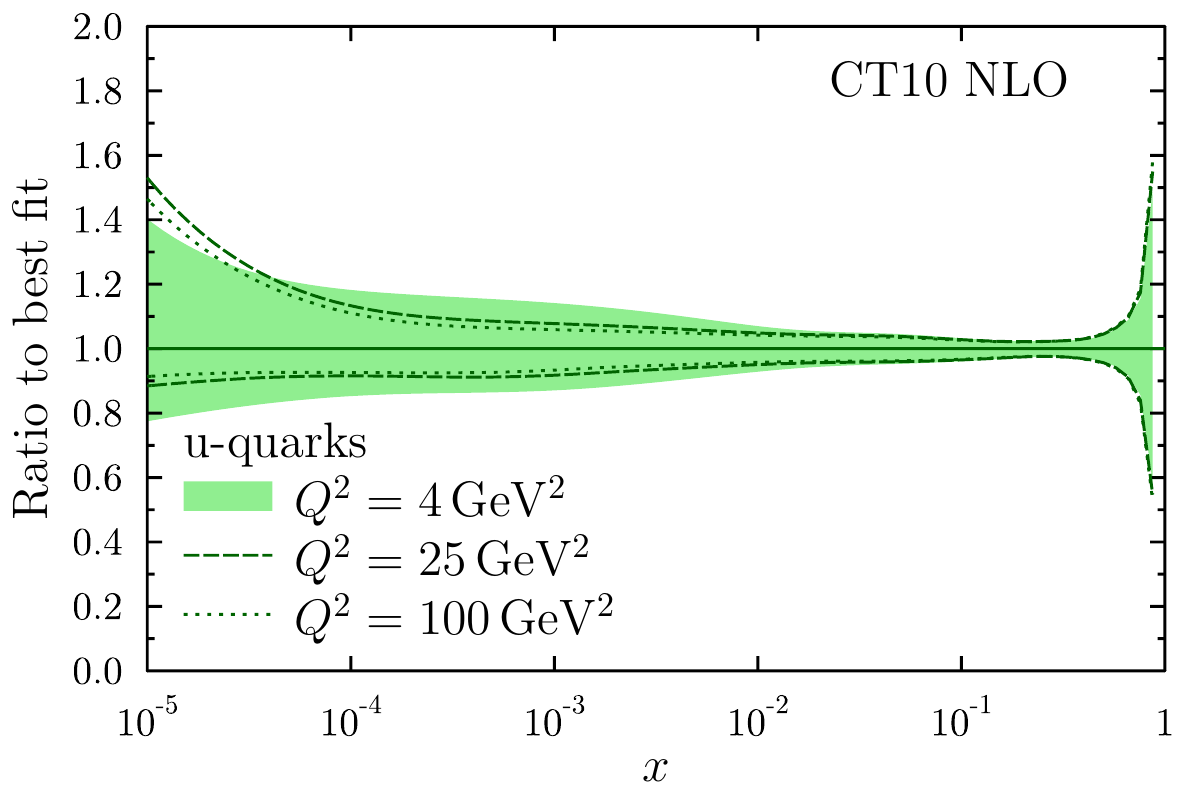}
\includegraphics[width=0.49\textwidth]{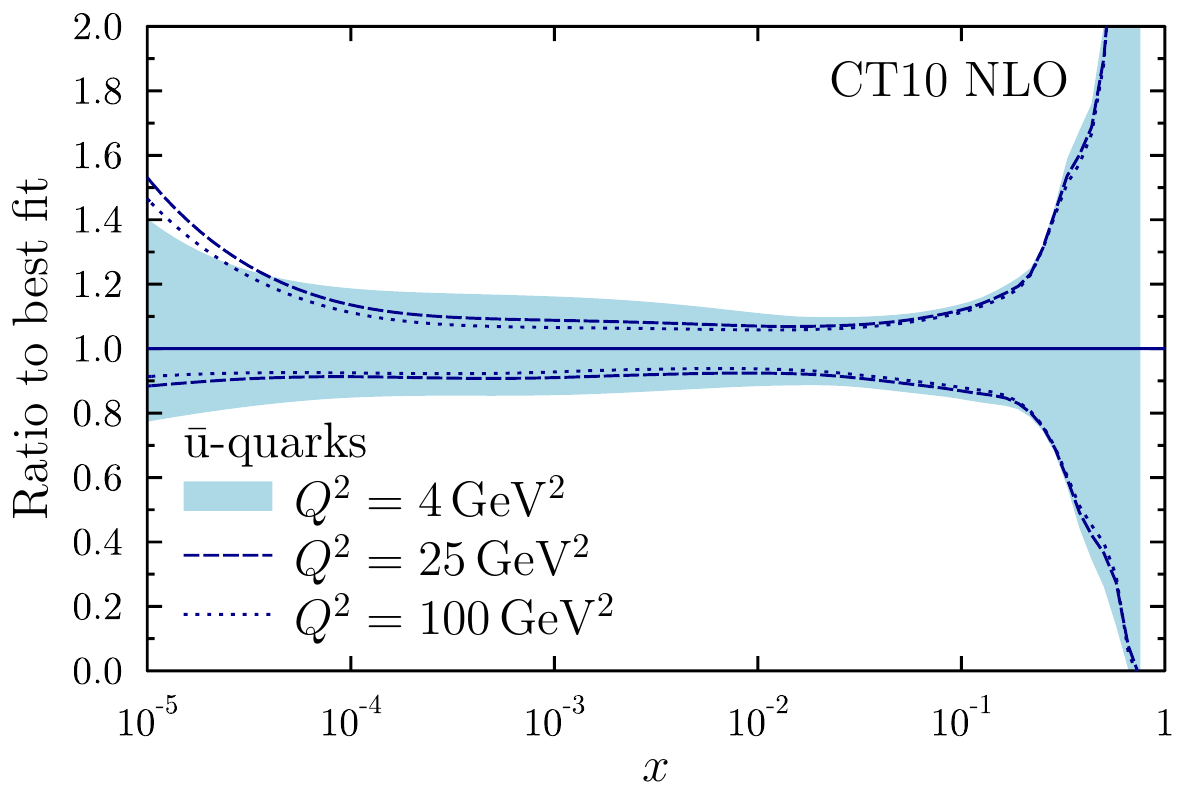}\\
\includegraphics[width=0.49\textwidth]{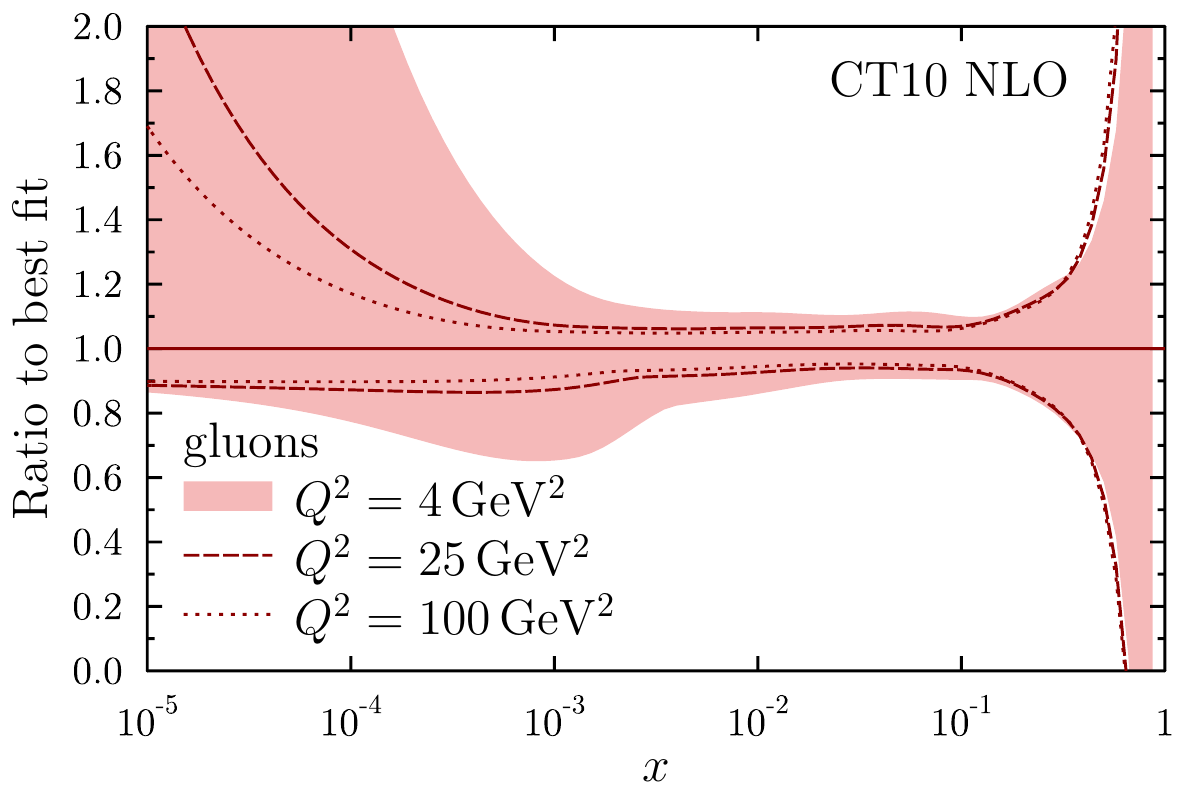}
\caption{The relative uncertainties of the u-quark (top left), $\rm \bar{u}$-quark (top right) and gluon (bottom) PDFs from the CT10 NLO set at scales $Q^2=4\,\mathrm{GeV^2}$ (solid), $25\,\mathrm{GeV^2}$ (dashed) and $100\,\mathrm{GeV^2}$ (dotted).}
\label{fig:ct10_err}
\end{figure}

The CT10 analysis provides also error sets that can be used to quantify how the uncertainties in the PDF analysis propagate to different observables. The relative uncertainties of the gluon, u-quark, and $\bar{\text{u}}$-quark distributions at different scales are shown in figure \ref{fig:ct10_err}. The relative uncertainties of the quark and antiquark distributions are very similar below $x\sim0.01$, of the order 20~\% except at the smallest values of $x$ but at $x>0.1$ the $\bar{\text{u}}$-uncertainties are much larger. This follows from the fact that at these values of $x$, the quark distributions dominate over the antiquark distributions so the quark-related cross sections typically are not very sensitive to the sea quark distributions here. The relative uncertainty of the gluon PDFs according to CT10 is rather large at $x<10^{-4}$ at small scales but due to the DGLAP evolution shrinks towards larger scales. For quarks the small-$x$ uncertainty actually increases slightly towards larger scales which is due to the connection of the quark and gluon DGLAP equations that transfers some amount of gluon uncertainty also to the quarks.

\section{Nuclear modifications of the PDFs}
\label{sec:nPDF}

The measurements in the lepton-nucleus DIS have indicated that there are non-trivial modifications in the nuclear structure functions $F^A_2(x,Q^2)$ relative to the $F_2(x,Q^2)$ in deuteron \cite{Amaudruz:1995tq, Arneodo:1995cs}. In the collinear factorization framework these non-trivial effects are absorbed into the nuclear PDFs $f_i^A(x,Q^2)$, which are assumed to obey the same DGLAP equations as the free nucleon PDFs $f_i(x,Q^2)$ but with a modified initial parametrization. A more trivial nuclear modification follows from the fact that nuclei consist of both protons and neutrons which have different valence quark distributions. Thus, the nPDF for an average nucleon bound to a nucleus with a mass number $A$, $Z$ protons and $N (=A-Z)$ neutrons can be written as
\begin{equation}
f_i^A(x,Q^2) = \frac{Z}{A}f_i^{{\rm p}/A}(x,Q^2) + \frac{N}{A}f_i^{{\rm n}/A}(x,Q^2),
\end{equation}
where the neutron PDFs are usually obtained from the proton PDFs by assuming {\em isospin symmetry}, i.e. that ${\rm {u}^{p}}(x,Q^2) = {\rm d^{n}}(x,Q^2)$ (${\rm \bar{u}^{p}}(x,Q^2) = {\rm \bar{d}^{n}}(x,Q^2)$) and vice versa. Thus, for the observables that have sensitivity also to electroweak coupling, some modifications are expected due to the modified u- and d-quark distributions relative to the protons only -case. This effect is referred to as the \emph{ isospin effect} and is present in observables that are sensitive to large values of $x$ in a nucleus (including deuterium) where the charge distributions are most modified due to the different valence quark distributions.

To quantify the nuclear modifications of the PDFs one can study the ratio between the bound  and free proton PDFs,
\begin{equation}
R_i^A(x,Q^2) = \frac{f_i^{{\rm p}/A}(x,Q^2)}{f_i^{\rm p}(x,Q^2)}.
\label{eq:ra}
\end{equation}
The nuclear PDFs can be obtained through a global analysis by introducing a parametrization for the absolute distributions $f_i^{{\rm p}/A}(x,Q^2)$ at some initial scale $Q_0$ or by using a well established free proton PDF set as a baseline and parametrize only the nuclear modifications $R_i^A(x,Q^2)$ at the initial scale. The latter have been more popular but also the former have been utilized e.g. by the nCTEQ collaboration \cite{Schienbein:2009kk}. As in the free proton case, also the parametrization of the $R_i^A(x,Q^2)$ must be flexible enough to accommodate all the relevant features seen in the data. Typically there are four distinct nuclear effects observed at different regions of $x$: shadowing at $x\lesssim 0.01$, anti-shadowing around $x\sim0.1$, EMC-effect\footnote{EMC stands for European Muon Collaboration, which provided the first experimental evidence for the nuclear modification of the structure functions \cite{Aubert:1983xm}} at $0.3\lesssim x \lesssim 0.7$, and Fermi-motion towards $x\rightarrow 1$. The origins of these effects are not discussed in this thesis but these conventional terms are used to identify the $x$ regions discussed. Currently there are several nuclear PDF sets available, e.g. DSSZ \cite{deFlorian:2011fp}, EPS09 \cite{Eskola:2009uj} and HKN07 \cite{Hirai:2007sx}, to name a few. The current status of the global nPDF analyses is reviewed in Refs.~\cite{Paukkunen:2014nqa, Eskola:2012rg}. In the articles of this thesis we have utilized the EPS09 and its ancestor EKS98 \cite{Eskola:1998df, Eskola:1998iy} nPDFs, the latter being a LO fit and the former including both the LO and NLO nPDFs.

\begin{figure}
\centering
\includegraphics[trim = 40pt 20pt 10pt 10pt, clip, width=0.75\textwidth]{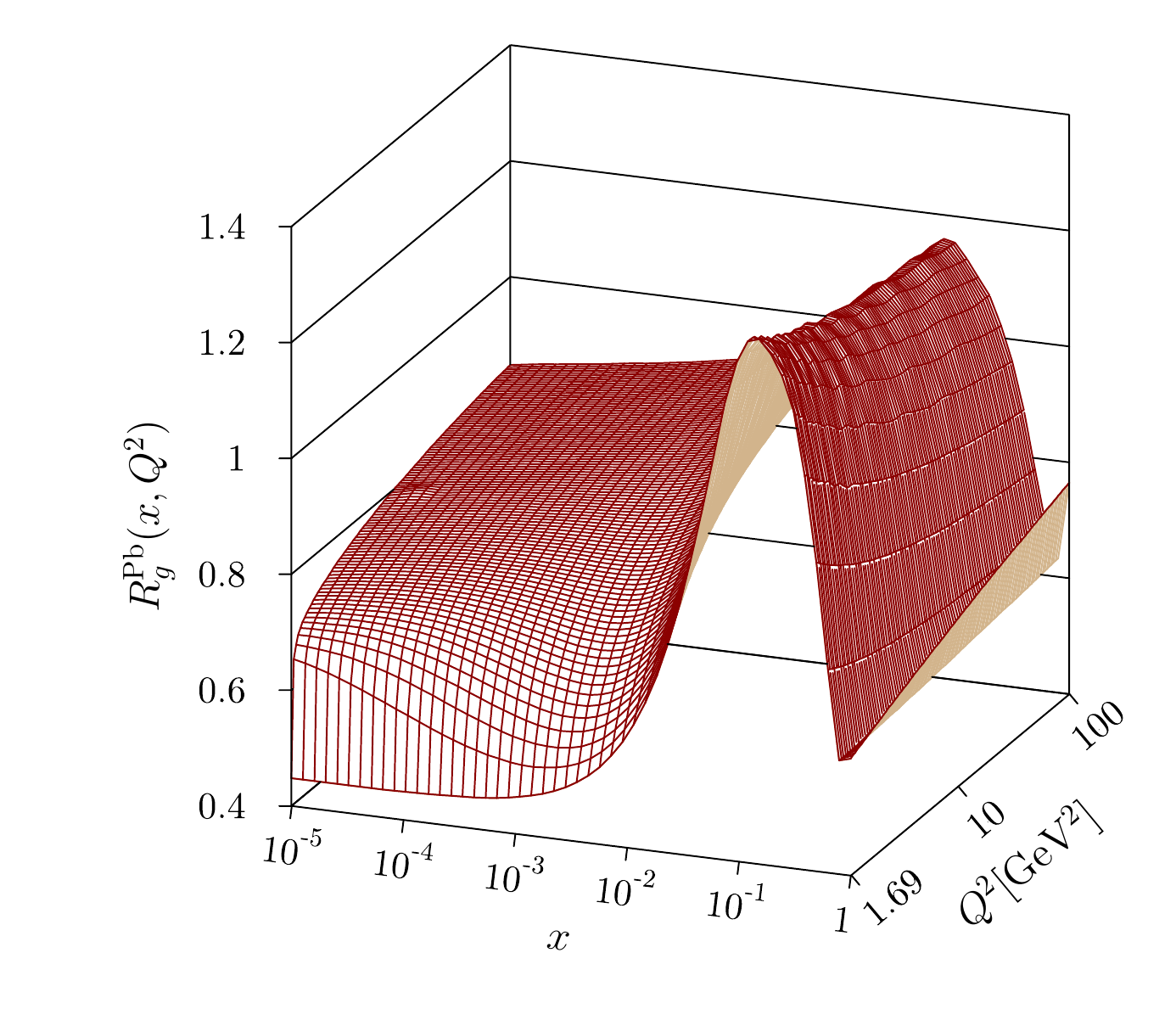}
\caption{The gluon nuclear modification $R_g^{\rm Pb}(x,Q^2)$ for the Pb-nucleus as a function of $x$ and $Q^2$ from EPS09 NLO.}
\label{fig:EPS09NLO_g}
\end{figure}

\begin{figure}
\centering
\includegraphics[trim = 40pt 20pt 10pt 10pt, clip, width=0.75\textwidth]{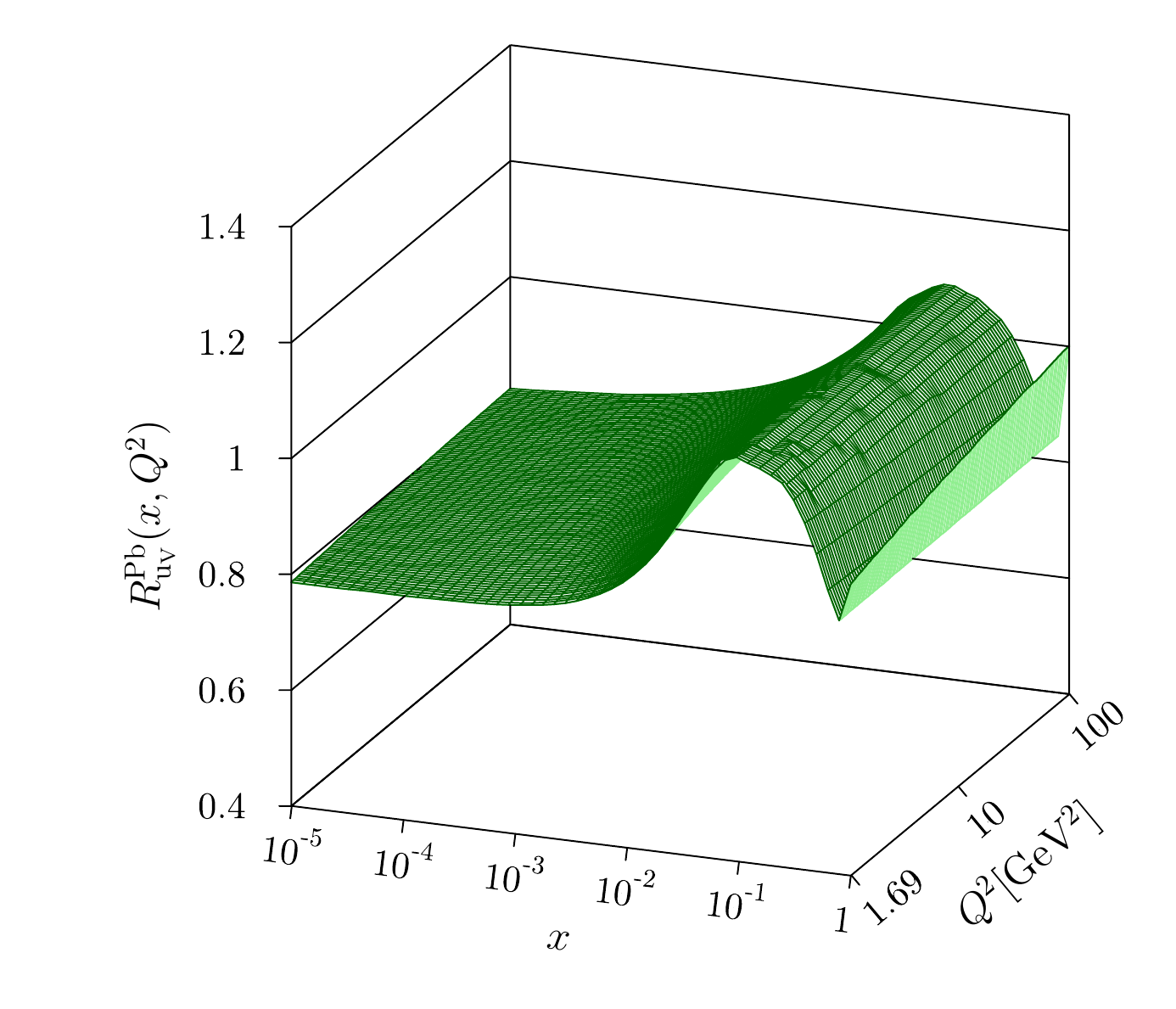}
\caption{Same as figure \ref{fig:EPS09NLO_g} but for $\rm u_V$-quarks}
\label{fig:EPS09NLO_uv}
\end{figure}

\begin{figure}
\centering
\includegraphics[trim = 40pt 20pt 10pt 10pt, clip, width=0.75\textwidth]{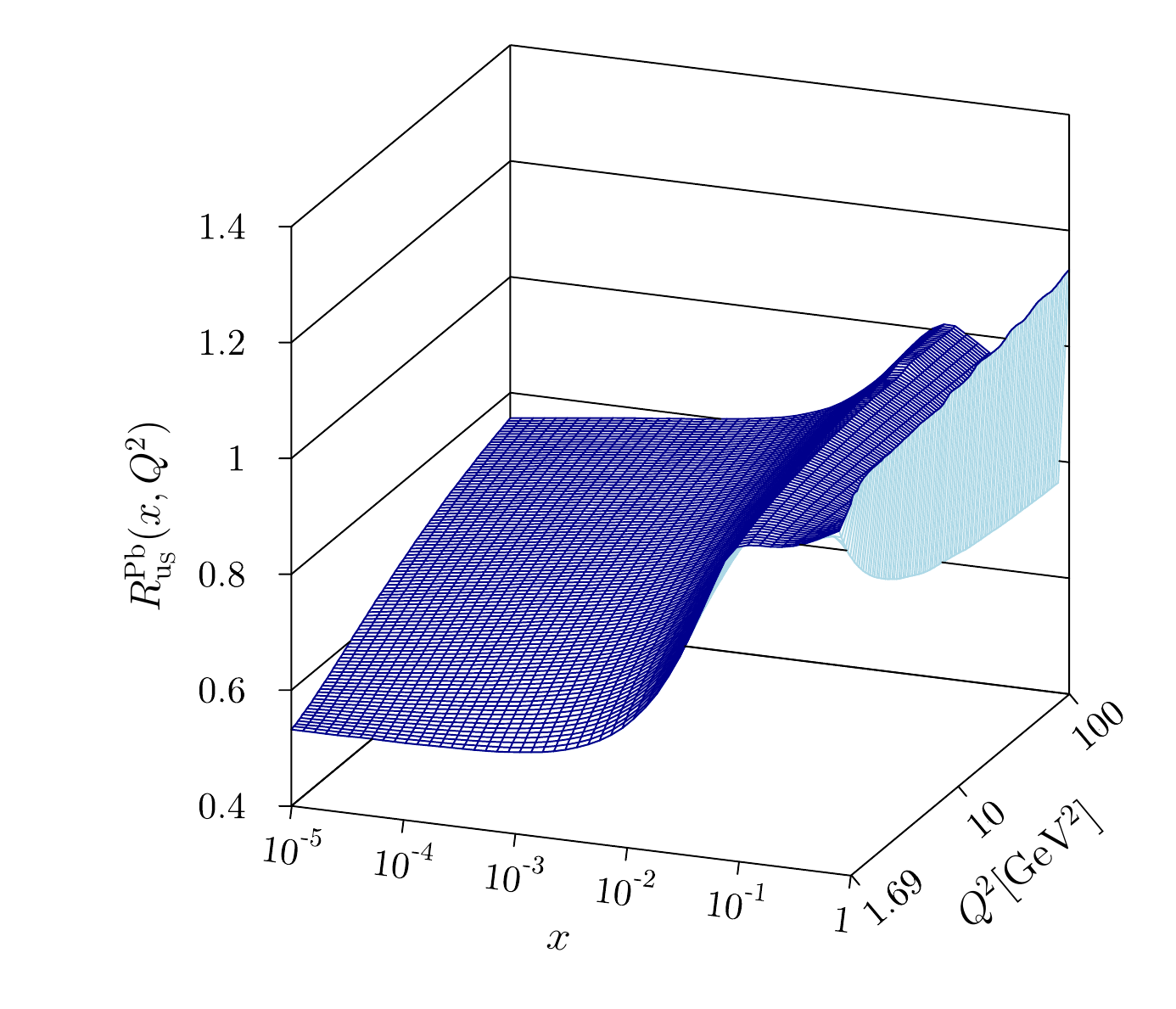}
\caption{Same as figure \ref{fig:EPS09NLO_g} but for $\rm u_S$-quarks}
\label{fig:EPS09NLO_us}
\end{figure}

Figures \ref{fig:EPS09NLO_g}, \ref{fig:EPS09NLO_uv}, and \ref{fig:EPS09NLO_us} show both the $x$ and the scale dependence of the gluon, $\rm u_V$, and $\rm u_S$ nuclear modification, respectively, from the NLO EPS09 fit. The common features are visible in all figures: a suppression at small $x$ due to the shadowing, an enhancement due to the antishadowing, a suppression again at the EMC-region and then a rapid rise due to the Fermi motion towards $x \rightarrow 1$, which is partly cut out from the figures for better readability. The scale evolution of the $R_i^A(x,Q^2)$ for valence quarks turns out be rather slow but for small-$x$ gluons a very rapid increase is present close to the initial scale. A very similar behaviour is seen also in the DSSZ nPDFs that have a different baseline PDF set and a different functional form for the initial parametrization. This hints that the rapid scale evolution of $R_g^A(x,Q^2)$ is rather a feature of the DGLAP equations than due to the details in the nPDF fit. For the sea quarks the scale evolution is somewhat stronger than for the valence quarks which follows partly from the connection of the quark and gluon DGLAP equations.

\begin{figure}
\centering
\includegraphics[width=\textwidth]{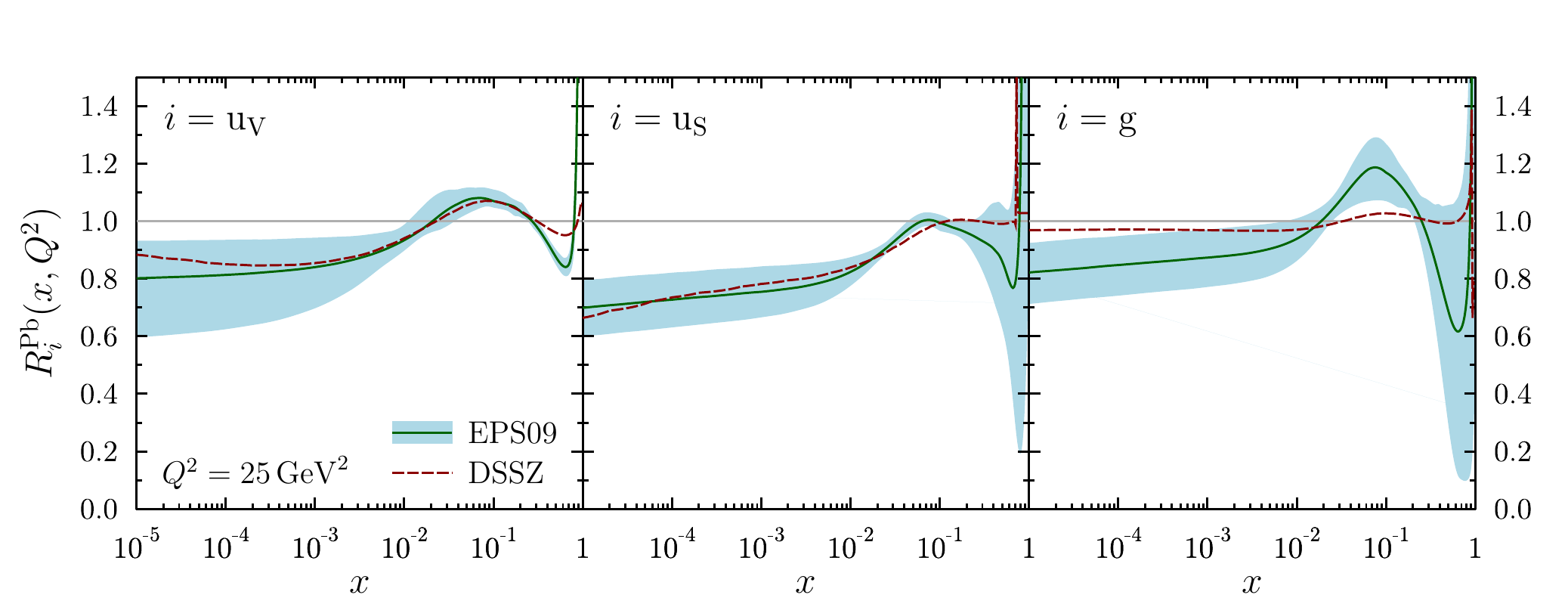}
\caption{The nuclear modification of Pb-nucleus for u-valence (left), u-sea (right), and gluons (right) from EPS09 (solid green) with uncertainties (blue band) and DSSZ (dashed brown) at scale $Q^2=25\,\mathrm{GeV^2}$.}
\label{fig:R_eps_dssz}
\end{figure}

Compared to the free proton PDFs, the amount of the data that can be used to constrain the nPDFs is much more limited. Most of the constraints come from nuclear DIS and Drell-Yan (DY) dilepton production in fixed target p+$A$ collisions which are primarily sensitive to (anti-)quark distributions. Some constraints for gluons for the present fits are provided by RHIC data for pion production in d+Au collisions, and DSSZ and nCTEQ have exploited also neutrino-nucleus DIS data which can be used also for flavor separation. Similarly as the CT10 PDFs for the free protons, the EPS09 (and DSSZ) analysis also provides error sets which can be used to quantify the uncertainty in the fit. These uncertainties are shown in figure \ref{fig:R_eps_dssz} for $\rm u_V$, $\rm u_S$ and gluons for the Pb-nucleus at a scale $Q^2=25\,\mathrm{GeV^2}$. The corresponding nuclear modifications from DSSZ are also plotted for comparison. Clearly the valence and sea quarks are well constrained by the DIS and DY data at $x>0.01$ (except at $x>0.3$ for sea quarks but this is not very relevant for the calculations in this thesis as the valence quarks dominate the considered cross sections in this region). However, the data included in the present fits do not give definite constraints for the gluon nuclear modification. This is reflected by the wide uncertainty band in the whole $x$ region considered and especially at large $x$. Some constraints are provided by the pion data from RHIC but the kinematic reach is limited to the antishadowing region. The lack of antishadowing in the DSSZ analysis was in Ref.~\cite{Eskola:2012rg} found to result from the use of nuclear fragmentation functions \cite{Sassot:2009sh} in the fit.

Currently the most promising source for further nPDF constraints are the p+Pb collisions at the LHC. The first data from these collisions were published already from the short pilot run which was carried out in September 2012. From the nPDF point of view, the most interesting observable was the minimum bias nuclear modification ratio $R^{ch}_{\rm pPb}(p_T,\eta)$ for inclusive charged hadron production at mid-rapidity measured by ALICE \cite{ALICE:2012mj}, which is discussed in section \ref{sec:FF_pA}. Also the data for the same observable from 2013 p+Pb run with increased statistics and a wider span in $p_T$ have recently become available from ALICE \cite{Abelev:2014dsa}, both being consistent with unity at $p_T > 7\,\mathrm{GeV/c}$. However, in the preliminary data from CMS \cite{CMS:2013cka} and ATLAS \cite{ATLAS_RpPb} one can notice an enhancement in charged hadron $R^{ch}_{\rm pPb}$ at $p_T>20\,\mathrm{GeV/c}$ that is not consistent with our NLO pQCD baseline calculation or with the ALICE data\footnote{This is one of the current experimental puzzles.}. However, to obtain constraints for the nPDFs, identified-hadron data (for pions in particular) would be preferable as there are hints of some non-perturbative or higher twist effects for baryon production around $p_T \sim 3 \,\mathrm{GeV/c}$ in the measurements --- these are discussed further in Chapter \ref{chap:FF}. 

In the article [V] of this thesis we performed a more detailed study about which values of $x$ are probed at different rapidities with different observables in the p+Pb collisions at the LHC. We found that to constrain the so far poorly known small-$x$ gluons an optimal observable would be direct photons at forward rapidities. The direct photon production will be discussed in detail in Chapter \ref{chap:direct_photon}. To constrain the nPDFs at larger values of $x$ one could use also the dijet measurements, as discussed in Ref.~\cite{Eskola:2013aya}. Recently published preliminary CMS data \cite{Chatrchyan:2014hqa} for the dijet rapidity distributions seem to favor the EPS09-style antishadowing for gluons that is not present e.g. in the DSSZ nPDFs.

\chapter{Parton-to-hadron fragmentation functions}
\label{chap:FF}

The high-energy hadrons that can be measured in the experiments are formed from the energetic partons created in the hard partonic scattering. As the transition from partons to hadrons happens at rather small virtuality scales it cannot be treated using pQCD but, similarly as for PDFs in the Chapter \ref{chap:PDF}, non-perturbative distributions can be used to describe this hadronization process in an inclusive way. These distributions are referred to as FFs, $D_k^h(z,Q_F^2)$, which give the probability density to generate a hadron $h$ from the parton $k$ with a momentum fraction $z$ of the parent parton. Analogously to the PDFs, the integral $\int^1_0\mathrm{d}z\,D_k^h(z,Q_F^2)$ gives the average number of hadrons $h$ formed from the parton $k$. The FFs can be obtained by a global analysis applying the QCD evolution equations. The general form of the evolution equations is the same for PDFs and FFs, but the NLO splitting functions \cite{Binnewies:1997gz} differ from the PDF case, resulting in a more singular behaviour at small values of $z$ \cite{deFlorian:1997zj}. This makes the FFs less reliable at very small $z$ but as discussed in [IV], the contribution from $z<0.1$ to single-inclusive hadron cross-sections is negligible.

In general, the single-inclusive hadron production in three different types of collisions have been used to constrain the FFs in the present fits:
\begin{itemize}
\item Electron-positron annihilations (\epem)
\item Semi-inclusive deep inelastic scatterings (SIDIS)
\item Proton-(anti-)proton collisions (p+p($\rm\bar{p}$))
\end{itemize}
The data from \epem collisions are very precise and their interpretation is unambiguous as the kinematics are entirely fixed by the measurement and as there are no parton distributions for the colliding particles involved. However, as the LO process produces only quark-antiquark pairs and the gluon radiation is an NLO correction, the gluon-to-hadron FFs are not stringently constrained by these data. The inclusion of the SIDIS data can help to separate the quark and antiquark FFs but as in the \epem collisions, the gluon production is again an NLO effect. Thus only the purely hadronic collisions, like the p+p($\rm\bar{p}$) collisions, are more directly sensitive to the gluonic FFs.

\section[Inclusive charged hadron production in hadronic collisions]{Inclusive charged hadron production in\\ hadronic collisions}

The cross section of single inclusive hadron production in hadronic collisions is calculated as a convolution integral between the partonic spectra and the fragmentation functions,
\begin{equation}
\frac{\mathrm{d} \sigma^{h+X}}{\mathrm{d}p_T \mathrm{d} \eta}(\mu^2,Q^2,Q_F^2) = \sum_{k,X'} \int \frac{\mathrm{d}z}{z}  D_k^h(z,Q_F^2) \left.\frac{\mathrm{d} \sigma^{k+X'}}{\mathrm{d}q_T \mathrm{d} \eta}(\mu^2,Q^2,Q_F^2)\right|_{q=p/z},
\end{equation}
where the partonic spectrum $\frac{\mathrm{d} \sigma^{k+X'}}{\mathrm{d}q_T \mathrm{d} \eta}$ includes the convolution between the PDFs and the partonic pQCD pieces as pointed out in equation (\ref{eq:coll_fact}), and scales are the renormalization scale ($\mu$), the factorization scale ($Q$) and the fragmentation scale ($Q_F$). Thus, compared to \epem collisions, there are two complications: First, due to the convolution with the PDFs there is no direct access to the partonic kinematics but only $z$-integrated observables can be studied. This could be improved by studying hadron-jet momentum correlations in p+p collisions as proposed in Ref.~\cite{Arleo:2013tya} but so far there are no data available. Second, there are three, in principle independent, hard scales related to the process that are not specified by the theory anymore. This results in a theoretical uncertainty that can be studied by varying the different scales in the calculation. It should be noted that this scale ambiguity is not a physical phenomenon but arises purely from the truncation of the perturbative series to a given order. In hadronic collisions there are potentially also some non-perturbative or higher twist effects due to multiparticle interactions or the underlying event which are not present in \epem annihilations. Thus one should consider only high enough energy scales where these effects are negligible.

\begin{figure}[htb]
\centering
\includegraphics[width=0.95\textwidth]{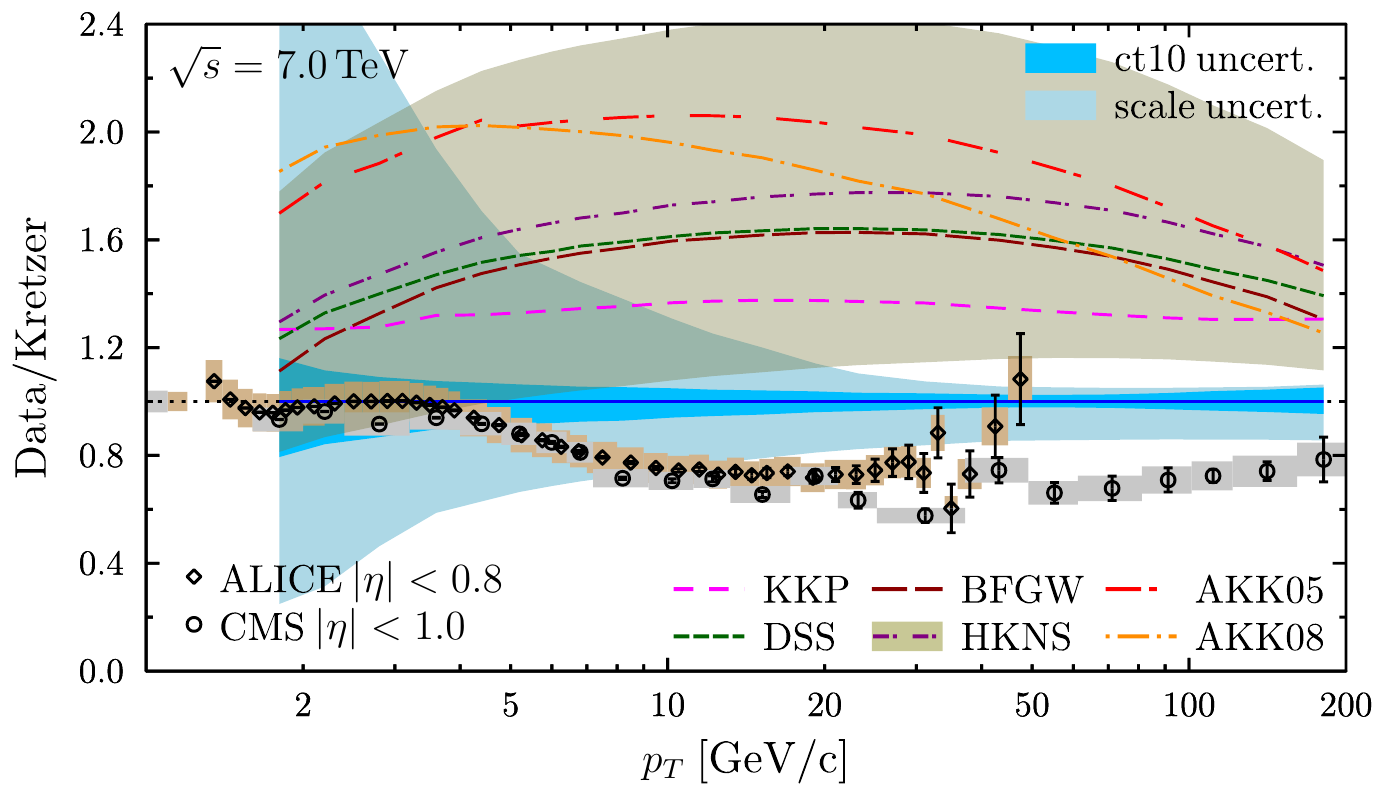}
\caption{The data-to-theory ratio for charged hadron production in p+p collisions at $\sqrt{s_{NN}}=7.0\,\mathrm{TeV}$. The data points are from CMS \cite{Chatrchyan:2011av} (circles) and ALICE \cite{Abelev:2013ala} (diamonds), and the baseline calculation is done with the Kretzer FFs and using \texttt{INCNLO} code (described in section \ref{subsec:incnlo}). Also the ratios between the calculations with different FF sets and the Kretzer set are presented, and in the case of HKNS also the uncertainty is shown (gray band). For the PDFs the CT10 set is used and its uncertainties are shown (dark blue band). The scale uncertainty (light blue band) is calculated as an envelope, explained in [IV]. Figure from [IV].}
\label{fig:pp7000_ch}
\end{figure}

At the moment there are only two charged hadron FF analyses that have exploited the data from hadronic collisions, DSS \cite{deFlorian:2007aj, deFlorian:2007hc} and AKK08 \cite{Albino:2008fy}, the former being the only one to include also SIDIS data. The earlier analyses HKNS \cite{Hirai:2007cx}, AKK05 \cite{Albino:2005me}, BFGW \cite{Bourhis:2000gs}, KKP \cite{Kniehl:2000fe} and Kretzer \cite{Kretzer:2000yf} are based purely on \epem data. This leads to large differences in gluon-to-hadron FFs between different FF sets as we discussed in the article [IV] of this thesis where we studied the charged hadron production in p+p collisions at the LHC energies. As the gluons dominate the parton spectra up to large values of $p_T$ ($\sim 20\,\mathrm{GeV/c}$ at $\sqrt{s_{NN}}=900\,\mathrm{GeV}$ and $\sim 100\,\mathrm{GeV/c}$ at $\sqrt{s_{NN}}=7.0\,\mathrm{TeV}$, see figure 3 in [IV]), also the calculated cross section is rather sensitive to the choice of the FF set. In the study [IV] we found that there are up to a factor of two differences between the inclusive hadron NLO cross sections calculated with different sets, and that for most of the FF sets the calculation clearly overshoots the experimental data at several center-of-mass (cms)-energies \cite{Chatrchyan:2011av, CMS:2012aa, Abelev:2013ala, Aaltonen:2009ne, Albajar:1989an}. As an example, the comparison between the data and the calculation with different FFs for the charged hadron production in p+p at $\sqrt{s_{NN}}=7.0\,\mathrm{TeV}$ is shown in figure \ref{fig:pp7000_ch}. Surprisingly, the calculations with the FFs that include also data from hadronic collisions are about a factor of two above the data. In the article [IV] we concluded that this follows from fitting to low $\sqrt{s}$, low $p_T$ p+p data\footnote{In e.g. the DSS analysis only 11 of 228 data points of the available charged hadron data from p+p($\rm\bar{p}$) collisions were above $p_T>10\,\mathrm{GeV/c}$} where the scale variations yield a wide uncertainty band which implies that the perturbative expansion is not yet well under control. To further elaborate this observation, the identified hadron production is discussed in the next section.

\subsection{Identified hadrons}
\label{subsec:inde_hadr}

\begin{figure}
\centering
\includegraphics[width=\textwidth]{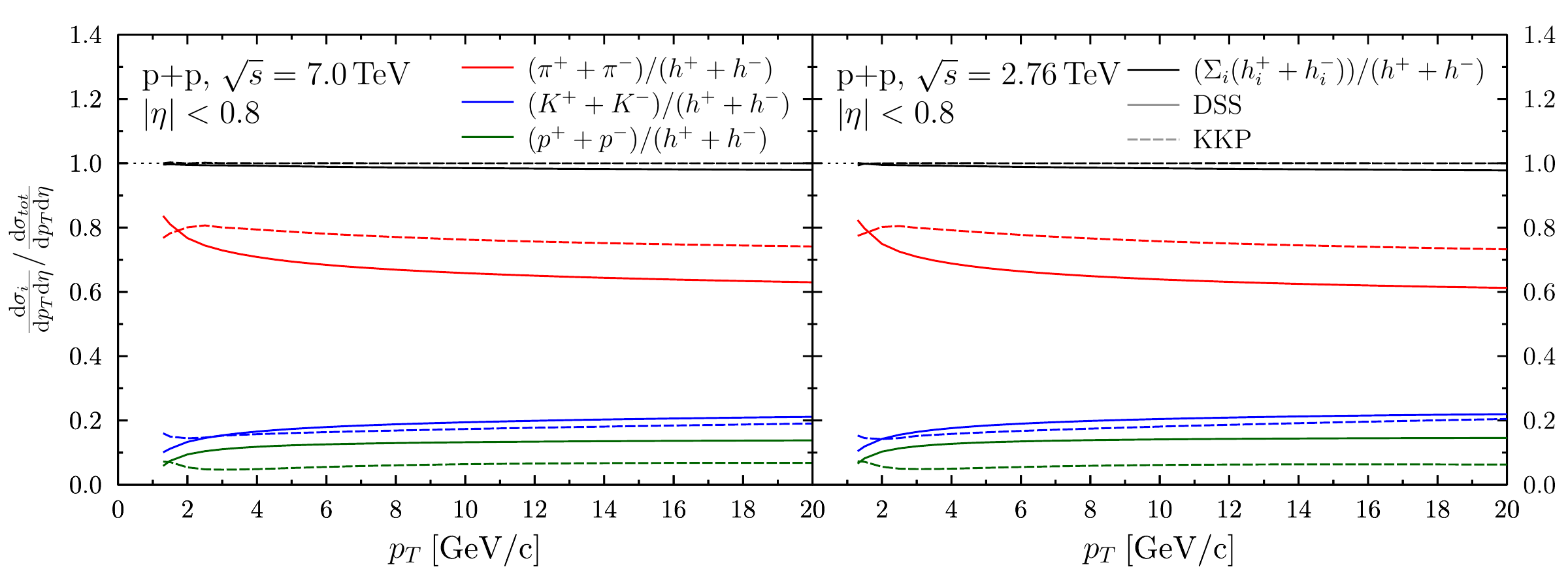}
\caption{The cross section ratio between identified charged hadrons and total charged hadrons using the DSS (solid) and KKP (dashed) FFs in p+p collisions at $\sqrt{s}=7.0\,\mathrm{TeV}$ (left) and $\sqrt{s}=2.76\,\mathrm{TeV}$ (right) at $|\eta|<0.8$.}
\label{fig:R_id2ch}
\end{figure}

From the experimental point of view, the sum of all charged hadrons is much simpler to measure than the identified hadrons as no challenging high-$p_T$ particle identification is required. However, the identified hadrons provide more detailed information of the particle production in hadronic collisions and can also help to interpret the total charged hadron data. The usual assumption is that the charged hadrons consist of charged pions, kaons and (anti-)protons. The DSS analysis, however, contains the FFs for charged pions, kaons, (anti-)protons, and also for the total charged hadrons separately. The resulting FFs support the canonical assumption as the residual charged hadron component after adding the pions, kaons and protons together gives only a few percent contribution to the total charged hadron cross section. This can be observed from figure \ref{fig:R_id2ch} which shows the ratio of cross sections between individual hadron species and the total charged hadrons with two different collision energies, $\sqrt{s}=7.0\,\mathrm{TeV}$ and $\sqrt{s}=2.76\,\mathrm{TeV}$ at $|\eta|<0.8$. The ratios with the KKP FFs are also shown for comparison. The cross sections are here calculated at NLO using the \texttt{INCNLO}-program \cite{Aversa:1988vb, Aurenche:1999nz}, discussed in more detail in section \ref{subsec:incnlo}, with the CT10 PDFs \cite{Lai:2010vv} and fixing all scales to the hadron $p_T$.
According to the calculations, roughly $70\,\%$ of the charged hadrons are pions, $20\,\%$ kaons, and $10\,\%$ (anti-)protons at the LHC energies.

\begin{figure}
\centering
\includegraphics[width=\textwidth]{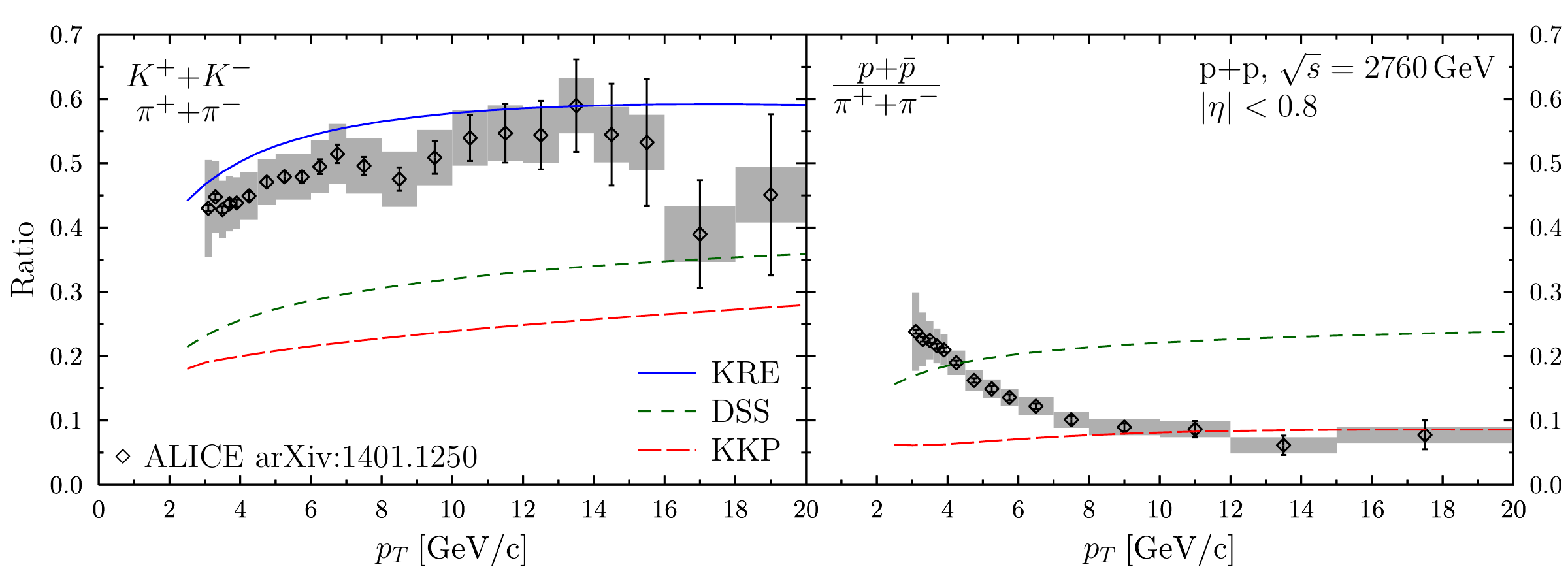}
\caption{The kaon-to-pion (left) and proton-to-pion ratio (right) in p+p collisions at $\sqrt{s}=2.76\,\mathrm{GeV}$. The data are from ALICE \cite{Abelev:2014laa} and the NLO calculations are done with the CT10 PDFs and three different FFs: Kretzer (solid blue), DSS (short-dashed green) and KKP (long-dashed red). The kretzer analysis does not provide parton-to-proton FFs, so the proton-to pion ratio is shown only for the DSS and KKP FFs.}
\label{fig:R_pi2p}
\end{figure}

Experimentally the relative contributions from different hadron species are usually studied using charged pions as a baseline. Figure \ref{fig:R_pi2p} shows a comparison of kaon-to-pion and proton-to-pion -ratios between the ALICE data \cite{Abelev:2014laa} and NLO calculation with the KKP, Kretzer, and DSS FFs at $\sqrt{s}=2.76\,\mathrm{TeV}$. There are large differences in the kaon-to-pion ratio between the different FF sets but all of them can qualitatively reproduce the slow increase with increasing $p_T$ seen in the data. However, the $p_T$ dependence of the proton-to-pion ratio is very different in the data and calculation: the measured ratio has a clear enhancement at $p_T\sim3\,\mathrm{GeV/c}$ and a clear decrease until it flattens out around $p_T=8\,\mathrm{GeV/c}$. The flat behavior at higher values of $p_T$ is well reproduced by the calculations, with the KKP FFs even quantitatively. The disagreement between the pQCD calculation and the data hints that at $p_T<8\,\mathrm{GeV/c}$ there is some non-perturbative or higher-twist component in the proton production, which supports our main conclusion of the article [IV] of this thesis: In future FF analyses one should use only $p_T\gtrsim10\,\mathrm<{GeV/c}$ data, which should be theoretically under a better control and free from non-perturbative effects. The bump in the measured proton-to-pion ratio at small $p_T$ actually seems very similar to what was observed in the data/NLO ratio in figure \ref{fig:pp7000_ch} and in figure 5 of the article [IV] in this thesis.

\begin{figure}[htb]
\centering
\includegraphics[width=\textwidth]{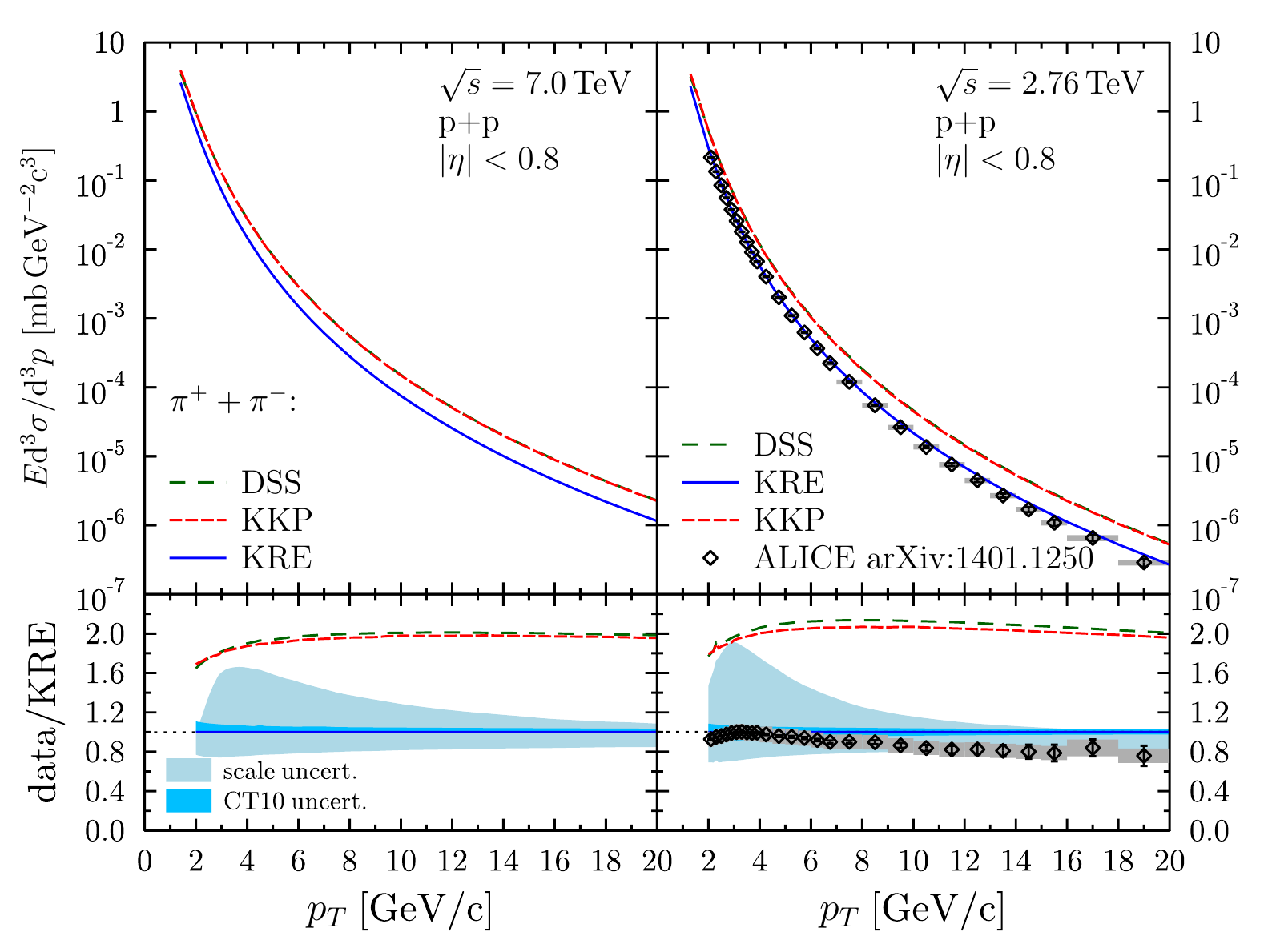}
\caption{The inclusive charged pion cross section in p+p collisions at $|\eta|<0.8$ for $\sqrt{s}=7.0\,\mathrm{TeV}$ (left) and $\sqrt{s}=2.76\,\mathrm{TeV}$ (right). The NLO calculations are done with the DSS, KKP and Kretzer FFs with $\mu=Q=Q_F=p_T$ and CT10 PDFs. The data at $\sqrt{s}=2.76\,\mathrm{TeV}$ at $|\eta|<0.8$ is from ALICE \cite{Abelev:2014laa}. The lower panels show the data/theory ratio using the Kretzer FFs including also the theoretical uncertainties.}
\label{fig:dN_pi_ALICE}
\end{figure}

In order to have a conclusive comparison between different FF sets and the current data for identified case, one should consider also the absolute cross sections. Figure \ref{fig:dN_pi_ALICE} shows the invariant yield of charged pions at $\sqrt{s}=7.0\,\mathrm{TeV}$ and $\sqrt{s}=2.76\,\mathrm{TeV}$ at $|\eta|<0.8$ from our NLO calculations with the DSS, KKP and Kretzer FFs. For $\sqrt{s}=2.76\,\mathrm{TeV}$ the calculations are compared to data from the ALICE measurement \cite{Abelev:2014laa}. To convert the measured invariant yield to invariant cross section the $\sqrt{s}=2.76\,\mathrm{TeV}$ data have been multiplied by $\sigma^{NN}=55.4\,\mathrm{mb}$ as instructed in Ref.~\cite{Abelev:2013ala} for the total charged hadron data. The situation is very similar as observed for the total charged hadron production in the article [IV]: The calculations with the DSS and KKP FFs clearly overshoot the data but with the Kretzer FFs the description is much better. Still the calculated cross section tends to be rather above the data but when considering also theoretical uncertainties, the calculation with Kretzer FFs is nicely consistent with the data. The two sources for the shown theoretical uncertainties are the uncertainty in the CT10 PDFs and the scale ambiguities, which is here quantified by setting $\mu=Q=Q_F=2p_T$ and $=p_T/2$ for the lower and upper limit, respectively. This results in a slightly thinner uncertainty band than what is obtained by varying the scales independently but at higher values of $p_T$ the difference is small. In the data/theory ratio the NLO calculations are integrated over the $p_T$ bins to be consistent with the finite size $p_T$ bins in the measurement. As the bin size is rather small and the cross section is a smooth function, a simple Simpson's rule with three points was found to be sufficient for this purpose.

From the results above and from the article [IV] of this thesis, three main conclusions of the current FF sets can be drawn: 
\begin{itemize}
\item The more recent FF analyses tend to have too hard gluon-to-hadron FFs (see figure 1 of [IV]), which results in too hard hadron spectra for both the identified and unidentified hadrons at the LHC. 
\item The pQCD calculations do not give even a qualitative description for the measured proton-to-pion ratio at $p_T<10\,\mathrm{GeV/c}$, which hints to some non-perturbative or higher twist effects in baryon production at low $p_T$. 
\item Using the older Kretzer FFs, the identified and also unidentified charged hadron spectra can be described fairly well with NLO calculations. Rather than concluding that this is ``the correct'' FF set, this should be taken as an indication that it should be possible to obtain a FF set that can describe simultaneously the very clean and accurate data from \epem collisions and also the high-$\sqrt{s}$ and high-$p_T$ data from hadronic collisions. 
\end{itemize}
The only conclusive way to confirm this finding requires a global QCD reanalysis of the FFs using the \epem (and also SIDIS) data together with the new LHC data for inclusive charged hadron production, including a lower cut for the $p_T$ of the produced hadrons.

\section{Proton-nucleus collisions}
\label{sec:FF_pA}

\begin{figure}[htb]
\centering
\includegraphics[width=0.8\textwidth]{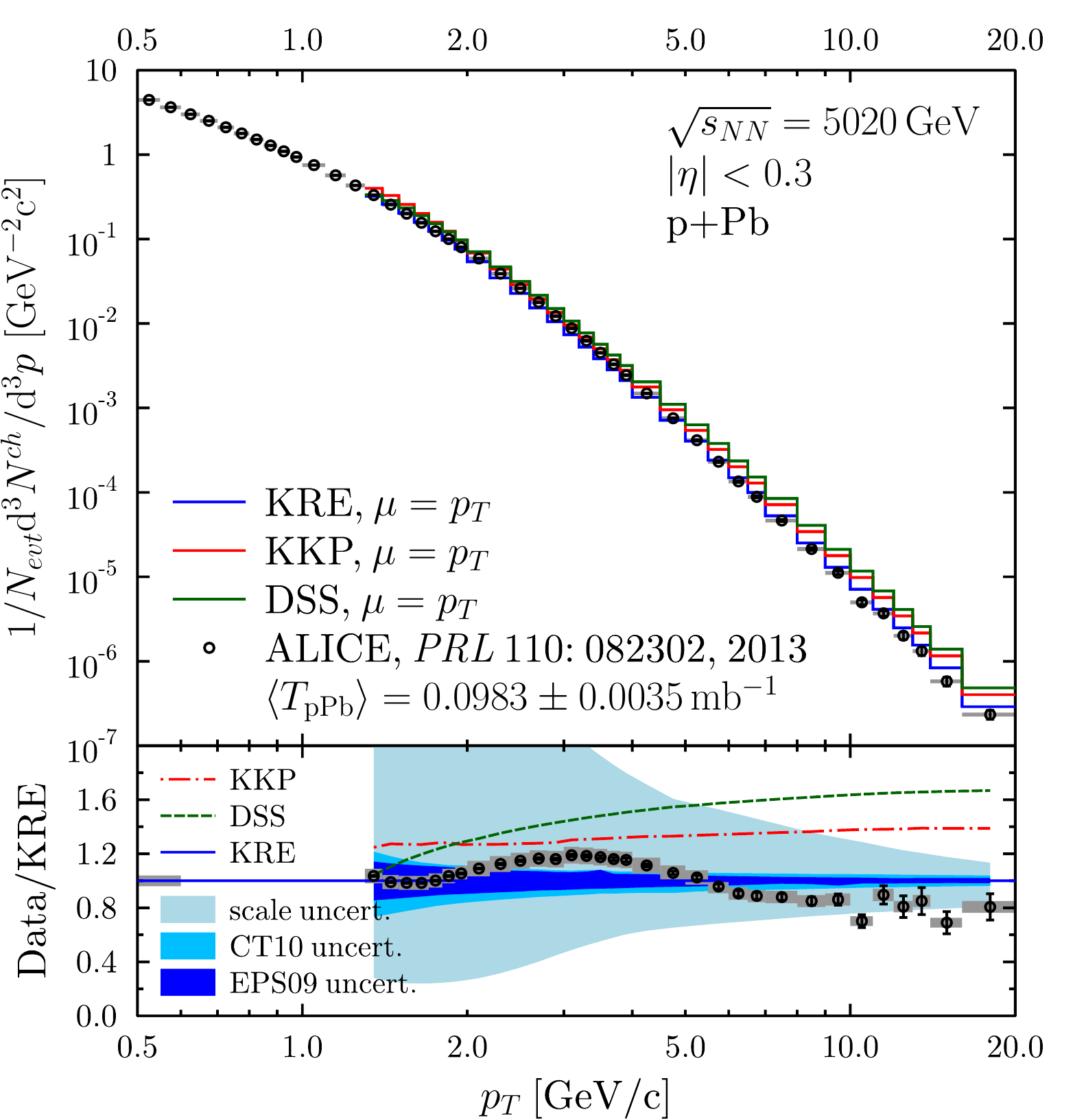}
\caption{Upper panel: The inclusive charged hadron cross section in p+Pb collisions at $\sqrt{s}=5.0\,\mathrm{TeV}$ and $|\eta|<0.3$. The NLO calculations are done with the DSS, KKP and Kretzer FFs with $\mu=Q=Q_F=p_T$ and CT10 PDFs with the EPS09 nuclear modifications. The data is from ALICE \cite{ALICE:2012mj}. Lower panel: The data/theory ratio using the Kretzer FFs and including also the theoretical uncertainties.}
\label{fig:dN_ch_ALICE}
\end{figure}

As discussed in the Chapter \ref{chap:PDF}, the inclusive charged hadron production measured in p+Pb collisions at the LHC could be used to constrain the nuclear modifications of the PDFs. Before including the measured nuclear modification factor $R_{\rm pPb}$ data into an nPDF analysis it should be ensured that the measured absolute $p_T$ spectra in p+Pb collisions are consistent with the pQCD framework. This can be done by calculating the invariant charged hadron cross section and comparing the calculation to the measured yield. Such a comparison with the ALICE data \cite{ALICE:2012mj} is shown in figure \ref{fig:dN_ch_ALICE}. To compare with the measured invariant yield, the calculated cross sections are multiplied by the average nuclear thickness function $\langle T_{\rm pPb} \rangle = 0.0983\,\mathrm{mb}^{-1}$ given by the experiment. The calculational framework is the same as for the pions above, but now for the lead nucleus the EPS09 nPDFs have been used and also their uncertainties are shown. The scale uncertainty band is calculated as in the article [IV].
The conclusions from the figure are very similar as for the p+p collisions studied in the article [IV]: The shape of the spectra is well reproduced with the NLO calculation with Kretzer FFs at $p_T>8\,\mathrm{GeV/c}$ but at smaller $p_T$ a bump in the data/theory ratio can be observed. Remarkably, the bump is again at the very same $p_T$ values as the observed excess in the proton-to-pion ratio in figure \ref{fig:R_pi2p}, supporting our conclusion of a non-perturbative baryonic contribution to the total charged hadron spectra around $p_T\sim3\,\mathrm{GeV/c}$. Also here we notice large differences between the calculations with the different FFs.

\begin{figure}[tbh]
\centering
\includegraphics[width=0.8\textwidth]{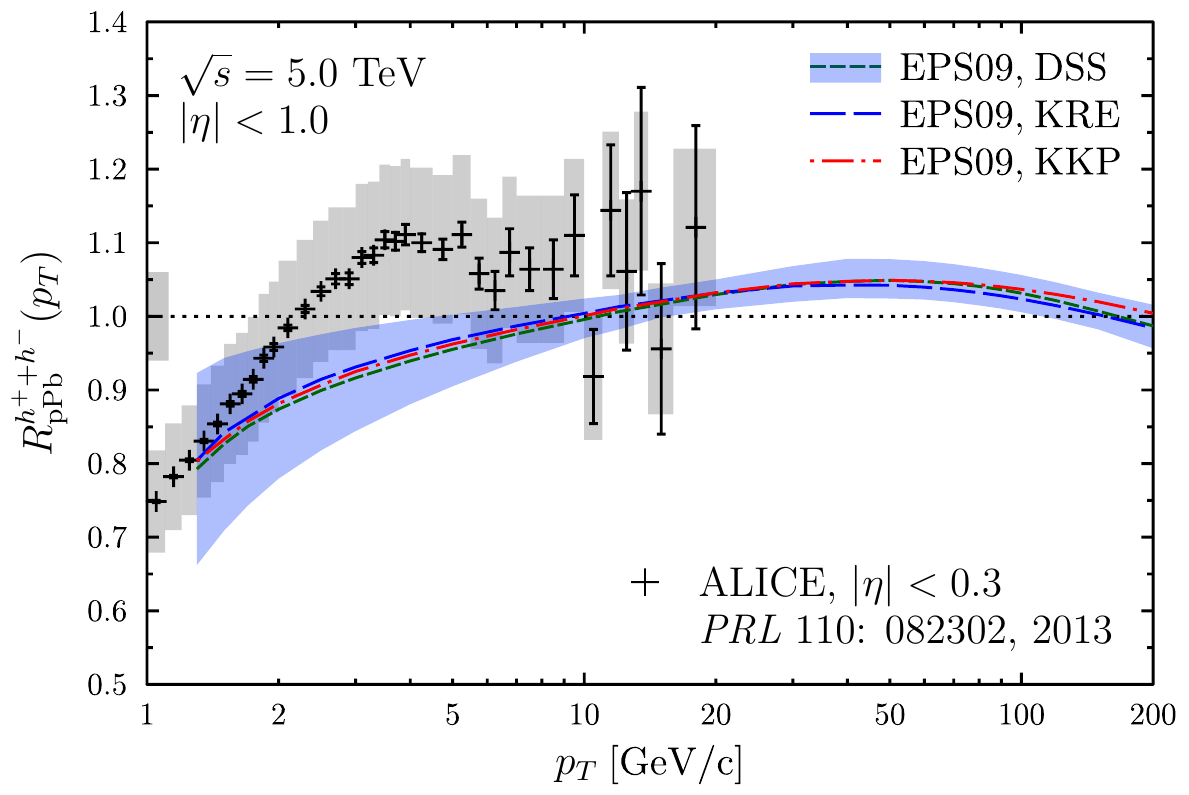}
\caption{The nuclear modification ratio for the inclusive charged hadron production in p+Pb collisions at $\sqrt{s_{NN}}=5.0\,\mathrm{TeV}$ with the EPS09 nPDFs and three parton-to-hadron FFs. The EPS09 uncertainty band is calculated with the DSS FFs and the data is from ALICE \cite{ALICE:2012mj}. The gray box on the left shows the additional $6\,\%$ overall normalization uncertainty of the measurement.}
\label{fig:R_pPb_ALICE}
\end{figure}

The minimum bias nuclear modification ratio for a hard process $k$ is defined as
\begin{equation}
R_{AB}^k(p_T,\eta) = \frac{1}{AB}\frac{\mathrm{d} \sigma_{AB}^k}{\mathrm{d} p_T \mathrm{d} \eta}\Big/ \frac{\mathrm{d} \sigma_{\rm pp}^k}{\mathrm{d} p_T \mathrm{d} \eta},
\label{eq:R_AB}
\end{equation}
where $\mathrm{d} \sigma_{AB}^k/\mathrm{d} p_T \mathrm{d} \eta$ is the cross section for the given process in an $A$+$B$ collision and $\mathrm{d} \sigma_{\rm pp}^k/\mathrm{d} p_T \mathrm{d} \eta$ the corresponding cross section in proton-proton collision. Figure \ref{fig:R_pPb_ALICE} shows a comparison of $R_{\rm pPb}^{ch}(p_T,\eta)$ between the ALICE data \cite{ALICE:2012mj} and the NLO calculation with the EPS09 nPDFs and different FFs for the charged hadron production in p+Pb collisions. The NLO calculation framework is otherwise the same as in the article [I] of this thesis but the pion FFs have been here replaced with the charged hadron FFs. The large differences between the results with different FFs in figure \ref{fig:dN_ch_ALICE} cancel out very efficiently in the ratio $R_{\rm pPb}$. The same holds also to the scale variations, which makes the nuclear modification ratio a conveniently robust observable, even though there are some theoretical uncertainties in the absolute cross section. 

The charged hadron $R^{ch}_{\rm pPb}(p_T,\eta)$ from ALICE show a $\sim 10\,\%$ enhancement around $\sim 3 \,\mathrm{GeV/c}$ where some non-perturbative effects were seen in the proton-to-pion ratio shown in figure \ref{fig:R_pi2p}. However, the enhancement is missing from the preliminary ALICE data for the charged pion $R^{\pi}_{\rm pPb}(p_T,\eta)$ shown recently in the ``Quark Matter 2014'' -conference \cite{ALICE_RpPb_pi} which agrees very nicely with our prediction in [I]. However, the preliminary proton $R^{p}_{\rm pPb}(p_T,\eta)$ by ALICE, shown also in \cite{ALICE_RpPb_pi}, features an even larger enhancement around the same $p_T$ region which confirms that, indeed, the observed enhancement in the charged hadron $R^{ch}_{\rm pPb}(p_T,\eta)$ is caused by the protons. Similar behaviour is observed also in the identified hadron $R_{\rm dAu}$ measured by PHENIX \cite{Adare:2013esx} and the conclusion is supported also by the ALICE proton-to-pion ratio in p+Pb collisions \cite{Abelev:2013haa} which is enhanced with respect to the p+p collisions around $\sim 3\,\mathrm{GeV/c}$ especially in high-multiplicity events. Thus, for the nPDF studies and for testing the universality of the factorization theorem, a nuclear modification factor for identified mesons would be preferred over the sum of all charged hadrons.

\chapter{Direct photon production in hadronic collisions}
\label{chap:direct_photon}

There is one more piece in the equation (\ref{eq:coll_fact}) that is not yet discussed in detail: the hard partonic piece $\mathrm{d}\hat{\sigma}^{ij\rightarrow k+X}$. In LO this corresponds to a partonic cross section but in NLO the interpretation of this term describing the partonic interactions is not so straightforward anymore. In this Chapter I discuss the partonic interactions taking place in direct photon production. 

There are two reasons why the direct photon production interesting for the nPDF studies. First, the direct photons provide a more direct access to the underlying partonic kinematics than hadrons, as part of the direct photons are formed directly at the hard scattering. Second, as the photons do not interact directly with the strongly interacting medium, the high-$p_T$ photons also in $A$+$A$ collisions could be used to study the modifications of the initial parton distributions. 

\section{Leading order}

\subsection{Prompt component}

\begin{figure}[hptb]
\centering
\includegraphics[width=0.6\textwidth]{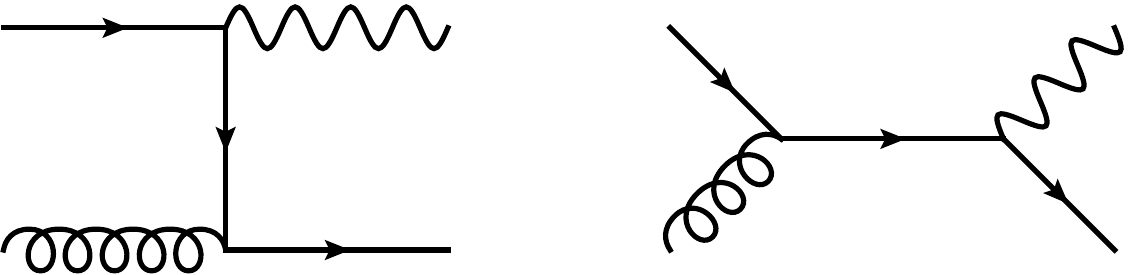}
\caption{Feynman diagrams for QCD Compton scattering.}
\label{fig:QCDCompton}
\end{figure}

\begin{figure}[hptb]
\centering
\includegraphics[width=0.6\textwidth]{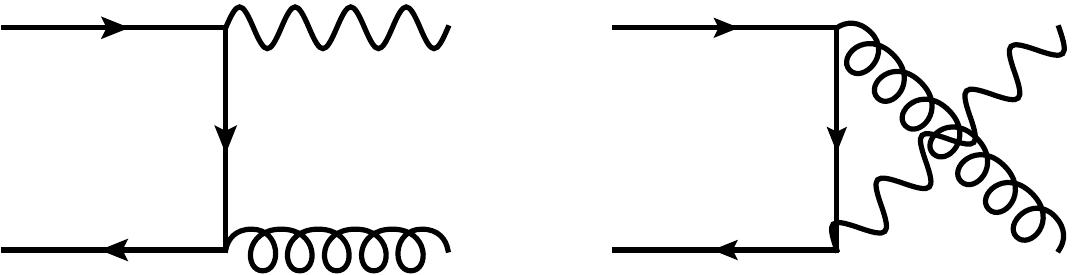}
\caption{Feynman diagrams for quark-antiquark annihilation process.}
\label{fig:LOannihilation}
\end{figure}

At LO in pQCD the photons can be produced directly in the hard scattering via two different processes, the QCD Compton scattering $q+g\rightarrow \gamma + q$ (figure \ref{fig:QCDCompton}) and the quark-antiquark annihilation $q+\bar{q}\rightarrow \gamma + g$ (figure \ref{fig:LOannihilation}), both including two diagrams that need to be taken into account.
Applying the Feynman rules to the diagrams one obtains the following partonic cross sections \cite{Gordon:1993qc}:
\begin{align}
\frac{\mathrm{d}\hat{\sigma}^{q+g\rightarrow \gamma + q}}{\mathrm{d}v} &= \frac{1}{N_C}\,\frac{\pi\alpha \alpha_s e_q^2}{\hat{s}}\, \frac{1+(1-v)^2}{1-v}\label{eq:LOCompton}\\
\frac{\mathrm{d}\hat{\sigma}^{q+\bar{q}\rightarrow \gamma + g}}{\mathrm{d}v} &= \frac{2 C_F}{N_C}\,\frac{\pi\alpha \alpha_s e_q^2}{\hat{s}}\, \frac{v^2+(1-v)^2}{v(1-v)},\label{eq:annihilation}
\end{align}
where $\alpha$ and $\alpha_s$ are the electromagnetic and strong coupling constants, respectively, $e_q$ the electric charge of the quark $q$ and $C_F=(N_C^2-1)/2N_C$, where $N_C$ is the number of colors. The partonic invariant $v$ is defined by $v=1+\hat{t}/\hat{s}$ where $\hat{s}$ and $\hat{t}$ are the usual partonic Mandelstam variables. The cross section in a hadronic collision is then calculated by convoluting the partonic cross sections with the PDFs:
\begin{equation}
\frac{\mathrm{d}\sigma_{\rm prompt}^{h_1+h_2\rightarrow \gamma + X}}{\mathrm{d}^2 p_T\mathrm{d}\eta}(\mu^2,Q^2)=\frac{1}{\pi p_T^4}\sum_{i,j} \int_{v_{min}}^{v_{max}}\mathrm{d}v\,x_1 f_i^{h_1}(x_1,Q^2)\,x_2 f_j^{h_2}(x_2,Q^2)\,v(1-v) \hat{s} \frac{\mathrm{d}\hat{\sigma}^{ij}}{\mathrm{d}v}.
\end{equation}
For the initial-state -crossed process ($\hat{t}\rightarrow \hat{u}$, where $\hat{s} + \hat{t} + \hat{u}=0$) the partonic invariant $v$ converts to $(1-v)$ so the cross section becomes
\begin{align}
&\frac{\mathrm{d}\sigma_{\rm prompt}^{h_1+h_2\rightarrow \gamma + X}}{\mathrm{d}^2 p_T\mathrm{d}\eta}(\mu^2,Q^2) = \frac{\alpha \alpha_s(\mu^2)}{N_C\, p_T^4} \int_{v_{min}}^{v_{max}}\mathrm{d}v \sum_q e_{q}^2 \notag\\
&\bigg\{ x_1 \left[f_{q}^{h_1}(x_1,Q^2)+f_{\bar{q}}^{h_1}(x_1,Q^2)\right]\,x_2 f_g^{h_2}(x_2,Q^2)\, v[1+(1-v)^2] \label{eq:LO_prompt_photon}\\
&\hspace{0pt} + x_1 f_g^{h_1}(x_1,Q^2)\,x_2 \left[f_{q}^{h_2}(x_2,Q^2)+f_{\bar{q}}^{h_2}(x_2,Q^2)\right](1-v)(1+v^2) \notag\\
+&\left[x_1 f_{q}^{h_1}(x_1,Q^2)\,x_2 f_{\bar{q}}^{h_2}(x_2,Q^2) + x_1 f_{\bar{q}}^{h_1}(x_1,Q^2)\,x_2 f_{q}^{h_2}(x_2,Q^2) \right] 2 C_F\left[ v^2+(1-v)^2\right]\bigg\}\notag
\end{align}
where the sum runs over the quark flavors. The momentum fractions can be expressed as $x_1=v_{min}/v$ and $x_2=(1-v_{max})/(1-v)$ where the kinematical limits are given by
\begin{equation}
v_{min}=p_T \mathrm{e}^{\eta}/\sqrt{s} \quad \text{and}\quad  v_{max}=1 - p_T \mathrm{e}^{-\eta}/\sqrt{s}
\end{equation}
in which the $p_T$ is the transverse momentum of the photon, $\eta$ its (pseudo)rapidity and $\sqrt{s}$ the hadronic collision energy. Using the partonic momenta the momentum fractions can be written as
\begin{equation}
x_1=\frac{p_T}{\sqrt{s}}(\mathrm{e}^{\eta} + \mathrm{e}^{\eta_{k}}) \quad \text{and}\quad  x_2=\frac{p_T}{\sqrt{s}}(\mathrm{e}^{-\eta} + \mathrm{e}^{-\eta_{k}}),
\label{eq:x1x2}
\end{equation}
where $\eta_k$ is the (pseudo)rapidity of the outgoing parton which is integrated over in the equation (\ref{eq:LO_prompt_photon}) for single inclusive photon cross section. 

The prompt photon LO cross section depends on two hard mass scales, the renormalization scale $\mu$ via the running of $\alpha_s$ (see equation~(\ref{eq:alpha_s})) and the factorization scale $Q$ via the scale dependent PDFs. Again, these scales are not specified unambiguously by the theory but are usually taken to be proportional to the photon $p_T$.

\subsection{Fragmentation component}
\label{subsec:photon_FFs}

In addition to the prompt photon production discussed above, the direct photons can be created also by fragmentation of energetic partons from the hard scattering into photons. Even though the partonic sub-process is of a higher order in $\alpha_s$, the fragmentation functions of the photon, $D_k^{\gamma}(z,Q_F^2)$, behave roughly as $\alpha/\alpha_s(Q_F^2)$ at large $Q_F$ \cite{Catani:2002ny,Owens:1986mp}, making the fragmentation component effectively of the same order as the prompt photon component.

The fragmentation component of the inclusive direct photon production is calculated in a similar manner as the hadron production, only the parton-to-hadron FFs are replaced with the parton-to-photon FFs. In the collinear factorization framework this can be written as
\begin{align}
&\frac{\mathrm{d}\sigma_{\rm frag.}^{h_1+h_2\rightarrow \gamma + X}}{\mathrm{d}p_T\mathrm{d}\eta}(\mu^2,Q^2,Q_F^2) = \\ 
&\sum_{i,j,k} \int \mathrm{d}x_1\,\mathrm{d}x_2\frac{\mathrm{d}z}{z}x_1 f_{i}^{h_1}(x_1,Q^2)\,x_2 f_j^{h_2}(x_2,Q^2) \left.\frac{\mathrm{d}\hat{\sigma}^{ij\rightarrow k}}{\mathrm{d}q_T\mathrm{d}\eta}\right|_{q_T=p_T/z} D_k^{\gamma}(z,Q_F^2).\notag
\end{align}
In LO there are eight different partonic $2\rightarrow 2$ sub-processes ${\mathrm{d}\hat{\sigma}^{ij\rightarrow k}}/{\mathrm{d}q_T\mathrm{d}\eta}$ contributing to the cross section which are listed e.g. in Ref.~\cite{Owens:1986mp} and a detailed discussion of how to combine these can be found in Ref.~\cite{Eskola:2002kv}. 

The QCD evolution equations for parton-to-photon fragmentation functions differ from the hadronic ones as they contain also an inhomogeneous term describing the splitting $q\rightarrow q \gamma$  \cite{Bourhis:1997yu}:
\begin{equation}
\frac{\partial D_i^{\gamma}}{\partial \mathrm{log}(Q^2)} = 1 \otimes P_{\gamma i} + \frac{\alpha_s(Q^2)}{2\pi} \sum_{j=q,g} P_{ji}\otimes D^{\gamma}_j,
\label{eq:photon_DGLAP}
\end{equation}
where the splitting functions $P_{ij}$ and $P_{\gamma i}$ can again be written as a perturbative expansion in $\alpha_s$ when higher orders are considered, and where the $1\otimes$ refers to the convolution with $\delta(1-z)$. The splitting of a gluon into a photon involves an intermediate quark, making the expansion of the splitting function to start from the order $\alpha_s$. Thus, in LO there is only the quark splitting function for the inhomogeneous term involved:
\begin{equation}
P_{\gamma q}(z) = \frac{\alpha}{2\pi} e_{q}^2 P_{\gamma q}^{LO}(z), \quad \text{where} \quad P_{\gamma q}^{LO}(z) = \frac{1+(1-z)^2}{z}.
\end{equation}
The NLO corrections to $P_{\gamma i}(z)$ can be found in Ref.~\cite{Aurenche:1992yc}.

Due to the presence of the inhomogeneous term in the equation (\ref{eq:photon_DGLAP}) there are two components in the full solution of the $D_{i}^{\gamma}(z,Q_F^2)$: a perturbative component, which in the literature \cite{Bourhis:1997yu,Aurenche:1992yc} is usually referred to as an anomalous component $D_{i,an}^{\gamma}(z,Q_F^2)$, and a non-perturbative component $D_{i,np}^{\gamma}(z,Q_F^2)$. The latter is the general solution of the homogeneous part of evolution equation (i.e. $P_{\gamma i}(z) = 0$), and the former is a solution of the full inhomogeneous set of equations. The perturbative component $D_{i,an}^{\gamma}(z,Q_F^2)$ is calculable in pQCD for large enough scales $Q_F$, and it is defined with an initial condition such as $D_{i,an}^{\gamma}(z,Q_{F,0}^2)=0$.

For the $D_{i,np}^{\gamma}(z,Q_{F,0}^2)$ component one needs to construct an ansatz and fix the parameters by fitting to the experimental data similarly as is done for the PDFs and hadronic FFs. In the BFG analysis \cite{Bourhis:1997yu} the form of the non-perturbative component is determined in the vector dominance model (VDM), in which the photon is described by a superposition of vector mesons ($\rho$, $\omega$, and $\phi$, neglecting the $J/\psi$) and the data constraints come from \epem annihilations at LEP and SLAC. Another study on the photonic fragmentation functions, the GRV analysis \cite{Gluck:1992zx}, used pre-determined LO parton-to-hadron FFs to estimate the effect from the non-perturbative component as inspired by the VDM. Recently there have not been considerable activity to reanalyze the photonic FFs.

In the leading order, the cross section of inclusive photon production in hadronic collisions is then simply the sum of the LO prompt and fragmentation component:
\begin{equation}
\frac{\mathrm{d}\sigma^{h_1+h_2\rightarrow \gamma + X}}{\mathrm{d}p_T\mathrm{d}\eta}(\mu^2,Q^2,Q^2_F) = \frac{\mathrm{d}\sigma_{\rm prompt}^{h_1+h_2\rightarrow \gamma + X}}{\mathrm{d}p_T\mathrm{d}\eta}(\mu^2,Q^2) + \frac{\mathrm{d}\sigma_{\rm frag.}^{h_1+h_2\rightarrow \gamma + X}}{\mathrm{d}p_T\mathrm{d}\eta}(\mu^2,Q^2,Q^2_F),
\label{eq:dsigma_dir_LO}
\end{equation}
where now the prompt component depends only on the renormalization scale $\mu$ and factorization scale $Q$, and the fragmentation component also on the fragmentation scale $Q_F$. In the literature what we here refer to as the ``prompt'' component is often referred to as the ``direct'' component and then the prompt photons are the ``fragmentation''+``direct''. This convention is also used in the article [II] of this thesis. In some earlier works, e.g. in Refs.~\cite{Aurenche:1992yc, Aurenche:1998gv}, the fragmentation component is referred also as ``bremsstrahlung''. However, as the experiments use the term ``direct'' for all but decay photons, we have chosen our naming convention to be consistent with this definition.

\section{NLO corrections}

In the next-to-leading order, $\mathcal{O}(\alpha \alpha_s^2)$, the inclusive direct photons are still the sum of the prompt and fragmentation components, but both components now receive corrections from a large number of new graphs. For the prompt components the corrections arise from 3 different contributions (see Ref.~\cite{Gordon:1993qc} for example graphs): (i) virtual (loop) corrections to $2\rightarrow 2$ process, ({ii}) gluon emissions from quarks, and (iii) photon emissions from quarks.

The NLO cross section for the prompt component can be written as
\begin{equation}
\begin{split}
\frac{\mathrm{d}\sigma_{\rm prompt}^{h_1+h_2\rightarrow \gamma + X}}{\mathrm{d}p_T\mathrm{d}\eta}(\mu^2,Q^2,Q^2_F) = \sum_{i,j} &\int \mathrm{d}x_1\,\mathrm{d}x_2\,x_1 f_{i}^{h_1}(x_1,Q^2)\,x_2 f_j^{h_2}(x_2,Q^2)\\
& \left[\frac{\mathrm{d}\hat{\sigma}^{ij\rightarrow \gamma + X}}{\mathrm{d}p_T\mathrm{d}\eta} + \frac{\alpha_s(\mu^2)}{2 \pi}K_{\rm prompt}^{ij}(\mu^2,Q^2,Q^2_F)\right],
\label{eq:dsigma_prompt_NLO}
\end{split}
\end{equation}
where now $K_{\rm prompt}^{ij}(\mu^2,Q^2,Q^2_F)$ stands for the NLO corrections. These are listed for all processes in the appendix D of Ref.~\cite{Gordon:1993qc}. Similarly for the fragmentation component the NLO cross section can be written as
\begin{equation}
\begin{split}
\frac{\mathrm{d}\sigma_{\rm frag}^{h_1+h_2\rightarrow \gamma + X}}{\mathrm{d}p_T\mathrm{d}\eta}&(\mu^2,Q^2,Q^2_F) = \sum_{i,j,k} \int \mathrm{d}x_1\,\mathrm{d}x_2 \frac{\mathrm{d}z}{z}\,x_1 f_{i}^{h_1}(x_1,Q^2)\,x_2 f_j^{h_2}(x_2,Q^2)\\ & D_k^{\gamma}(z,Q_F^2)
 \left[\left.\frac{\mathrm{d}\hat{\sigma}^{ij\rightarrow k + X}}{\mathrm{d}q_T\mathrm{d}\eta}\right|_{q_T=p_T/z} + \frac{\alpha_s(\mu^2)}{2 \pi}K_{\rm frag.}^{ij\rightarrow k}(\mu^2,Q^2,Q^2_F)\right].
\label{eq:dsigma_frag_NLO}
\end{split}
\end{equation}
The NLO corrections for the partonic scatterings $K_{\rm frag.}^{ij\rightarrow k}(\mu^2,Q^2,Q^2_F)$ using the $\overline{\rm MS}$ renormalization scheme are presented in Ref.~\cite{Aversa:1988vb}.

What deserves some attention is that now also the prompt component depends on the fragmentation scale $Q_F$ due to the NLO corrections. This follows from the singularity in collinear photon emission which needs to be regulated. According to Ref.~\cite{Catani:2002ny} (see also Ref.~\cite{Gordon:1994ut}), using dimensional regularization with $d=4-2\epsilon$ dimensions, the singular term in the collinear approximation (and restricting the integral to a $(\eta,\phi)$ cone of a radius $R$ around the photon) is proportional to
\begin{equation}
\frac{\alpha}{2 \pi}e_q^2(P_{\gamma q}^{LO}(z) - \epsilon z)\left(-\frac{1}{\epsilon}\right)\frac{\Gamma(1-\epsilon)}{\Gamma(1-2\epsilon)}\left(\frac{4 \pi \mu_{DR}^2}{R^2p_{T\,\gamma}^2(1-z)^2}\right)^{\epsilon},
\end{equation}
where the scale $\mu_{DR}$ arises from the dimensional regularization. Using the expansion
\begin{equation}
A^{\epsilon} = 1 + \epsilon \log(A) + \mathcal{O}(\epsilon^2)
\end{equation}
this can be written as (dropping terms $\mathcal{O}(\epsilon)$):
\begin{equation}
D_{q,an,LO}^{\gamma}(z,Q^2_F) + \frac{\alpha}{2 \pi}e_q^2 \left[ 2 \log \left( \frac{R\,p_{T\,\gamma}(1-z)}{Q_F} \right) P_{\gamma q}^{LO}(z) +z \right],
\label{eq:coll_sing_term}
\end{equation}
where $D_{q,an,LO}^{\gamma}(z,Q^2_F)$ is the LO part of the anomalous component in the quark-to-photon fragmentation function in the $\overline{\rm MS}$ scheme, to which the remaining $1/\epsilon$ singularity is absorbed:
\begin{equation}
\begin{split}
D_{q,an,LO}^{\gamma}(z,Q^2_F)&=-\frac{1}{\epsilon} \frac{\Gamma(1-\epsilon)}{\Gamma(1-2\epsilon)}\left( \frac{4\pi\mu^2}{Q_F^2} \right)^{\epsilon} \frac{\alpha}{2 \pi} e_q^2P_{\gamma q}^{LO}(z)\\
&=\frac{\alpha}{2\pi}e_q^2P_{\gamma q}^{LO}(z)\left(\log \frac{Q_F^2}{\mu^2} - \frac{1}{\hat{\epsilon}}\right) + \mathcal{O}(\epsilon).
\end{split}
\end{equation}
Combining this term with the partonic $2\rightarrow 2$ cross section and the phase-space integral shows that the term proportional to $D_{q,an,LO}^{\gamma}(z,Q_F)$ is already included into the fragmentation component and thus, to avoid double counting, is subtracted from the NLO prompt component. The remaining $\log(Q_F)$ dependence remains in the prompt component being responsible for the $Q_F$ dependence in $K_{\rm prompt}^{ij}$. The $R$ dependence cancels out in the full result where the integration over the region outside the cone is included.

\begin{figure}[hptb]
\centering
\includegraphics[width=0.6\textwidth]{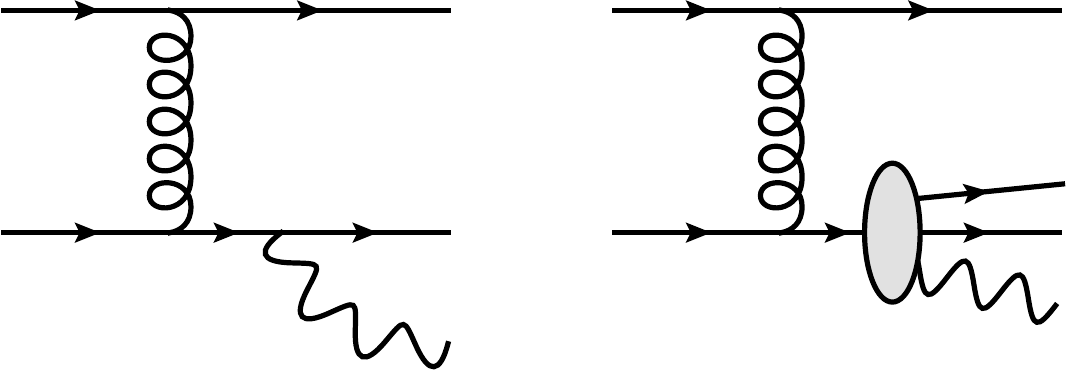}
\caption{{\bf Left}: Photon emission graph contributing to the NLO correction of prompt photon production. {\bf Right:} A LO graph contributing to the fragmentation component.}
\label{fig:PhotonEmission}
\end{figure}

The interplay between the prompt NLO component and the LO fragmentation component is not surprising as the collinear photon emissions are already included into the fragmentation functions. To demonstrate this the figure \ref{fig:PhotonEmission} shows two graphs, one contributing to the NLO prompt component and the other one to the LO fragmentation component. Although the $Q_F$ dependence of the prompt component can be large, it is partly compensated by the $Q_F$ dependence in fragmentation component. Thus, the scale ambiguity here can be interpreted as freedom to choose which processes are treated perturbatively and which are included to the non-perturbative fragmentation functions. By numerical studies it turns out that the total inclusive direct photon cross section is not very sensitive to the choice of the fragmentation scale $Q_F$. This emphasizes the fact that at NLO neither the prompt nor the fragmentation component alone is a valid physical observable but only the sum of these two is.

\section{Isolation cut}

In a hadronic collision there are several mechanisms that can produce photons. One major source is the photons that are produced via decays of unstable hadrons. Thus, to study the direct photons, experiments often introduce an isolation cut for the candidate photons (see e.g. Refs.~\cite{Chatrchyan:2012vq, Adare:2012yt, Abazov:2005wc}). As the photons from the hadronic decays are usually accompanied by hadrons nearby, the isolation requirement effectively cuts out these processes. Also the photons that are produced via fragmentation are surrounded by (nearly) collinear hadrons originating from the same parent parton, and this is essentially the case also for the prompt NLO photons with collinear emissions. Thus, to be able to consistently compare the pQCD predictions with the measurements, the isolation effects for the direct photons have to be studied.

\begin{figure}[hptb]
\centering
\includegraphics[width=0.7\textwidth]{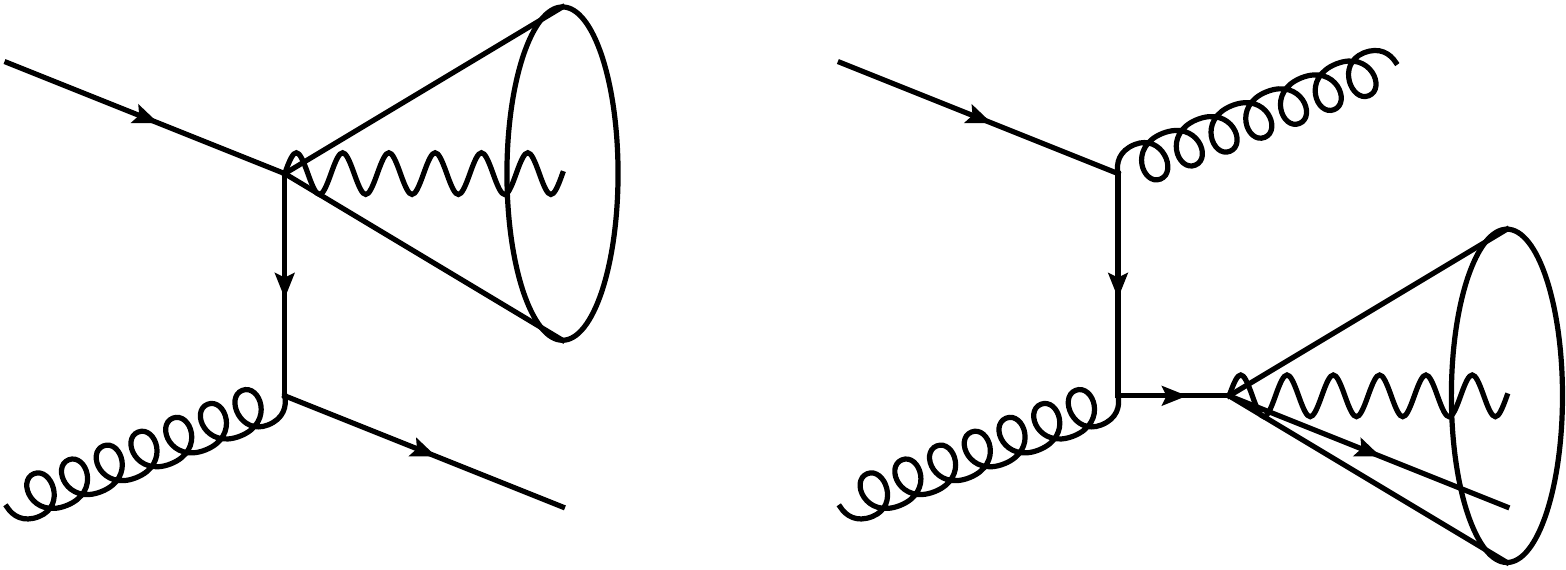}
\caption{{\bf Left}: Isolated photon. {\bf Right:} Non-isolated photon.}
\label{fig:IsolatedPhoton}
\end{figure}

The isolation cut in hadronic collisions is defined as follows: First one draws a cone with a radius $R$ around the photon in the $(\eta, \phi)$ space, as demonstrated in figure \ref{fig:IsolatedPhoton}. Then the transverse energy of the hadrons inside the cone is summed to $\Sigma E_T$:
\begin{equation}
\Sigma E_T=\sum_i E_{T,i}\theta(R-R_i),
\label{eq:transverseEnergy}
\end{equation}
where $E_{T,i}$ is the transverse energy of a hadron $i$ and $\theta(x)$ is the Heaviside step function. The distance $R_i$ between the photon and the hadron $i$ in the $(\eta, \phi)$ space is calculated from
\begin{equation}
R_i=\sqrt{(\eta_{\gamma}-\eta_i)^2+(\phi_{\gamma}-\phi_i)^2},
\label{eq:isolationCone}
\end{equation}
where $\eta_{\gamma}$ ($\eta_{i}$) is the pseudorapidity and $\phi_{\gamma}$ ($\phi_{i}$) the azimuthal angle of the photon candidate (hadron $i$). The photon is isolated if $\Sigma E_T < E_{T\,max}$, where the $E_{T,max}$ is chosen suitably. The $E_{T,max}$ can be either a fixed value, usually of the order few GeV's, or it can be chosen to be proportional to the photon $p_T$. In the calculations this corresponds to cutting the phase space in the $z$ integration in both the prompt and fragmentation contributions. For the prompt component this affects only the NLO contribution via the collinear emissions as in equation (\ref{eq:coll_sing_term}), resulting actually in an {\em increased} contribution because the isolation cut suppresses the part that gives a negative contribution to the prompt photon cross section. For the fragmentation contribution the isolation cut suppresses both the LO and NLO terms so that the total cross section, which is the physical observable, is always reduced when an isolation cut is imposed. 

In Ref.~\cite{Frixione:1998jh} also a modified isolation criteria was proposed. There the cone size $R$ is not a constant but a continuous variable and the photon is then isolated if
\begin{equation}
\sum_i E_{T,i}\theta(R-R_i)\le F( R) \quad \forall R \le R_0,
\end{equation}
where $F( R)$ depends on the photon transverse energy $E^{\gamma}_{T}$ and $F\rightarrow 0$ when $R\rightarrow 0$. It is claimed that this kind of an isolation criteria would remove {\em all} fragmentation photons but as this has not yet been implemented into the measurements, we have not considered this criterion in our studies.

In the experiments the isolation cut might also reject the photons which would be isolated from the theoretical point of view but which are accompanied by hadrons from the underlying event. Corrections for this should be taken care of especially in collisions where the multiplicity is large, e.g. in p+p collisions with high pile-up or in heavy-ion collisions. Experimentally one can try to estimate the contribution from the underlying event and subtract the part from the $\Sigma E_T$ as done e.g. in Ref.~\cite{Chatrchyan:2012vq} for the Pb+Pb collisions. Theoretical studies of this would require a full event simulation with an event generator.

\section{Numerical implementations}

\subsection{\texttt{INCNLO} code}
\label{subsec:incnlo}

In the articles included in this thesis the NLO single-inclusive direct photon cross sections are calculated using the public \texttt{INCNLO} fortran code\footnote{\url{http://lapth.cnrs.fr/PHOX_FAMILY/readme_inc.html}}. The code combines the NLO prompt photon calculations from Refs.~\cite{Aurenche:1998gv, Aurenche:1987fs} with the NLO fragmentation calculations from Refs.~\cite{Aversa:1988vb, Aurenche:1999nz}. The program can be used also to compute the cross section of inclusive hadron production as the partonic sub-processes are the same for the hadrons and photons, only the fragmentation functions are different for each case.

The aim of the code is to efficiently calculate single inclusive cross sections. There is a 2-dimensional integral for the LO and virtual corrections and a 3-dimensional integral for the $2\rightarrow 3$ processes for the fragmentation component with fixed $p_T$ and $\eta$. For the prompt component there is one dimension less and integration over an $\eta$ interval contains one dimension more. The multidimensional numerical integrations for the fragmentation contribution are by default done using the \texttt{BASES} -Monte Carlo (MC) routine (V5.1) from Ref.~\cite{Kawabata:1995th}. However, this default routine is rather old and in some cases does not provide the required numerical accuracy even with enhanced number of sampling points. To overcome this I have replaced the default MC routine by a Vegas -type MC routine provided in the open source GNU Scientific Library (GSL)\footnote{\url{http://www.gnu.org/software/gsl/}}. The GSL routines are written in C, but can be straightforwardly implemented also to a \texttt{FORTRAN} code. If performance is the priority and the dimension of the integral is not too high, also the \texttt{CERNLIB D120} routine produced reliable results with reduced computing time.

When preparing results for the article [V] of this thesis a more severe problem with the \texttt{INCNLO} code occurred: When considering the phase space region with a large $\sqrt{s}$ ($\sim\rm{TeV}$), large $\eta$ ($\sim3$) and $p_T<10\,\mathrm{GeV/c}$ the phase space integration of the fragmentation component does not converge as the numerical precision is not sufficient. This limited also the kinematic reach of the calculations in the article [I] of this thesis and earlier also e.g. in Ref.~\cite{Arleo:2011gc}. We traced the problem to arise from certain gluonic sub-processes that involve divisions between small numerical values which causes some numerical instabilities. In the following I will briefly explain how we cured these problems.

For a $2\rightarrow 3$ processes the integration variables in the program are $z$, $v$ and $w$, where $z$ is the momentum fraction in the fragmentation functions and $v$ and $w$ are related to the momentum fractions $x_1$ and $x_2$ in the PDFs. The integration limits are
\begin{align}
z_{min} &= 1 - V + VW &  z_{max} &= 1 \\
v_{min} &= VW/z & v_{max} &= 1- (1-V)/z \\
w_{min} &= VW/(zv) & w_{max} &= 1,
\end{align}
where the hadronic variables $V$ and $W$ can be calculated from
\begin{equation}
V=1-\frac{p_T}{\sqrt{s}}\mathrm{e}^{-y} \quad \text{and} \quad W = \frac{p_T \mathrm{e}^{y}}{\sqrt{s}-p_T \mathrm{e}^{-y}}.
\end{equation}
The cross section then involves terms which in the original code contain large powers of $(1-v)$ and $(1-vw)$ both in the numerator and in the denominator that become very small when $v\rightarrow 1$, see the example code below (the integration variables $v$ and $w$ are here written in capital letters):
\begin{lstlisting}
LVW =(-2*N**4*(V-1)*VC*(V**5*W**5-5*V**4*W**4+10*V**3*W**3
  -10*V**2*W**2+5*V*W-1)*(V**5*W**5-5*(V-1)*V**4*W**4+10*(
  V-1)**2*V**3*W**3-10*(V-1)**3*V**2*W**2+5*(V-1)**4*V*W-(
  V-1)**5)*(16*V**5*W**5-16*V**4*(V+1)*W**4+V**3*(32*V**2-
  43*V+43)*W**3-V**2*(10*V**2-7*V+13)*W**2-(6*V**4-21*V**3
  +12*V**2+3*V-4)*W-4*(V-1)*(2*V**2-2*V+1))+2*N**2*(V-1)*V
  C*(V**3*(V+7)*W**3-V**2*(2*V**2+V+5)*W**2+(2*V**4+V**3-V
  +2)*W-4*(V-1)*(2*V**2-2*V+1))*(V**5*W**5-5*V**4*W**4+10*
  V**3*W**3-10*V**2*W**2+5*V*W-1)*(V**5*W**5-5*(V-1)*V**4*
  W**4+10*(V-1)**2*V**3*W**3-10*(V-1)**3*V**2*W**2+5*(V-1)
  **4*V*W-(V-1)**5)-2*(V-1)**2*VC*W*(V**3*W**2-2*V**3*W+(2
  *V-1)*(V**2-V+2))*(V**5*W**5-5*V**4*W**4+10*V**3*W**3-10
  *V**2*W**2+5*V*W-1)*(V**5*W**5-5*(V-1)*V**4*W**4+10*(V-1
  )**2*V**3*W**3-10*(V-1)**3*V**2*W**2+5*(V-1)**4*V*W-(V-1
  )**5))*LOG(1-V*W)/(N**2*(V-1)**3*V**2*W**2*(V*W-1)**5*(
  V*W-V+1)**5)
\end{lstlisting}
However, these expressions can be further simplified. For example, the code above after simplifying becomes:
\begin{lstlisting}
LVW =LOG(1-V*W)*(2*VC*(N**2*(2*(2 + W) + 
    V*(-12 - W + V*
    (16 - 8*V + V*(1 + 2*V)*W - (5 + V + 2*V**2)*W**2 + 
    V*(7 + V)*W**3))) - 
    (-1 + V)*W*(-2 + V*(5 + V*(-3 + V*(2 + (-2 + W)*W)))) + 
    N**4*(-4*(1 + W) + 
    V*(3*(4 + W) + 
    V*(-16 + W*(12 + 13*W) - 
    16*V**3*W**3*(2 + (-1 + W)*W) + 
    V**2*W*(6 + W*(10 + W*(43 + 16*W))) - 
    V*(-8 + W*(21 + W*(7 + 43*W))))))))/
    (N**2*(-1 + V)**2*V**2*W**2)

\end{lstlisting}
where now the powers of the $(1-v)$ and $(1-vw)$ are significantly reduced most notably in the denominator, which makes the expression numerically more stable. This simplifying procedure needs to be done for the functions \texttt{STRUV13-16} in the \texttt{hadlib.f} -file, where each function contains nine terms similar to the example above. Due to the number and length of the terms the simplifications should be automated rather than doing all these by hand. This can be done using the \texttt{FullSimplify} command in \texttt{Mathematica} together with some scripts converting the expressions from \texttt{FORTRAN} form to a \texttt{Mathematica} form and back. This, however, is still not quite enough for all the kinematic regions under consideration but also the numerical precision needs to be increased. As this is rather straightforward and depends on the used compiler, I will not discuss this in more detail here. 

\begin{figure}[hptb]
\centering
\includegraphics[width=0.8\textwidth]{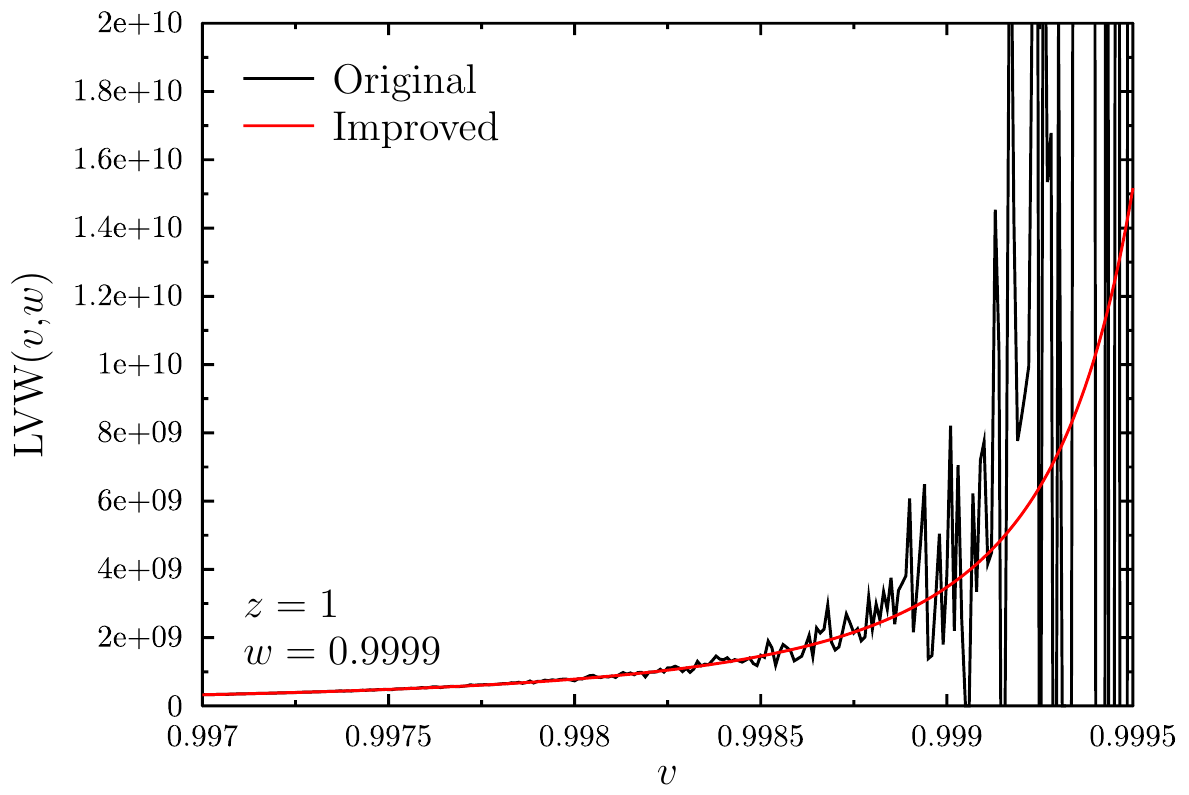}
\caption{A part of the cross section integrand from the original INCNLO code (black) and a improved version of the same part (red).}
\label{fig:hadlib_ex}
\end{figure}

Taking now e.g. $\sqrt{s}=5.0\,\mathrm{TeV}$, $p_T=5.0\,\mathrm{GeV/c}$ and $y=3$ gives $v_{max}=0.99995$, where the results from original code already diverge as shown in figure \ref{fig:hadlib_ex}. However, the improved version of the same code with the above simplification produces reliable results and the cross section integral converges. All the modified parts of the code have gone through an extensive number of cross checks with the original version\footnote{The author wishes to thank Hannu Paukkunen for collaboration in these details.}. After these measures, we were able to perform the novel study of forward photon production at the LHC also at small values of $p_T$, see the article [V] of this thesis.

\subsection{\texttt{JETPHOX} code}
\label{subsec:jetphox}

The \texttt{INCNLO} program described above is very efficient for single-inclusive cross section calculations as all possible phase space integrations are done analytically. However, this means that it is not possible to consider different kinematic cuts, such as the isolation cut discussed above, or to study two-particle correlations, with these kind of codes. For such a purpose a Monte Carlo (MC) approach is more suitable. In the MC approach the phase space is left unintegrated and the 4-momenta of the final state particles are sampled according to the probability distribution given by the differential cross section (see e.g. Ref.~\cite{Sjostrand:2006za} for details). This way one can then implement diverse kinematic cuts to study various measurable observables and also reproduce the inclusive cross sections for comparison if needed. The price to pay for the extended flexibility is the enhanced computing time as one typically needs a large number MC-events to reduce the numerical fluctuations due to the sampling with finite statistics. The event generation is of course trivial to parallelize after the generator is initialized if a computing server with a large number of cores is available.

\begin{figure}[htb]
\centering
\includegraphics[width=0.8\textwidth]{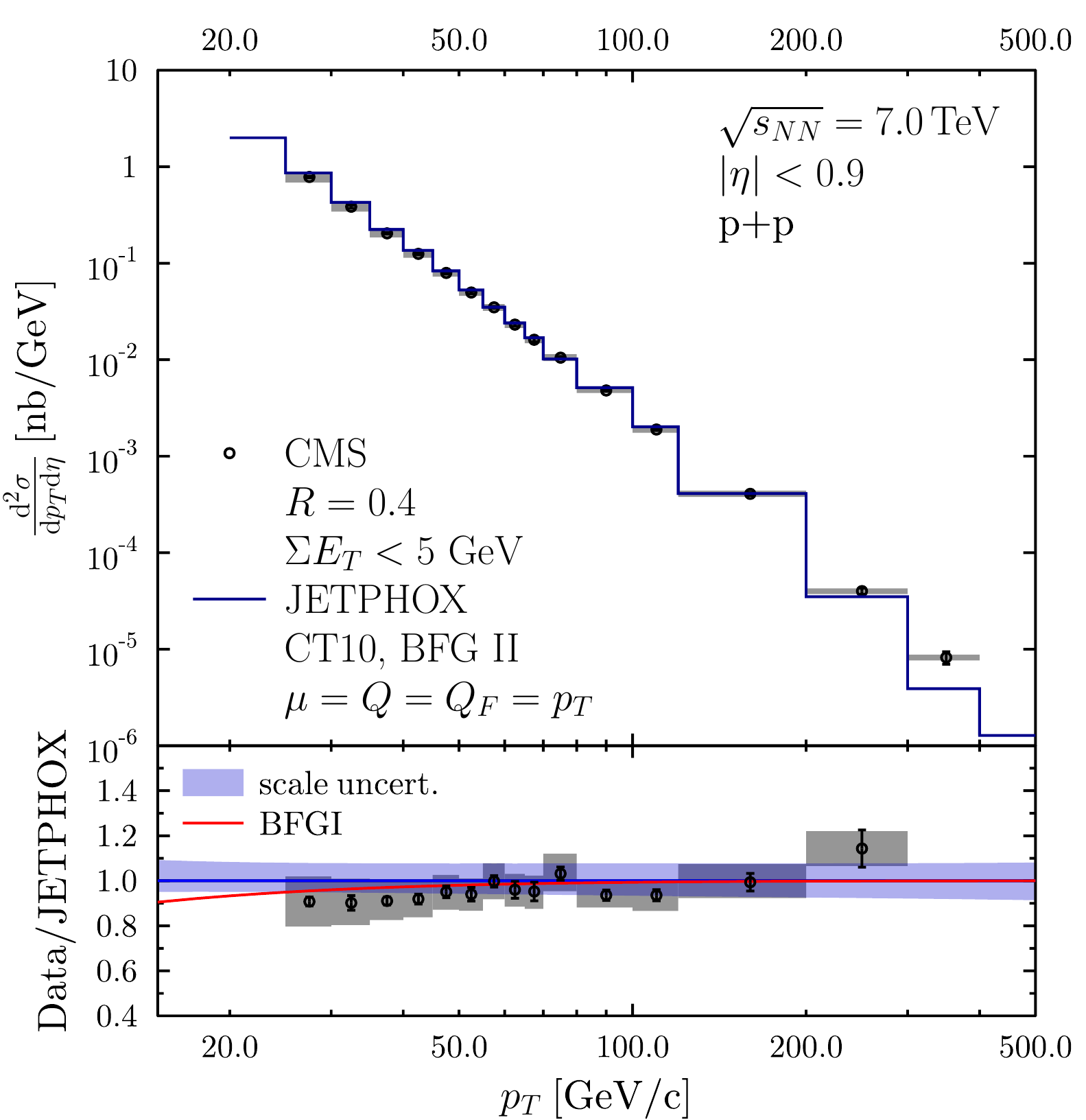}
\caption{{\bf Top:} The isolated photon cross section in p+p collisions at $\sqrt{s}=7.0\,\mathrm{TeV}$ and $|\eta|<0.9$. The data is from CMS \cite{Chatrchyan:2011ue} and the NLO calculations are performed using \texttt{JETPHOX}, applying the CT10 PDFs and BFG II FFs. {\bf Bottom:} The data-to-theory ratio with BFG II FFs. The scale uncertainties and the BFG~I result are calculated for inclusive photons with \texttt{INCNLO} (see the text for the reasoning).}
\label{fig:dsigma_CMS_pp}
\end{figure}

A suitable tool to study the direct photon production in hadronic collisions with the Monte Carlo approach at NLO accuracy is the \texttt{JETPHOX} code\footnote{\url{http://lapth.cnrs.fr/PHOX_FAMILY/jetphox.html}} \cite{Catani:2002ny}. The code is shown to reproduce accurately the isolated photon data from different experiments with different collision energies \cite{Aurenche:2006vj} and has been already used to study free the proton PDFs with isolated photon data in Ref.~\cite{d'Enterria:2012yj}. Also the CMS data \cite{Chatrchyan:2011ue} for the isolated photon cross section in p+p collisions at $\sqrt{s}=7.0\,\mathrm{TeV}$ and $|\eta|<0.9$ is well reproduced as shown in figure \ref{fig:dsigma_CMS_pp}. For the calculation the CT10 PDFs and BFG II FFs are used and all the scales have been set to the photon $p_T$. The isolation cut is defined by requiring $\Sigma E_T<5\,\mathrm{GeV}$ inside $R=0.4$ to match to the criterion used in the measurement. The scale uncertainties have been quantified by varying all the scales from $p_T/2$ to $2p_T$, and also the result with the BFG I FFs is shown. Both of these are calculated with the \texttt{INCNLO} code without isolation to avoid the statistical uncertainty due to the numerical fluctuations. In doing this, we should keep in mind that as the isolation cut suppresses mostly the fragmentation component, the BFG I result likely overestimates the difference to the BFG II set for the isolated case by some amount. Also, as the variation of the fragmentation scale modifies the relative contributions from prompt and fragmentation component, also the resulting scale uncertainty for isolated photons might slightly differ from the presented inclusive result. In any case, as the size of these effects is $<10\,\%$, a detailed MC study of these uncertainties for the isolated case would require a huge number of generated events to overcome the fluctuations arising from the MC sampling.

\begin{figure}[htb]
\centering
\includegraphics[width=0.8\textwidth]{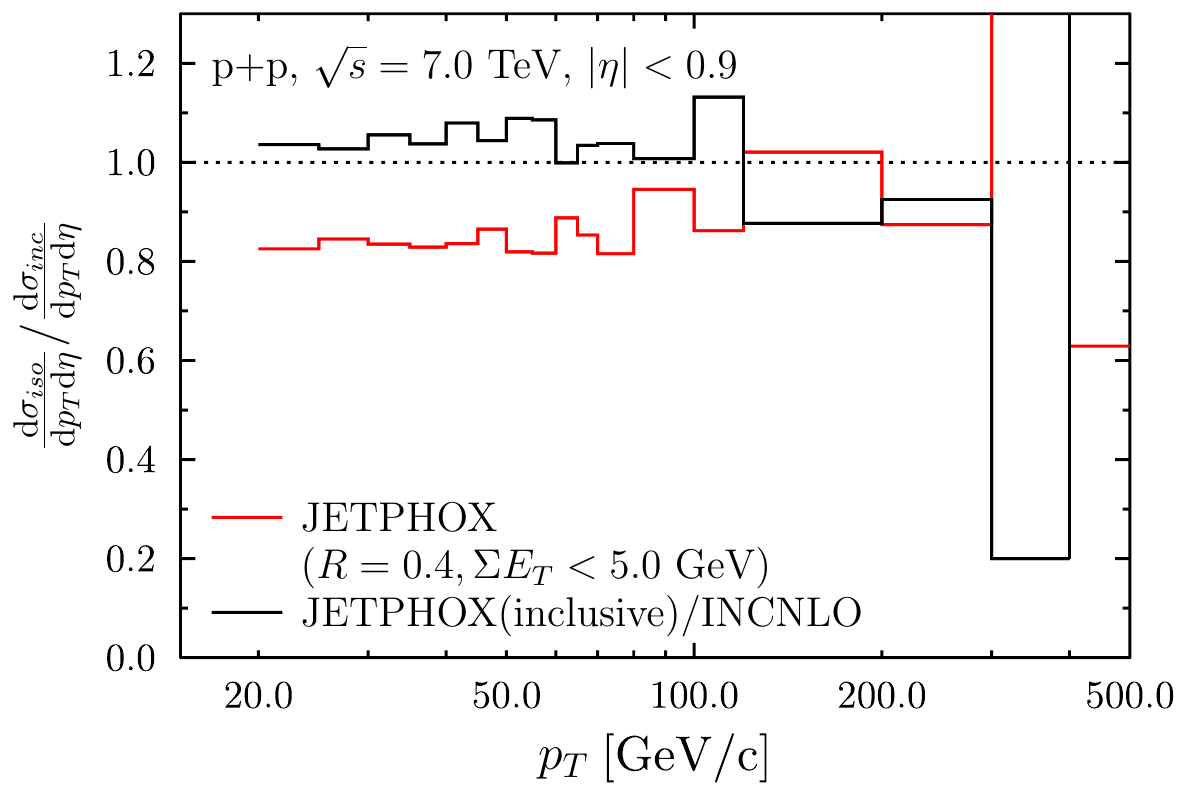}
\caption{The ratio between the isolated and inclusive photon cross sections (red) and the ratio between the inclusive direct photon cross section from \texttt{JETPHOX} and \texttt{INCNLO} (black) in p+p collisions at the LHC.}
\label{fig:iso_vs_inc}
\end{figure}

Figure \ref{fig:iso_vs_inc} shows the effect of the isolation cut to the direct photon cross section with the above kinematics. Here for both the prompt and fragmentation component 40 M events were generated. Between $p_T\in[20,100]\,\mathrm{GeV/c}$ the isolation cut reduces the direct photon cross section by $\sim 20\,\%$. Above that, the numerical fluctuations start to be of the same order and shroud the effect from the isolation. To cross check between the two programs and to estimate the size of numerical fluctuations, figure \ref{fig:iso_vs_inc} shows also the ratio between the \texttt{JETPHOX} and \texttt{INCNLO} results for the inclusive direct photon cross section. The agreement between the two NLO calculations seems very good as expected and the numerical fluctuations here are tolerable below $p_T=300\,\mathrm{GeV/c}$. Above that the greatly reduced cross section gives only a few events to the bins which can be observed as a very large fluctuations in both ratios. 

To study how the isolation cut affects each component in the direct photon cross section, the ratios between the prompt and total (prompt + fragmentation) cross section are presented in figure \ref{fig:prom_frag} for inclusive and isolated photons. Again, the ratio of inclusive photons is calculated from both the \texttt{JETPHOX} and \texttt{INCNLO} codes to cross check the result. The figure shows that the reduction of the cross section observed in figure \ref{fig:iso_vs_inc} is indeed due to the suppression of the fragmentation component as the relative contribution of the prompt component increases about the same amount as the isolation cut reduces the total cross section. At this high values of $p_T$ the prompt component dominates over the fragmentation component and after the isolation cut the contribution from fragmentation is $20\,\%$ or less. One should however keep in mind that at the NLO level the division into two separate contributions is ambiguous due to the fragmentation scale dependence as discussed earlier and thus the results in the figure \label{fig:prom_frag} should be considered suggestive only. The figures 4 and 5 in the article [V] show how the relative contributions alter when all the scales are varied. These figures also demonstrate how the fragmentation component becomes dominant towards smaller $p_T$, which largely follows from the rapidly increasing small-$x$ gluon luminosity.
\begin{figure}[hptb]
\centering
\includegraphics[width=0.8\textwidth]{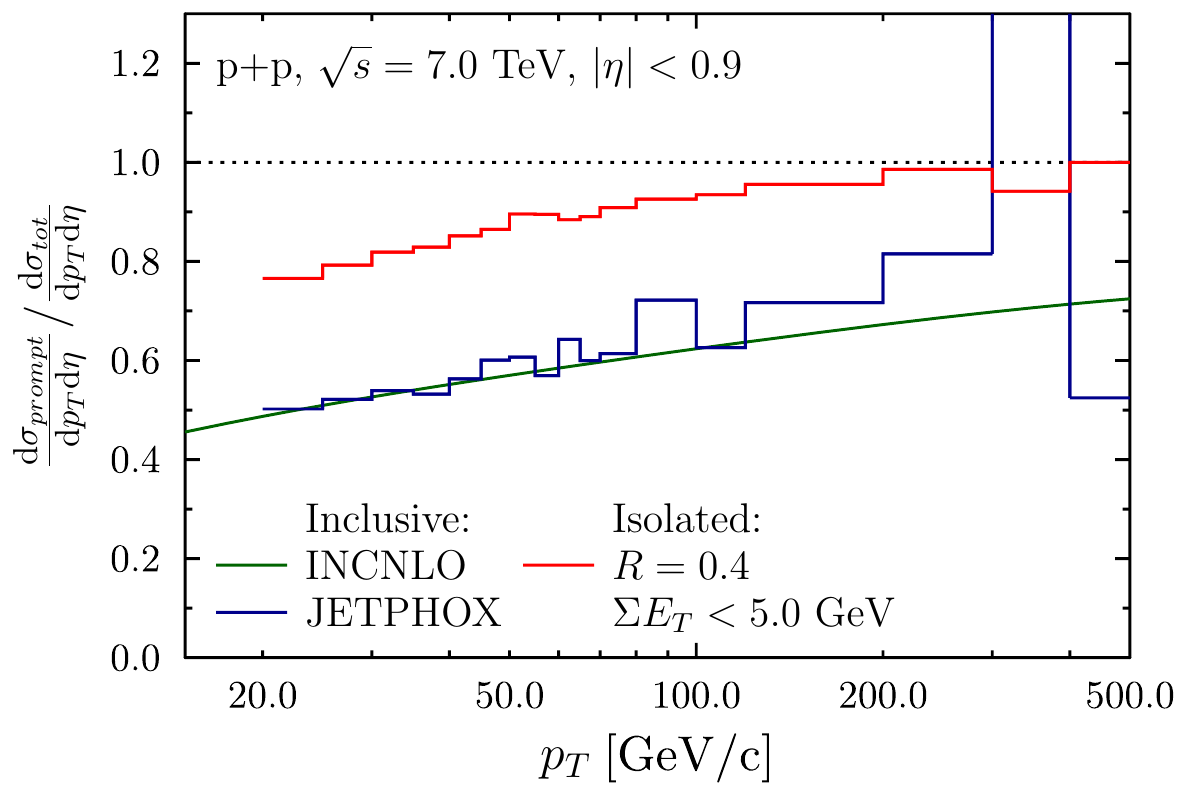}
\caption{The ratio between the prompt and total cross section for inclusive (\texttt{JETPHOX}: blue, \texttt{INCNLO}: green) and isolated direct photons (red) in p+p collisions at the LHC.}
\label{fig:prom_frag}
\end{figure}

\section{Sensitivity to gluon nPDFs}

\begin{figure}[hptb]
\centering
\includegraphics[width=0.8\textwidth]{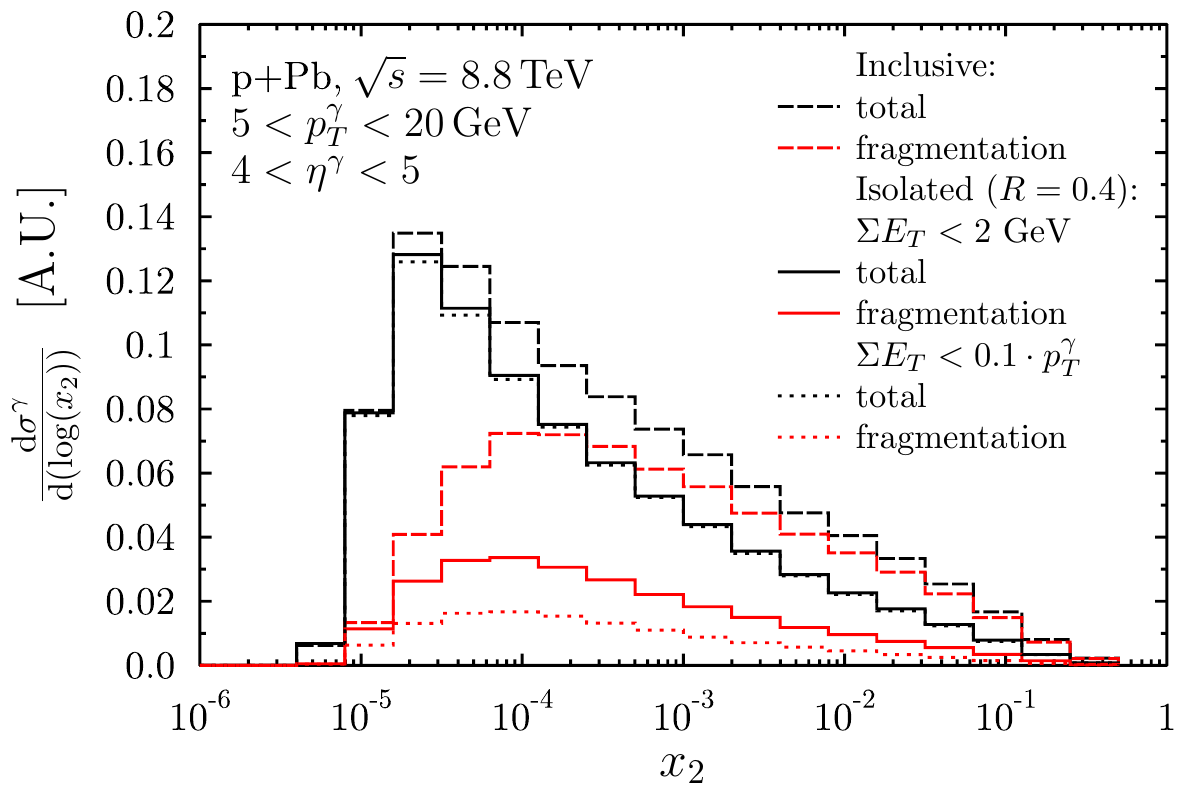}
\caption{The direct photon cross section as a function of $\log x_2$ with isolation cuts $\Sigma E_T<2\,\mathrm{GeV}$ (solid) and $\Sigma E_T<0.1\cdot p_T^{\gamma}$ (dotted), and without the isolation cut (dashed) in p+Pb collisions at the LHC. Also the fragmentation contribution is plotted for each case (red).}
\label{fig:dsigma_x2_y45}
\end{figure}

In the article [V] of this thesis we used the \texttt{JETPHOX} code to calculate the differential cross section of the isolated photon production at forward rapidities in p+Pb collisions at the LHC. The goal was to study in detail which regions of $x$ of the gluon nuclear PDFs are probed at different rapidities. As the cross sections at forward rapidities are somewhat lower than at mid-rapidities and as the main interest is at small  values of $x$ (where the gluons are badly known), we concentrated on lower values of $p_T$, where the fragmentation component is more pronounced. Also there we noticed that the isolation cut reduces the contribution from the fragmentation component, which makes the cross sections more sensitive to smaller values of $x_2$ in the nucleus as the $p_T$ of a fragmentation photon is always smaller than the $p_T$ of the parent parton. However, although the fragmentation component is reduced more than by a factor of two when requiring $\Sigma E_T<2\,\mathrm{GeV}$, the moderate growth of the prompt component's NLO correction partly compensates the reduction. This can be seen from figure \ref{fig:dsigma_x2_y45} showing the differential cross section (in $\log x_2$) of the inclusive and isolated photon production together with the fragmentation contribution alone. The interplay between the different components is visible especially at $x_2>10^{-3}$ where the total cross section clearly does not reduce as much as the fragmentation component. This indicates that similarly as the fragmentation scale, the isolation cut also affects the division to prompt and fragmentation contributions at NLO, again underlining that only the sum of these two is a meaningful physical quantity. Indeed, by choosing an isolation criterion $\Sigma E_T<0.1\cdot p_T^{\gamma}$ instead of $\Sigma E_T<2\,\mathrm{GeV}$ the fragmentation component can be reduced even further but still the two components together give a very similar $x_2$ distribution.

The $x_2$ distributions in the figure \ref{fig:dsigma_x2_y45} and in the article [V] of this thesis are calculated by a ``brute force'' method, in which the PDFs are modified so that they are zero everywhere but in a specific $x$ bin and the cross section integrals are calculated over each bin separately. The method is not very elegant or efficient but a very robust one. It would be advantageous if the $x_1$ and $x_2$ would be saved for each event similarly as the 4-momenta of the final state particles as this would enable one to study the influence of the PDFs without generating a new set events\footnote{Such tools already exists, e.g. \texttt{APPLGRID} \cite{Carli:2010rw} for general NLO cross sections and \texttt{FASTNLO} \cite{Wobisch:2011ij} for jets.}. Especially this would be useful for the global PDF analysis if also the flavors of the initial partons are known.

In the article [V] we calculated also the nuclear modification factor for the isolated photons at forward rapidities in p+Pb collisions. As was shown in figure \ref{fig:iso_vs_inc}, the MC sampling generates numerical fluctuations easily of the order $10\,\%$ to the cross section. When studying ratios of cross sections with the effects of the similar size, e.g. the nuclear modifications of the PDFs, accurate predictions might be hard to obtain. Of course one could just generate billions of events to overcome this but luckily there is also a more handy way. When generating events the PDF uncertainties can be studied by calculating the fully differential cross section at the generated phase-space point with each error set in the PDFs at hand. Then the cross section ratios between the central PDF set and the error sets are saved to the each event as weighting factors from which one can then calculate the uncertainty band after a sufficient number of events is generated. If one is interested in the nPDF effects one can also replace the error sets with the nPDFs. This way the nuclear modifications in each phase-space point are exact and to obtain an accurate $R_{AB}(p_T,\eta)$ one needs to generate only one set of events that gives a reasonable spectrum with the given binning. The only drawback is that the event generation is optimized for the p+p collisions but as the nPDF effects are rather small the optimization works well also for the nucleus case.

\begin{figure}[hptb]
\centering
\includegraphics[width=0.75\textwidth]{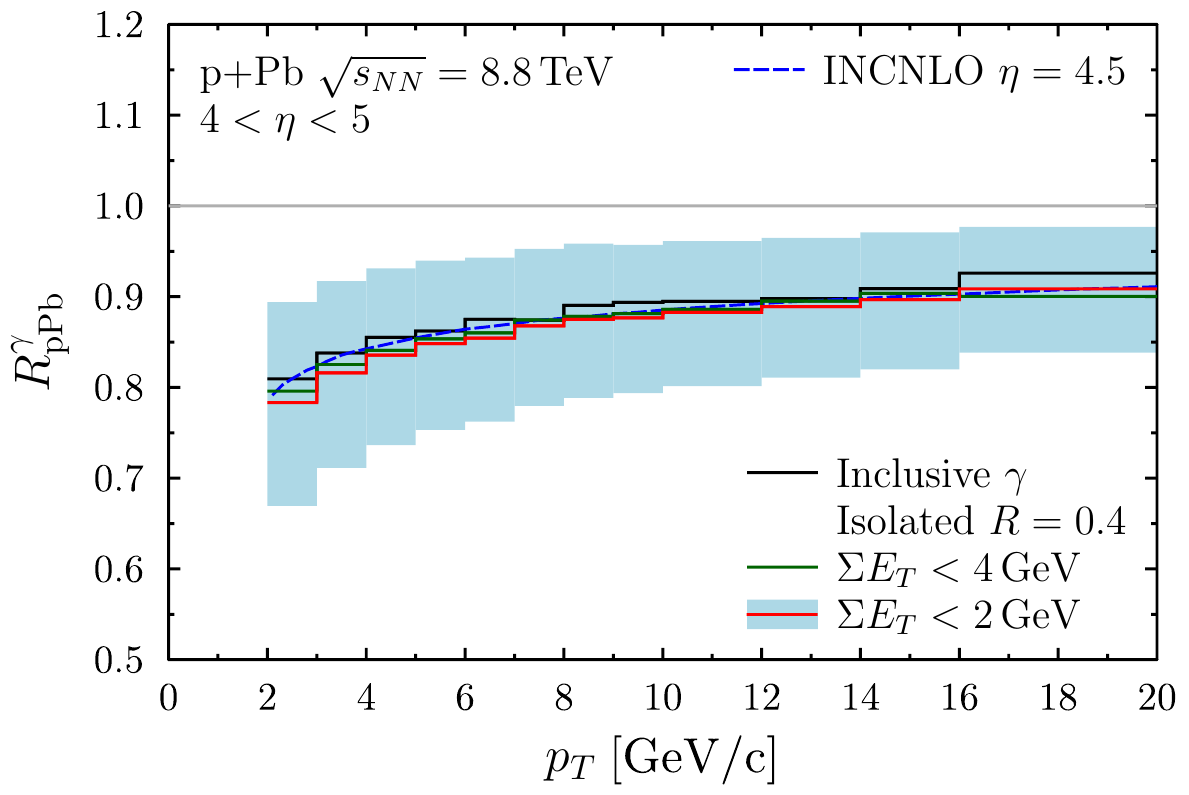}
\caption{The nuclear modification factor $R_{\rm pPb}^{\gamma}$ obtained with \texttt{JETPHOX} for inclusive (black) and isolated photons with two isolation criteria, $\Sigma E_T < 4\,\mathrm{GeV}$ (green) and $\Sigma E_T < 2\,\mathrm{GeV}$ (red). Also the inclusive photon $R_{\rm pPb}^{\gamma}$ with \texttt{INCNLO} is plotted for comparison (blue dashed). Figure from [V].}
\label{fig:R_pPb_iso}
\end{figure}

\begin{figure}[hptb]
\centering
\includegraphics[width=0.8\textwidth]{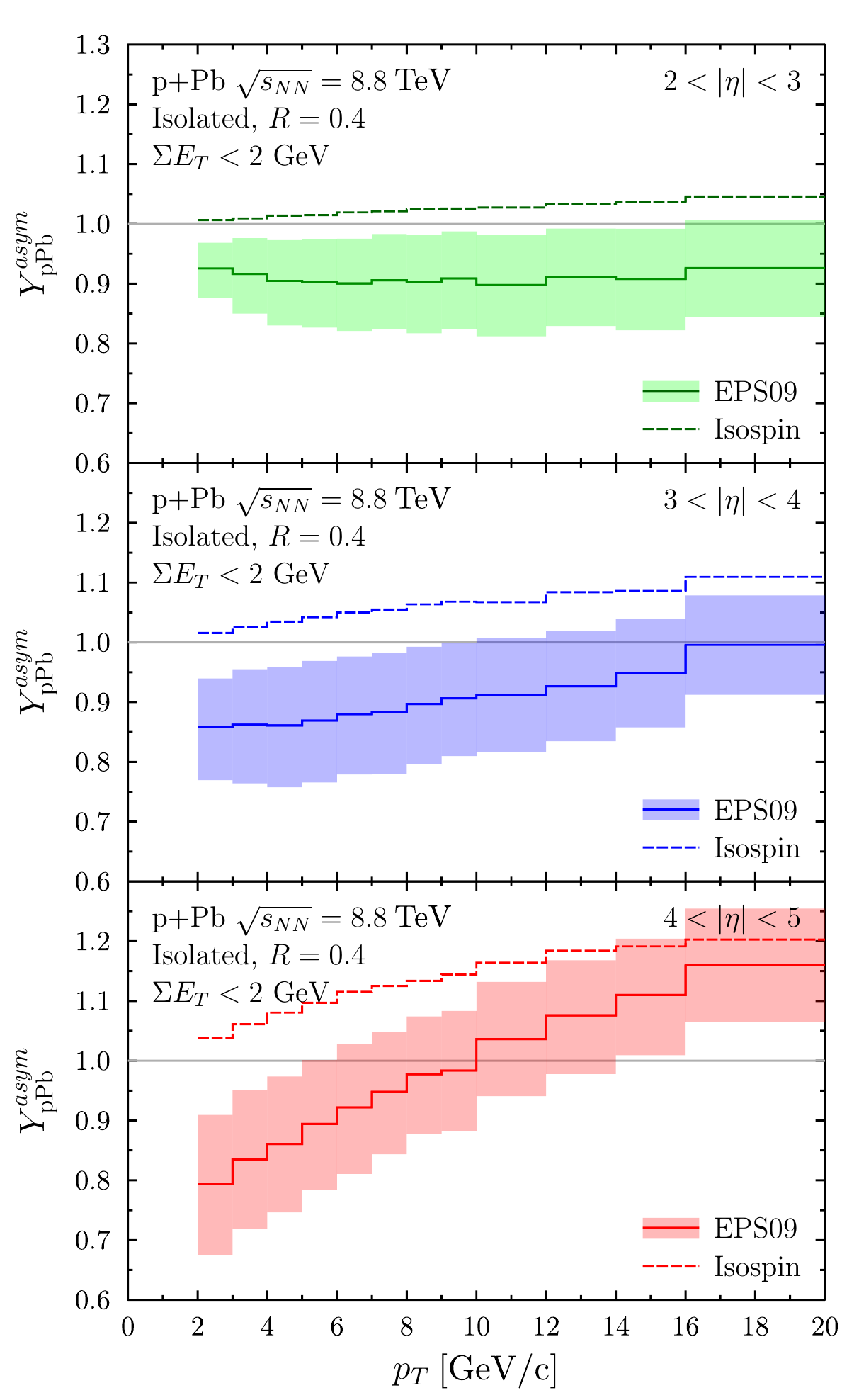}
\caption{The yield asymmetry $Y_{AB}^{\rm asym}(p_T,\eta)$ at rapidity bins $|\eta| \in ]2,3[$ (top), $]3,4[$ (middle), and $]4,5[$ (bottom) of isolated photons with $\Sigma E_T<2\,\mathrm{GeV}$ and $R=0.4$ calculated with \texttt{JETPHOX} using the BFG II FFs and CT10 PDFs with EPS09 nPDFs (solid line with EPS09 uncertainty band). Also the isospin effect is quantified by calculating the same observable without the nuclear modification of the PDFs (dashed line).}
\label{fig:R_fb_pPb}
\end{figure}

Figure \ref{fig:R_pPb_iso} shows the nuclear modification factor for isolated and inclusive photons in p+Pb collisions at $\sqrt{s_{NN}}=8.8\,\mathrm{TeV}$ and $4<\eta<5$. The inclusive photon $R_{\rm pPb}^{\gamma}$ is calculated using both numerical codes discussed above to cross check the results. To obtain the \texttt{INCNLO} result at this kinematic region the improved numerical stability was necessary and the \texttt{JETPHOX} result was computed by generating the p+p spectra using the nPDFs as a weight as discussed above. Both codes yield the same result to a very good accuracy and the numerical fluctuations due to the MC sampling in the \texttt{JETPHOX} case is found to be $\lesssim 2\,\%$ in the whole kinematic region considered. For the isolated $R_{\rm pPb}^{\gamma}$ two different criteria are used, $\Sigma E_T < 4\,\mathrm{GeV}$ and $\Sigma E_T < 2\,\mathrm{GeV}$. The increased sensitivity to the smaller $x_2$ region due to the isolation yields, however, only a slightly more pronounced suppression in $R_{\rm pPb}^{\gamma}$ at $p_T<7\,\mathrm{GeV/c}$. This can be understood by studying the $x$ dependence of the gluon nuclear modification which was shown in figure \ref{fig:EPS09NLO_g}. Due to the rapid DGLAP evolution the originally strong shadowing is rapidly reduced, resulting in a rather mild $x$ dependence in the small-$x$ region probed by the direct photons at forward rapidities. 
As the probed values of $x_2$ at forward rapidities in p+Pb collisions are very small the isospin effect here is negligible. However, as can be seen in figure \ref{fig:R_fb_pPb}, if considering p+Pb collisions at backward rapidities the isospin effect is expected to play a role due to large values of $x_2$ probed.

The main results of article [V] can be summarized as follows
\begin{itemize}
\item The inclusive direct photon cross section at forward rapidity is more sensitive to smaller values of $x_2$ than the inclusive hadron cross section at the same values of $p_T$ and $\eta$. This is due to the presence of the prompt component in the direct photon production that has a more direct connection to the partonic kinematics. Imposing an isolation criterion for the direct photons cuts the fragmentation contribution and thus further increases the sensitivity to small-$x$ physics.
\item Although the small-$x$ sensitivity is increased when moving towards more forward rapidities (or by imposing an isolation cut), the effect to the predicted $R_{\rm pPb}^{\gamma}$ is very modest. This suggests that measurements already at $2 < \eta < 3$ would provide  significant constraints for gluon nPDFs as the $x$ dependence in this region is weak due to the DGLAP evolution. Measurements at larger rapidities would of course serve as an important test of factorization and DGLAP dynamics in general.
\item As the measured nuclear modification ratio often suffers from $\sim 10\,\%$ normalization uncertainty (see e.g. figures \ref{fig:R_pPb_ALICE} and \ref{fig:R_AuAu_gamma_y0}) we proposed also an alternative observable that could be used to study the small-$x$ effects: The yield asymmetry between the forward and backward rapidities, defined as
\begin{equation}
Y_{AB}^{\rm asym}(p_T,\eta) \equiv \left.\frac{\mathrm{d}^2\sigma_{AB}}{\mathrm{d}p_T \mathrm{d}\eta}\right|_{\eta\in[\eta_1,\eta_2]}\bigg/ \left.\frac{\mathrm{d}^2\sigma_{AB}}{\mathrm{d}p_T \mathrm{d}\eta}\right|_{\eta\in[-\eta_2,-\eta_1]}.
\label{eq:Yasym}
\end{equation}
In this kind of measurement a large part of the systematic uncertainties would cancel out as they are the same for the nominator and denominator. Our predictions for rapidity intervals $|\eta| \in [2,3]$, $\in [3,4]$, and $\in [4,5]$ are shown in figure~\ref{fig:R_fb_pPb}. 
\end{itemize}

The same problem with the MC statistics arises when calculating other cross section ratios than the nuclear modification ratio, e.g. the yield asymmetry $Y_{AB}^{\rm asym}(p_T,\eta)$ above. However, here one can utilize the same trick as above for the $R_{AB}(p_T,\eta)$. First, as the p+p collision is symmetric, the cross sections are equal in forward and backward rapidities, i.e.
\begin{equation}
\left.\frac{\mathrm{d}^2\sigma_{\rm pp}}{\mathrm{d}p_T \mathrm{d}\eta}\right|_{\eta\in[\eta_1,\eta_2]} = \left.\frac{\mathrm{d}^2\sigma_{\rm pp}}{\mathrm{d}p_T \mathrm{d}\eta}\right|_{\eta\in[-\eta_2,-\eta_1]}.
\label{eq:dsigma_pp_eta}
\end{equation}
Thus the yield asymmetry can be written as
\begin{equation}
Y_{AB}^{\rm asym}(p_T,\eta) = \frac{\left.\left(\frac{1}{AB}\frac{\mathrm{d}^2\sigma_{AB}}{\mathrm{d}p_T \mathrm{d}\eta}\Big/ \frac{\mathrm{d}^2\sigma_{pp}}{\mathrm{d}p_T \mathrm{d}\eta}\right)\right|_{\eta\in[\eta_1,\eta_2]}}{\left.\left(\frac{1}{AB}\frac{\mathrm{d}^2\sigma_{AB}}{\mathrm{d}p_T \mathrm{d}\eta}\Big/ \frac{\mathrm{d}^2\sigma_{pp}}{\mathrm{d}p_T \mathrm{d}\eta}\right)\right|_{\eta\in[-\eta_2,-\eta_1]}}=\frac{R_{AB}(p_T,\eta)}{R_{AB}(p_T,-\eta)},
\label{eq:Yasym2}
\end{equation}
where now the $R_{AB}(p_T,\pm\eta)$ can be accurately (with small statistical errors) calculated using the method discussed earlier. The $Y^{\gamma}_{\rm pPb}(p_T, \eta)$ presented in the figure \ref{fig:R_fb_pPb} (and in the article [V]) have been calculated in this manner using the \texttt{JETPHOX} code.

\chapter{Centrality in heavy-ion collisions}
\label{chap:centrality}

\begin{figure}[hptb]
\centering
\includegraphics[width=0.4\textwidth]{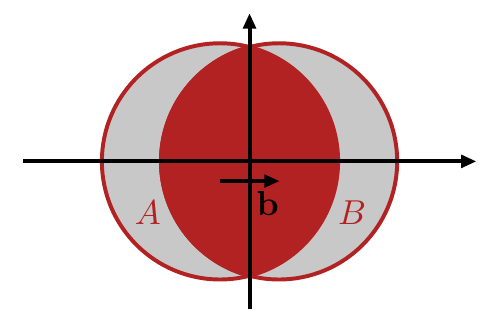}
\hspace{0.05\textwidth} \includegraphics[width=0.4\textwidth]{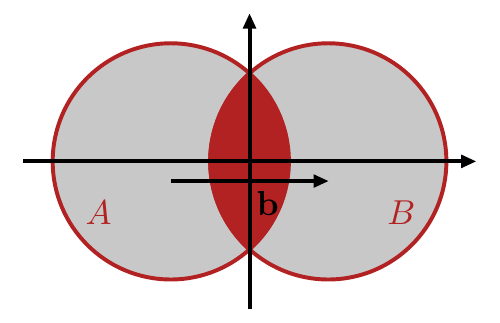}
\caption{{\bf Left}: A central collision with a small impact parameter $b$. {\bf Right:} A peripheral collision with a large impact parameter $b$.}
\label{fig:coll_geom}
\end{figure}

A collision between two heavy nuclei takes place at a specific impact parameter $\mathbf{b}$ whose length $b\equiv |\mathbf{b}|$ defines the distance between the centers of the colliding nuclei $A$ and $B$ in the transverse plane. Instead of the impact parameter, the collision geometry is often referred to as the centrality. A collision with a small impact parameter is referred to as a central collision and a collision with a large impact parameter to as a peripheral collision. Two different collision geometries are presented in figure \ref{fig:coll_geom}.

The impact parameter, however, cannot be directly measured in the collision experiments but a correlation between the collision geometry and an extensive observable such as multiplicity or transverse energy in the collisions have been proposed. Using the multiplicity, the events can be classified into centrality classes, where each class contains all the events in a given multiplicity interval. In this chapter I discuss two theoretical frameworks which can be used to relate the collision geometry and the experimental centrality classes in $A$+$A$ collisions.

\section{Optical Glauber model}

\subsection{Nuclear overlap function}
\label{subsec:T_AB}

\begin{figure}[hptb]
\centering
\includegraphics[width=0.6\textwidth]{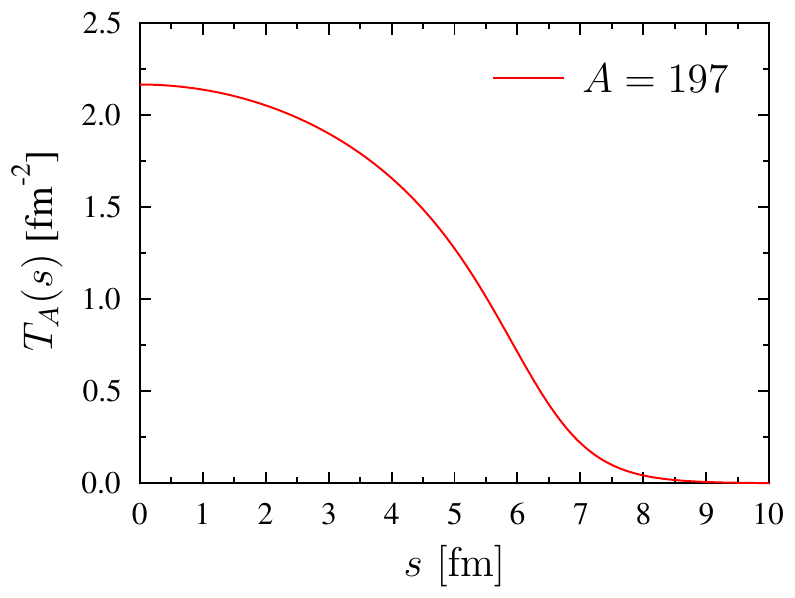}
\caption{The nuclear thickness function $T_A(s)$ for Au as a function of transverse distance $s$.}
\label{fig:T_Au}
\end{figure}

In the Optical Glauber model, extensively reviewed in Ref.~\cite{Miller:2007ri}, the colliding nuclei are described as a smooth density distributions. The nucleon density of the nucleus at a mass number $A$ can be described with a spherically symmetric two parameter Woods-Saxon distribution:
\begin{equation}
\rho_A(\mathbf{s},z)=\frac{n_0}{1+\exp\left[ \frac{\sqrt{\mathbf{s}^2 + z^2} - R_A }{d} \right]},
\label{eq:woodsSaxon}
\end{equation}
where $\mathbf{s}$ is the transverse position vector, $z$ the longitudinal position, $R_A$ the effective radius of the nucleus $A$, $d$ a parameter controlling the width (diffusiveness) of the edge, and $n_0$ defines the central density. In this thesis we use $d=0.54 \,\mathrm{fm}$ and $n_0=0.17 \,\mathrm{fm^{-3}}$. The nuclear thickness at a given $\mathbf{s}$ along the beam axis can be obtained by integrating the equation (\ref{eq:woodsSaxon}) over the longitudinal $z$-coordinate:
\begin{equation}
T_A(\mathbf{s})=\int_{-\infty}^{\infty}\mathrm{d}z\rho_A(\mathbf{s},z),
\label{eq:T_A}
\end{equation}
where the integral has to be done numerically. The parameters $n_0$ and $d$ are related to $R_A$ via the normalization condition $\int \mathrm{d}^2\mathbf{s}\,T_A(\mathbf{s})\equiv A$, which gives (see e.g. Ref.~\cite{Eskola:1988yh})
\begin{equation}
n_0 = \frac{3}{4}\frac{A}{\pi R_A^3}\frac{1}{\big(1+ (\frac{\pi d}{R_A})^2\big)}.
\end{equation}
From this, the radius of a nucleus can be accurately approximated as
\begin{equation}
R_A = 1.12 A^{1/3}-0.86 A^{-1/3}\,\mathrm{fm}.
\label{eq:RA}
\end{equation}
As an example, the $T_A(\mathbf{s})$ for a gold nucleus is shown in figure \ref{fig:T_Au} as a function of the transverse distance from the center $s=|\mathbf{s}|$.

Using the nuclear thickness function, one can construct the so called nuclear overlap function $T_{AB}(\mathbf{b})$, which describes the total amount of interacting nuclear matter at an impact parameter $\mathbf{b}$. This is defined as
\begin{equation}
T_{AB}(\mathbf{b})=\int\mathrm{d}^2\mathbf{s}\,T_A(\mathbf{s_1})\,T_B(\mathbf{s_2}),
\end{equation}
where $\mathbf{s_1}$ and $\mathbf{s_2}$ are defined as 
$\mathbf{s_1}=\mathbf{s}+\mathbf{b}/2$ and $\mathbf{s_2}=\mathbf{s}-\mathbf{b}/2$. The above normalization condition gives then
\begin{equation}
\int\mathrm{d}^2\mathbf{b}\,T_{AB}(\mathbf{b}) = AB,
\end{equation}
when integrating over the whole $2$-dimensional impact parameter space. 

\subsection{Centrality classes}

\begin{figure}[hptb]
\centering
\includegraphics[width=0.7\textwidth]{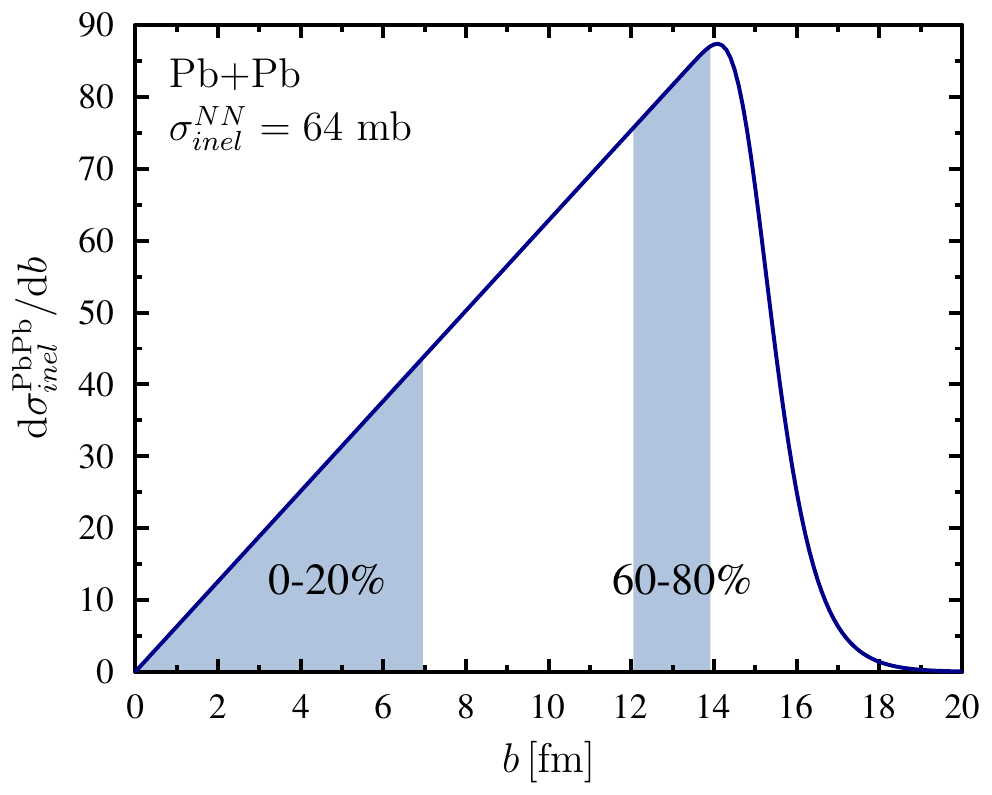}
\caption{The inelastic cross section $\mathrm{d}\sigma_{inel}^{AB}/\mathrm{d} b$ from equation \ref{eq:sigma_inel_ab} as a function of impact parameter for a Pb+Pb collision at $\sqrt{s}=2.76\,\mathrm{TeV}$. The centrality classes $0-20\,\%$ and $60-80\,\%$ are shown by the blue regions.}
\label{fig:sigma_inel_pa}
\end{figure}

As discussed in the Appendix of article [I] of this thesis (see also Ref.~\cite{wong1994intr}), the inelastic cross section for an impact parameter interval $b\in[b_1,b_2]$ ($b=|\mathbf{b}|$) can be calculated from
\begin{equation}
\sigma_{inel}^{AB}(b_1,b_2) = \int_{b_1}^{b_2}\mathrm{d}b\,2\pi b (1-\mathrm{e}^{-T_{AB}(b)\sigma_{inel}^{NN}}),
\label{eq:sigma_inel_ab}
\end{equation}
where $\sigma_{inel}^{NN}$ is the inelastic nucleon-nucleon cross section which depends on the collision energy and can be measured in the experiments, see e.g. Refs.~\cite{Antchev:2011vs, Antchev:2013iaa}.	The centrality classes in the Optical Glauber model are defined by requiring that the integration over a given impact parameter interval generates a certain fraction of the total inelastic cross section $\sigma_{inel}^{AB} = \sigma_{inel}^{AB}(0,\infty)$. For example then the impact parameters corresponding to the $0-20\,\%$ centrality are obtained from
\begin{equation}
\int_{b_1}^{b_2}\mathrm{d}b\,2\pi b (1-\mathrm{e}^{-T_{AB}(b)\sigma_{inel}^{NN}}) = 0.2 \int_{0}^{\infty}\mathrm{d}b\,2\pi b (1-\mathrm{e}^{-T_{AB}(b)\sigma_{inel}^{NN}}),
\end{equation}
where $b_1=0$. In practice the value for $b_2$ here is calculated iteratively by integrating over different impact parameter values until the desired cross section is obtained. A useful approximation here is
\begin{equation}
\int_{0}^{b_2}\mathrm{d}b\,2\pi b (1-\mathrm{e}^{-T_{AB}(b)\sigma_{inel}^{NN}}) \approx \int_{0}^{b_2}\mathrm{d}b\,2\pi b = \pi b_2^2,
\end{equation}
from which one can start iterating until the required accuracy is obtained. The procedure is illustrated in figure \ref{fig:sigma_inel_pa} which shows the inelastic cross section for a Pb+Pb collision as a function of $b$ and impact parameter values that correspond to the centrality classes $0-20\,\%$ and $60-80\,\%$.

Another important centrality-related quantity is the number of binary collisions $N^{AB}_{bin}(\mathbf{b})$. This is required to get the correct normalization of hard processes when comparing the $A$+$B$ collisions to p+p collisions. In the Optical Glauber model at a given impact parameter this can be calculated from
\begin{equation}
N^{AB}_{bin}(\mathbf{b}) = \sigma_{inel}^{NN}T_{AB}(\mathbf{b}).
\label{eq:N_bin}
\end{equation}
The average number of binary collisions in inelastic collisions at a given centrality class can then be obtained from 
\begin{equation}
\langle N_{bin}^{AB} \rangle_{b_1,b_2} = \frac{\int_{b_1}^{b_2} \mathrm{d}^2\mathbf{b}\, N^{AB}_{bin}(\mathbf{b})}{\sigma_{inel}^{AB}(b_1,b_2)}.
\label{eq:N_bin_ave}
\end{equation}

The Optical Glauber model has been used to determine the centrality classes in the articles [I-III] of this thesis. However, if one is interested in studying e.g. the effects of the fluctuations in the initial nucleon configurations (e.g. the event-by-event studies in Ref~\cite{Holopainen:2010gz}), the Optical Glauber model is not sufficient. Thus, I will briefly discuss also another way, which was used in the thermal photon study [III], to translate the collision geometry into the centrality classes.

\section{Monte-Carlo Glauber model}
\label{sec:MCGlauber}

In the Monte Carlo Glauber model the colliding nuclei are no longer treated as smooth distributions but they consist of individual nucleons. The nucleons are randomly positioned (hence the name ``Monte Carlo'') according to the Woods-Saxon density profile (equation (\ref{eq:woodsSaxon})). The sampling procedure generates fluctuations to the nucleon configuration which are found to be necessary to explain the observed triangular flow in heavy-ion collisions \cite{Alver:2010gr,ALICE:2011ab,Alver:2010dn}. Before the introduction of the actual MC Glauber model I will briefly discuss the MC sampling in general. Further discussion on the MC techniques can be found in Refs.~\cite{Beringer:1900zz,Sjostrand:2006za}.

\subsection{Sampling a distribution}

Let $f(x)$ be a positive function from which we want to select a random value $x$ from an interval $x\in[x_{min},x_{max}]$. If the function $f(x)$ is integrable so that $\int \mathrm{d}x\,f(x) = F(x)$ and $F(x)$ has an inverse function $F^{-1}(x)$, the $x$ can be obtained from
\begin{equation}
x=F^{-1}\left\{ F(x_{min}) + R\left[F(x_{max})-F(x_{min})\right]\right\},
\end{equation}
where $R$ is a random number between 0 and 1. However, in most cases the distributions of interest do not have an invertible integral but advanced methods are needed. Often one can find a well behaving function $g(x)$ for which $f(x)\ge g(x)$ holds, where $g(x)$ can be constructed also piecewise. Then $x$ can be obtained by selecting first $x$ from
\begin{equation}
x=G^{-1}\left\{ G(x_{min}) + R\left[G(x_{max})-G(x_{min})\right]\right\},
\end{equation}
where now $\int\mathrm{d}x\,g(x)=G(x)$ and $G^{-1}(x)$ its inverse. Then one takes a new random number $R'$ and compares this to the ratio $f(x)/g(x)$. If the ratio is larger than $R'$, then the sampled $x$ is the result. If $f(x)/g(x)\le R'$, the $x$ is rejected and the process is repeated. For efficient sampling one should try to find a function which is close to $f(x)$ to reduce the number of reruns. If the approximative function is piecewise, i.e. $g(x)=\sum_i g_i(x)$, one needs first to pick $i$ randomly using the integrals $G_i(x_{max})-G_i(x_{min})$ as weights.

\subsection{Sampling the Woods-Saxon distribution}

\begin{figure}[htb]
\centering
\includegraphics[width=0.7\textwidth]{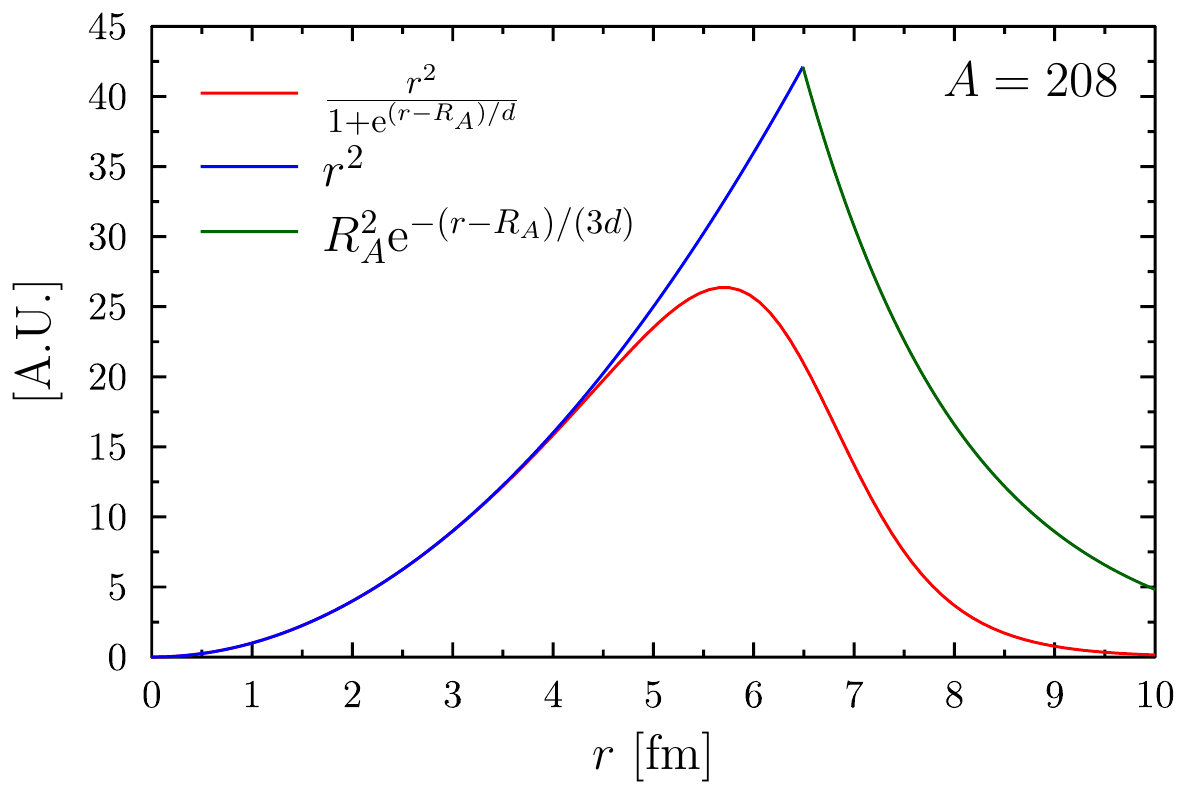}
\caption{The radial part of the Woods-Saxon density (red) and the approximative functions $g_1(r )$ (blue) $g_2(r )$ (green) for Pb nucleus.}
\label{fig:woodssaxon_approx}
\end{figure}

To efficiently sample the 3-dimensional density distribution one can make use of the radial symmetry. In spherical coordinates the differential distribution takes the form (normalization is not relevant for the sampling)
\begin{equation}
\frac{\mathrm{d}x\,\mathrm{d}y\,\mathrm{d}z}{1+\exp\Big[\frac{\sqrt{x^2+y^2+z^2}-R_A}{d} \Big]} = \frac{\mathrm{d}r\,\mathrm{d}\theta\,\mathrm{d}\phi\, r^2 \sin \theta}{1+\exp\left[\frac{r-R_A}{d} \right]},
\end{equation}
where the variables $r$, $\theta$ and $\phi$ can now be sampled independently. The sampling of both angles is now trivial as the $\phi$ distribution is flat and the integral of $\sin \theta$ has an inverse ($\cos^{-1} \theta$). However, for the effective sampling of the radial part of the Woods-Saxon distribution one needs to use the method introduced in the previous section. As there are two clearly distinct behaviours in the radial part, the quadratic rise at small $r$ and the exponential tail at large $r$, it is useful to divide the radial distribution into two separate regions:
\begin{itemize}
\item $r\le R_A$: At small $r$ the exponential term in the denominator is rather small so a good choice for $g_1(r)$ can be found by estimating
\begin{equation}
\frac{r^2}{1+\exp[(r-R_A)/d]} < r^2 \equiv g_1(r).
\end{equation}
This choice clearly has a known primitive function $(G_1(r)=r^3/3)$ which has also an inverse $(G^{-1}_1(r)=3\, r^{1/3})$ from which the values for $r$ can be sampled for $0\le r \le R_A$.
\item $r > R_A$: This part is a bit trickier. As the dominant behaviour here is the exponential fall in $r$ one can first notice that
\begin{equation}
\frac{r^2}{1+\exp[(r-R_A)/d]} < r^2 \exp[-(r-R_A)/d].
\end{equation}
Also this has a primitive function but not an invertible one. However, by using the  equation (\ref{eq:RA}) for the $R_A$ and $d=0.54\,\mathrm{fm}$, one can find that
\begin{equation}
\frac{r^2}{1+\exp[(r-R_A)/d]} < R_A^2 \exp[-(r-R_A)/(3\,d)] \equiv g_2(r)
\label{eq:ws_g2}
\end{equation}
holds for $A\ge3$. As this have now an invertible primitive function, it can be used for sampling at $r > R_A$.
\end{itemize}
Figure \ref{fig:woodssaxon_approx} shows the chosen $g_1(r)$ and $g_2(r)$ together with the exact radial Woods-Saxon distribution for $R_A=6.49\,\mathrm{fm}$, corresponding the radius of the Pb nucleus. One can notice that the sampling could be further optimized for the given nucleus by taking e.g. $g_2(r) = R_A^2 \exp[-(r-R_A)/(2\,d)]$, but as this would violate the $g(r)\ge f(r)$ condition at small values of $A$, the $g_2(r)$ defined in equation \ref{eq:ws_g2} is used for the studies in this thesis.

The sampling algorithm above does allow the nucleons to overlap with each other, which introduces potentially larger thickness fluctuations that are realistic. A simple rejection of nucleons that overlap with the nucleons generated earlier would introduce some bias to the configuration and the original Woods-Saxon distribution would not be reproduced. In Ref.~\cite{Alvioli:2009ab} this problem was addressed by taking into account the nucleon-nucleon correlations. However, as the effects due to correlations were found to be small in Ref.~\cite{Alvioli:2011sk} for the quantities studied here, the issue is not addressed further in this thesis.

\subsection{Centrality classes from MC Glauber}

\begin{figure}[htb]
\centering
\includegraphics[width=0.8\textwidth]{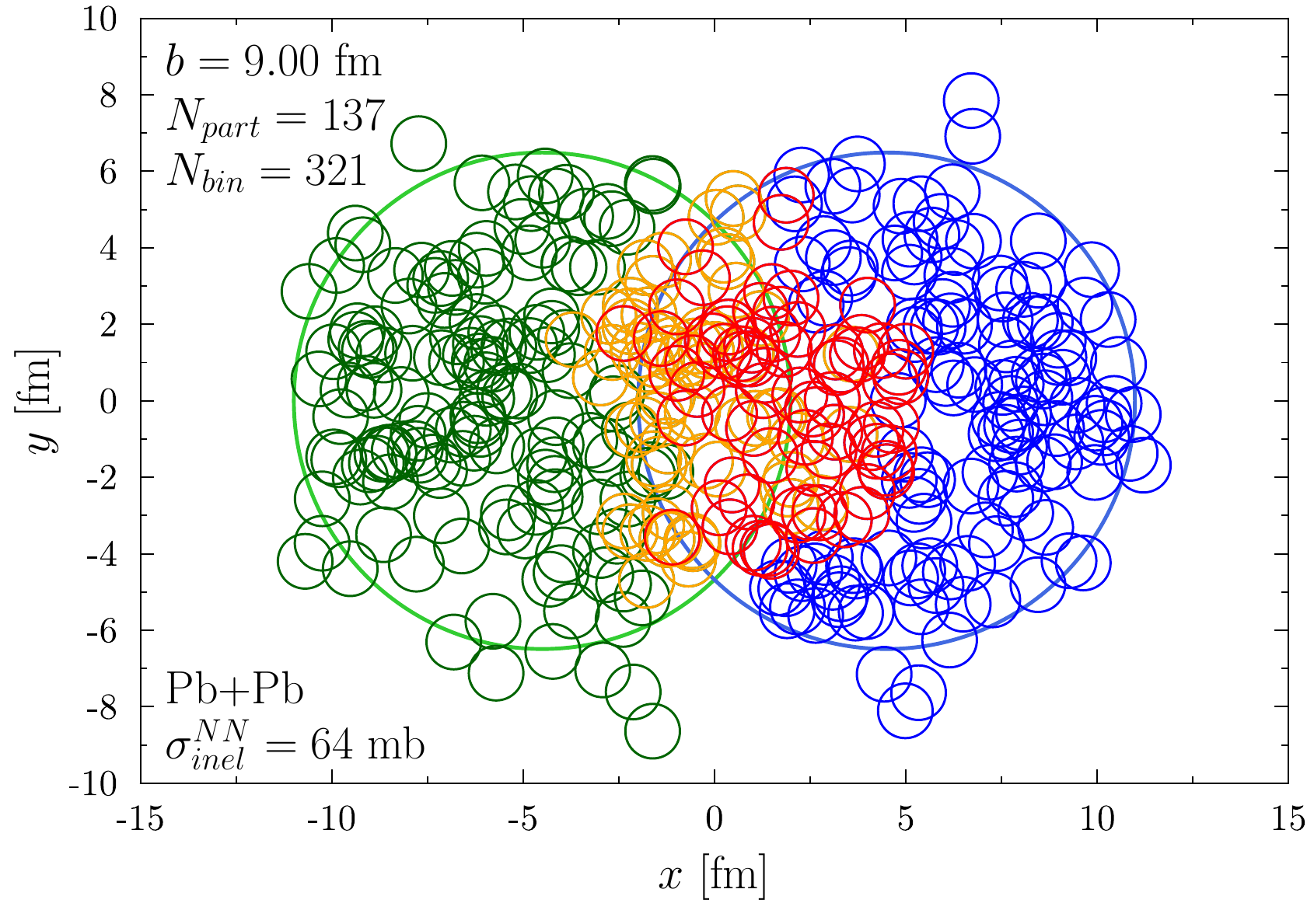}
\caption{A Pb+Pb collision at $\sqrt{s}=2.76\,\mathrm{TeV}$ with $b=9.00\,\mathrm{fm}$. The small green (blue) circles are the nucleons in the target (projectile) nucleus that did not collide and the yellow (red) circles are the collided nucleons. The large circles express the size of the nuclei.}
\label{fig:PbPb_coll}
\end{figure}

As the impact parameter in the MC Glauber model does not define the collision geometry unambiguously due to the fluctuations in the nucleon configurations, the centrality classification should be based on some other quantity. The two frequently used possibilities here that are related to the measured event multiplicity are the number of binary collisions $N_{bin}$ and the total number of participating nucleons $N_{part}$, which both can be calculated in each event after the collision is generated. To generate a collision one needs first to generate the colliding nuclei using e.g. the sampling procedure introduced above, and then generate the impact parameter for the collision. The impact parameter is sampled from $\mathrm{d}\sigma/\mathrm{d}b = 2 \pi b$ which, in principle, does not have an upper limit. However, as the tail of the Woods-Saxon distribution falls exponentially, the correction from limiting the impact parameter space to some finite value is negligible when the upper limit is sufficiently high. The adequate value for the upper limit depends on the considered nuclei and also slightly on the collision energy. When the nuclei and the impact parameter are generated one goes through all the nucleons in the projectile nucleus and counts the number of collisions with the nucleons in the target nucleus. In the black disc approximation a nucleon-nucleon collision takes place if
\begin{equation}
d_{ij}\le \sqrt{\frac{\sigma_{inel}^{NN}}{\pi}},
\end{equation}
where $d_{ij}$ is the transverse distance between the centers of the nucleons $i$ and $j$. The procedure is illustrated in figure \ref{fig:PbPb_coll} which shows a Pb+Pb collision at $b=9.0\,\mathrm{fm}$ using $\sigma_{inel}^{NN}=64\,\mathrm{mb}$, which corresponds to the LHC collision energy $\sqrt{s_{NN}}=2.76\,\mathrm{TeV}$. 

To define the centrality classes in the MC Glauber model one needs to generate an ensemble of collisions with the given kinematics and calculate the $N_{bin}$ and $N_{part}$ for each collision. Then one sorts the collisions based on the chosen criterion ($N_{bin}$ or $N_{part}$) and classifies the collisions according to the desired centrality bins. To compare the different centrality definitions table \ref{tab:PbPb_centralities} shows the average impact parameters and numbers of binary collisions for different centrality classes in Pb+Pb collisions at $\sqrt{s_{NN}}=2.76\,\mathrm{TeV}$ defined with the Optical Glauber model and with the MC Glauber model using $N_{bin}$ and $N_{part}$.
\begin{table}[tbh]
\caption{The average impact parameters and numbers of binary collisions from the Optical Glauber model and MC Glauber (MCG) model using $N_{part}$ and $N_{bin}$ to define the centrality classes. Five centrality classes in Pb+Pb collisions at $\sqrt{s_{NN}}=2.76\,\mathrm{TeV}$ are listed.}
\begin{center}
\begin{tabular}{ccccccc}
\hline
Centrality & \multicolumn{2}{c}{Optical Glauber} & \multicolumn{2}{c}{MCG $(N_{bin})$} & \multicolumn{2}{c}{MCG $(N_{part})$}\\
& $\langle b \rangle \,\mathrm{[fm]}$ & $\langle N_{bin} \rangle$  
& $\langle b \rangle \,\mathrm{[fm]}$ & $\langle N_{bin} \rangle$ 
& $\langle b \rangle \,\mathrm{[fm]}$ & $\langle N_{bin} \rangle$ \\
\hline
$0-20~\%$   & 4.637 & 1255  & 4.595 & 1261  & 4.589 & 1257  \\
$20-40~\%$  & 8.478 & 431.0 & 8.347 & 457.2 & 8.352 & 458.8 \\
$40-60~\%$  & 10.98 & 113.3 & 10.81 & 133.7 & 10.81 & 135.5 \\
$60-80~\%$  & 13.00 & 19.08 & 12.85 & 27.28 & 12.84 & 27.94 \\
$80-100~\%$ & 15.00 & 2.417 & 14.66 & 3.532 & 14.66 & 3.596 \\
\hline
\end{tabular}
\end{center}
\label{tab:PbPb_centralities}
\end{table}
The comparison shows that with the MC Glauber model the difference between the centrality classes defined using $N_{bin}$ or $N_{part}$ is very small for the considered quantities. This is true also for the comparison between the Optical and MC Glauber in central collisions but for more peripheral collisions the difference in the $\langle N_{bin} \rangle$ becomes notable. This suggests that at large $b$ there are some differences between these two approaches.




\chapter{Spatial dependence of nuclear PDFs}
\label{chap:spatial_nPDFs}

The present nuclear PDF fits have so far considered only centrality averaged, minimum bias, data. In fact the purely inclusive DIS data, which make most of the data points in the present global analyses, is not sensitive to the collision geometry as only the scattered lepton is measured. However, as the nuclear thickness depends on the transverse position in the nucleus it could be expected that also the nuclear modification of the PDF depends on the point in which it is probed. As the measured centrality dependence of different observables in heavy-ion collisions can provide important constraints for different theoretical models, the lack of spatial dependence in the nPDFs global fits is clearly a shortcoming. In the article [I] in this thesis, with the motivation of studying hard process cross sections in different centrality classes in more detail, we introduced two spatially dependent nPDF sets which are reviewed in this Chapter. 

\section{Previous works}

There have been several attempts to model the spatial transverse coordinate dependence of the nPDFs. In Ref.~\cite{Frankfurt:2011cs} (and references therein) the spatial dependence of the small $x$ shadowing was obtained using the PDFs derived from diffractive HERA data and a generalized Gribov-Glauber theory. In spite of being so far the only dynamically derived spatially dependent nPDFs, the FGS10 nPDFs \cite{Frankfurt:2011cs} have not been subjected to a full global analysis (such as in e.g. EPS09 and DSSZ). For studies of the nPDFs in the longitudinal spatial direction, see e.g. Ref. \cite{Vanttinen:1998iz}.

In somewhat simpler (more phenomenological) approaches in Refs.~\cite{Eskola:1991ec, Emel'yanov:1999bn, Klein:2003dj, Vogt:2004hf}, motivated by the small-$x$ studies referred to above, the authors postulated that the nuclear modification depends linearly on the nuclear thickness:
\begin{equation}
r_i^A(x,Q^2,\mathbf{s}) = 1 + c_i^A(x,Q^2)\, T_A(\mathbf{s}),
\end{equation}
where the $T_A(\mathbf{s})$ is calculated as in equation (\ref{eq:T_A}). The value for the parameter $c_i^A(x,Q^2)$ at given $x$ and $Q^2$ can in this case be obtained from the condition
\begin{equation}
\frac{1}{A}\int \mathrm{d}\mathbf{s}\, T_A(\mathbf{s})r^A_i(x,Q^2,\mathbf{s}) \equiv R^A_i(x,Q^2),
\label{eq:spatial_average}
\end{equation}
where the minimum bias (transversely averaged) nuclear modification $R^A_i(x,Q^2)$ is obtained for each parton type $i$ and nucleus $A$ from a chosen global analysis (e.g. EKS98 was used in Ref.~\cite{Emel'yanov:1999bn}). The condition (\ref{eq:spatial_average}) ensures that the spatially averaged quantities are correctly reproduced and the parameter $c_i^A(x,Q^2)$ is then given by
\begin{equation}
c_i^A(x,Q^2) = \frac{A}{T_{AA}(0)}\left[ R_i^A(x,Q^2) - 1 \right].
\end{equation}
In Ref.~\cite{Nagle:2010ix} the authors studied also a quadratic and exponential dependence on the nuclear thickness function for the nPDFs in the case of $J/\Psi$ production in d+Au collisions. A linear dependence on the nuclear thickness is included also into the \texttt{HIJING} event generator \cite{Gyulassy:1994ew, Li:2001xa} but there the nucleus is described by a hard sphere.

The problem with these effective approaches is that the parameter $c_i^A(x,Q^2)$ remains dependent on the mass number $A$ as shown in figure 2 in the article [I]. This implies that a simple linear dependence on $T_A(\mathbf{s})$ is not enough to capture the $A$ dependence obtained in the global analyses with nuclear data but a more involved form for the spatial dependence is required.

\section{The framework}
\label{sec:nPDF_framewrok}

Motivated by the earlier studies, we assumed in [I] that the spatial dependence is related to the thickness function $T_A(\mathbf{s})$ through a power series form
\begin{equation}
r_i^A(x,Q^2,\mathbf{s}) = 1 + \sum_{j=1}^N c^j_i(x,Q^2)\, [T_A(\mathbf{s})]^j.
\label{eq:r_a}
\end{equation}
Now the parameters $c^j_i(x,Q^2)$ are $A$ independent and their values can be obtained for each $x$ and $Q^2$ by fitting to the $A$ dependence of the minimum bias modification $R_i^A(x,Q^2)$, i.e. by minimizing the $\chi^2$ defined as
\begin{equation}
\chi^2(x,Q^2)=\sum_A \left[\frac{R_i^A(x,Q^2)-\frac{1}{A}\int\mathrm{d}^2 \mathbf{s}\,T_A(\mathbf{s})r_i^A(x,Q^2,\mathbf{s})}{W_i^A(x,Q^2)}\right]^2,
\label{eq:fitting}
\end{equation}
where $W_i^A(x,Q^2)$ is a weight factor, an artificial error, that is chosen to obtain an optimal fit. The required number of terms $N$ in the power series (\ref{eq:r_a}) can be obtained by studying the $\chi^2$ values with varying $N$.

\section{Results}

In the article [I] we used two globally analyzed nPDF sets for which we performed the fitting (\ref{eq:fitting}), EPS09 \cite{Eskola:2009uj} and EKS98 \cite{Eskola:1998df, Eskola:1998iy}. For these sets we found that with $N=4$ we can reproduce the $A$ dependence (see figure 4 in [I]) of both sets accurately for $A\ge 16$ using weights $W_i^A(x,Q^2) = R_A^i(x,Q^2)-1$ for EKS98 and $W_i^A(x,Q^2) = 1$ for EPS09. The weights were chosen to produce optimal fits in each case. The $(x,Q^2)$ grid for which the fitting was performed and the kinematic reach in $x$ and $Q^2$ was chosen to match those in the original EPS09 and EKS98 analyses, respectively. For the EPS09 case the fit was performed for both the LO and NLO sets including also the 30 error sets in both cases.

\begin{figure}[htb]
\begin{center}
\includegraphics[width=0.49\textwidth]{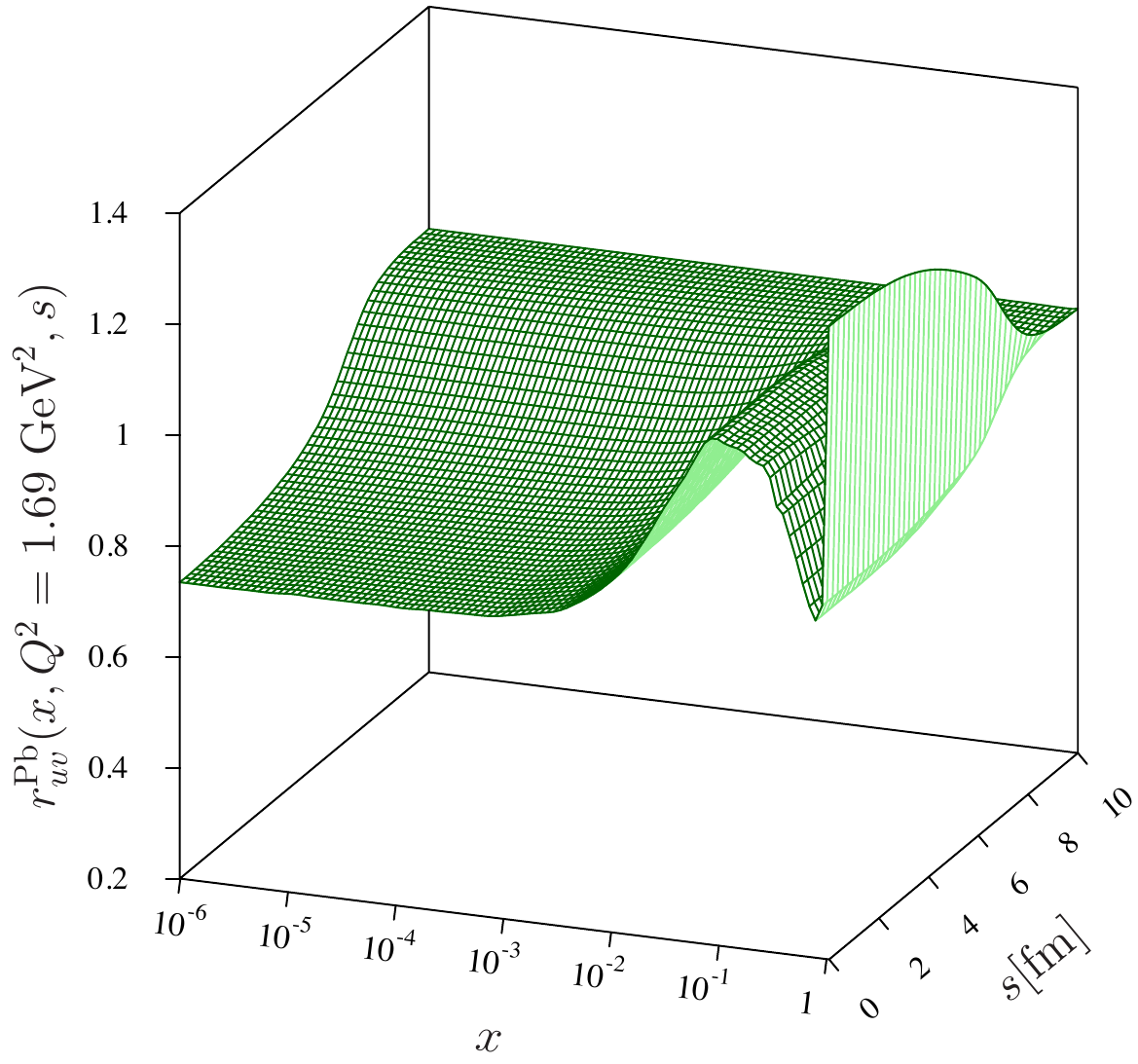}
\includegraphics[width=0.49\textwidth]{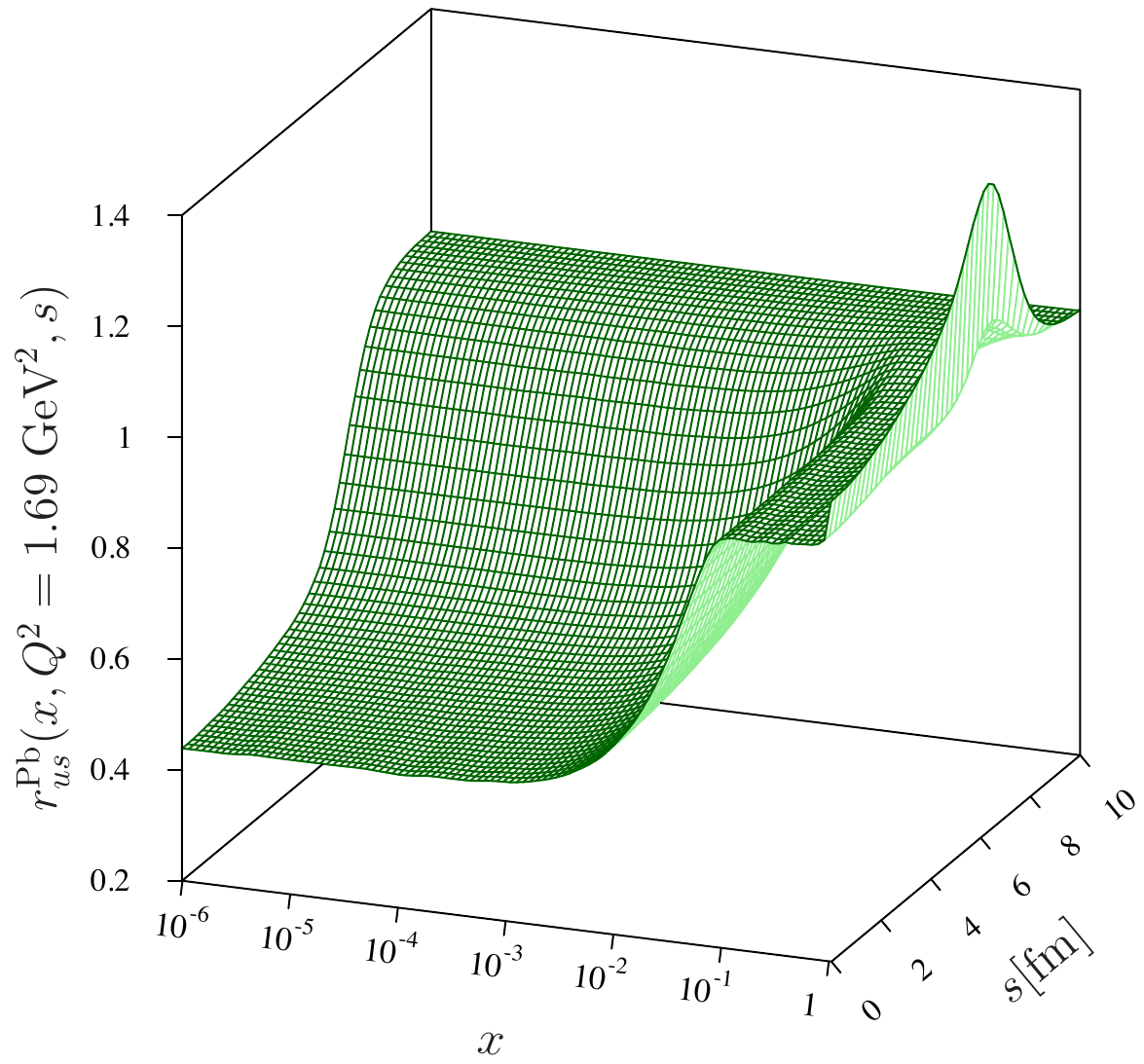}
\includegraphics[width=0.49\textwidth]{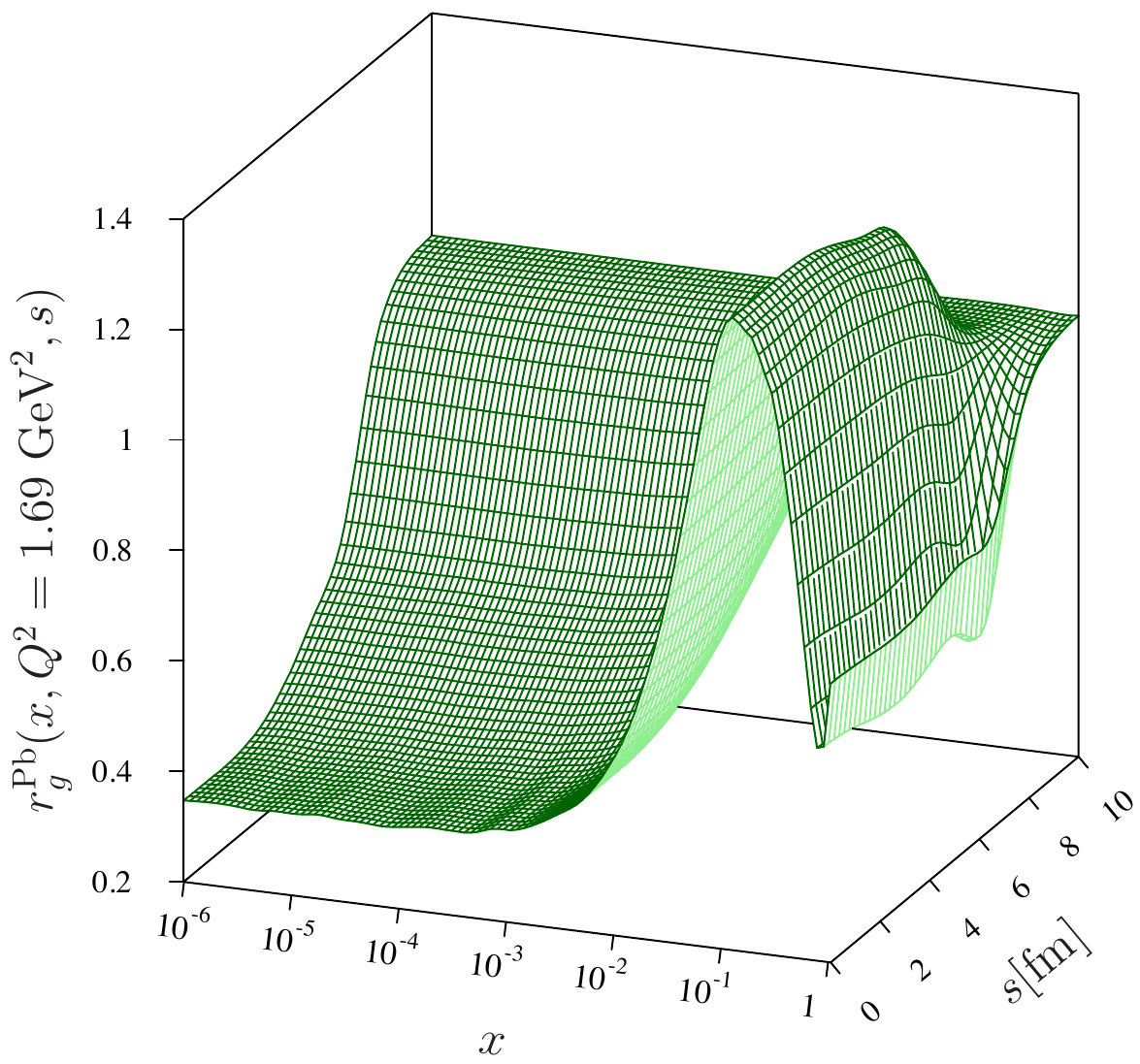}
\caption{The $x$ and $s$ dependence of the EPS09s NLO nuclear modification for $\rm u_V$ (upper left), $\rm u_S$ (upper right) and gluons (lower plot) at scale $Q^2=1.69\,\mathrm{GeV^2}$ for $A=208$. Figures from [I].}
\label{fig:rsx3d}
\end{center}
\end{figure}

As an outcome of the fitting procedure we obtained the spatially dependent nPDF sets, named as \texttt{EPS09s} and \texttt{EKS98s}, where the ``\texttt{s}'' refers to ``spatial''. These sets are one of the main results of this thesis and they are available for public use\footnote{\url{https://www.jyu.fi/fysiikka/en/research/highenergy/urhic/nPDFs}}. As an example, the figure \ref{fig:rsx3d} shows the spatial dependence of the nuclear modification of the Pb-nucleus in NLO \texttt{EPS09s} for $\rm u_V$, $\rm u_S$ and gluons at the initial scale $Q^2=1.69\,\mathrm{GeV^2}$. For the corresponding figures for EPS09sLO and EKS98s, see [I]. As can be seen in the figures, the spatial dependence turns out as expected: The nuclear effects are larger at the center of the nucleus and disappear towards the edge. An interesting feature, which follows from the behaviour of the thickness function is that the nuclear effects remain significant over several fm's (until $s\sim R_A$) and then they rapidly disappear towards the edge of the nucleus.

\section{Applications}

We have applied the new spatially dependent nuclear PDFs in the articles [I], [II], and [III] of this thesis to calculate the centrality dependence of different hard process observables in nuclear collisions. The applications are discussed also in various conference proceedings \cite{Helenius:2012wk, Helenius:2012zd, Helenius:2012ny, Helenius:2014oea, Helenius:2014cva}.

\subsection{Centrality dependent nuclear modification ratio}

When we replace the minimum bias nPDFs with the spatially dependent ones, the invariant yield of hard particles $k$ at a given impact parameter can be calculated from (see the article [I] for more details)
\begin{align}
\mathrm{d} N&^{AB\rightarrow k + X}(\mathbf{b}) = \sum\limits_{i,j,X'}   \int \mathrm{d}^2\mathbf{s_1} \, T_A(\mathbf{s_1}) \, r_i^A(x_1,Q^2,\mathbf{s_1}) \, f_i(x_1,Q^2) \, \label{eq:dN_AB} \\ & \otimes \int \mathrm{d}^2\mathbf{s_2} \, T_B(\mathbf{s_2}) \, r_j^B(x_2,Q^2,\mathbf{s_2}) \, f_{j}(x_2,Q^2) \otimes  \mathrm{d}\hat{\sigma}^{ij\rightarrow k + X'}
\delta(\mathbf{s_2} - \mathbf{s_1} - \mathbf{b}),\notag 
\end{align}
where $\mathrm{d}\hat{\sigma}^{ij\rightarrow k + X'}$ is the partonic piece containing also the convolution with the fragmentation functions if required for the process, and the nuclear thickness functions are calculated according to equation (\ref{eq:T_A}).

When the centrality classes are defined in terms of impact parameters using the Optical Glauber model (see tables 1, 2, and 3 in [I] and table 1 in [II]), one can define the nuclear modification ratio at a given centrality class $b\in[b_1,b_2]$ as
\begin{equation}
R_{AB}^{k}(p_T,\eta; b_1,b_2) \equiv \dfrac{\left\langle\dfrac{\mathrm{d}^2 N_{AB}^{k}}{\mathrm{d}p_T \mathrm{d}\eta}\right\rangle_{b_1,b_2}}{\langle N_{bin}^{AB} \rangle_{b_1,b_2} \dfrac{1} {\sigma^{NN}_{inel}}\dfrac{\mathrm{d}^2\sigma_{\rm pp}^{k}}{\mathrm{d}p_T \mathrm{d}\eta}},
\label{eq:R_AB_b1b2}
\end{equation}
where $\langle N_{bin}^{AB} \rangle_{b_1,b_2}$ is calculated as in equation (\ref{eq:N_bin_ave}) and the averaging for the differential yield can be done similarly. Working out the averages gives a simple ratio for the nuclear modification
\begin{equation}
R_{AB}^{k}(p_T,\eta; b_1,b_2) = \dfrac{\int_{b_1}^{b_2} \mathrm{d}^2 \mathbf{b} \dfrac{\mathrm{d}^2 N_{AB}^{k}(\mathbf{b})}{\mathrm{d}p_T \mathrm{d}\eta} }{ \int_{b_1}^{b_2} \mathrm{d}^2 \mathbf{b} \,T_{AA}(\mathbf{b})\dfrac{\mathrm{d}^2\sigma_{\rm pp}^{k}}{\mathrm{d}p_T \mathrm{d}\eta}}
\label{eq:R_AB_b1b2_2}
\end{equation}
from which the minimum bias nuclear modification ratio can be recovered by integrating over whole impact parameter space, $b_1\rightarrow 0$ and $b_2\rightarrow \infty$.

\subsection{Hadron production in p(d)+$A$}

In the article [I] we studied the LO inclusive jet production in Au+Au collisions at RHIC and Pb+Pb collisions at the LHC at different centralities, which could serve as a baseline for energy loss studies. Also the centrality dependence of the nuclear modification factor for inclusive pion production in d+Au collisions at RHIC and p+Pb collisions at the LHC was calculated at NLO using the \texttt{INCNLO} code discussed in section \ref{subsec:incnlo}. 

\begin{figure}[htb]
\begin{center}
\includegraphics[width=\textwidth]{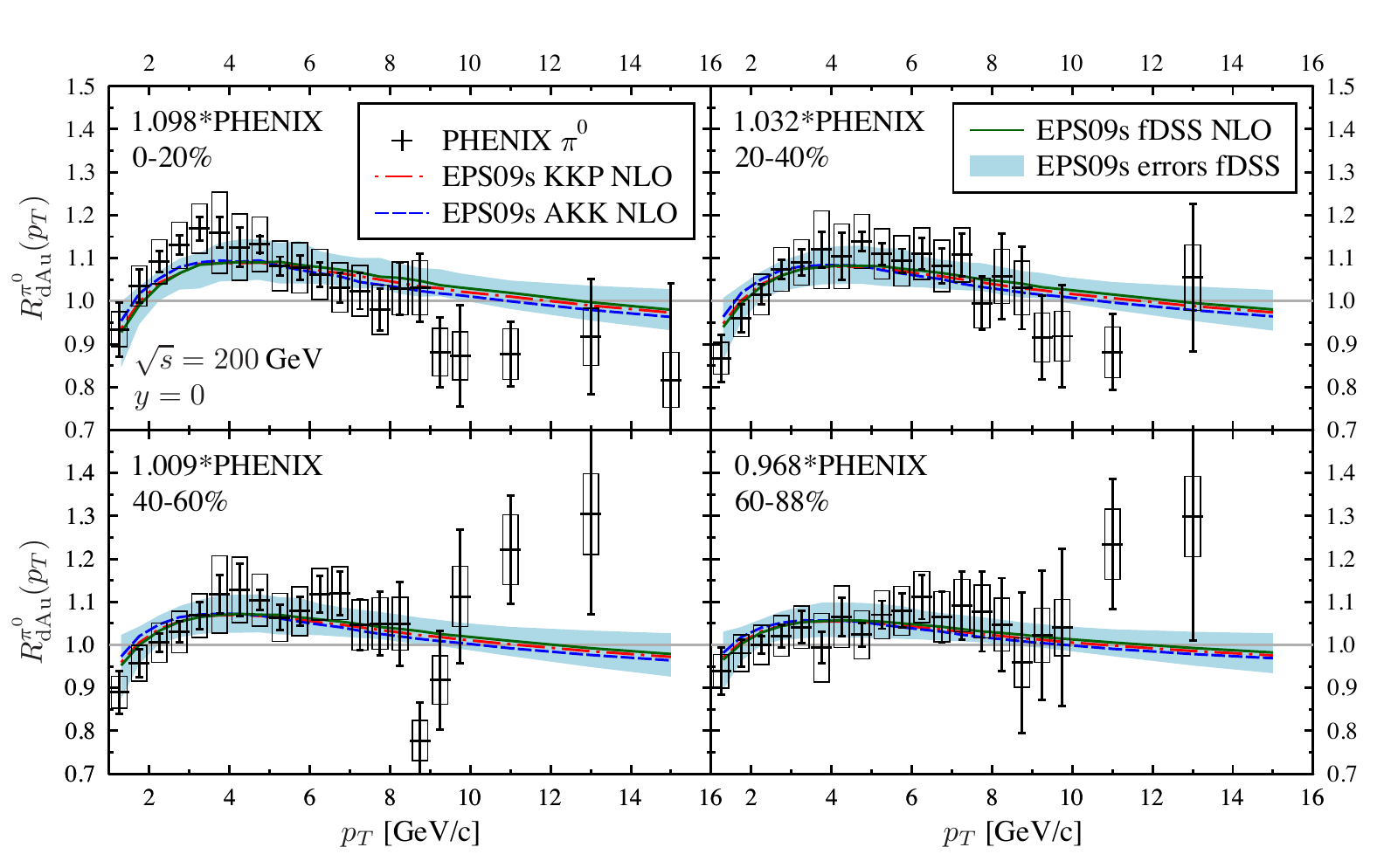}
\caption{The nuclear modification ratio for neutral pion production in d+Au collisions at $y=0$ and $\sqrt{s_{NN}}=200\,\mathrm{GeV}$ in four centrality classes (different panels). The NLO calculations are done with three FFs, KKP (red dot-dashed), AKK05 (blue dashed) and DSS (green solid), the blue band quantifies the uncertainty from EPS09s, and the data points are from PHENIX \cite{Adler:2006wg} at $|\eta|<0.35$. Note also that the data points in each panel have been shifted vertically by the indicated amount within the normalization uncertainties. Figure from [I].}
\label{fig:R_dAu_y0}
\end{center}
\end{figure}

The first main result was the comparison between PHENIX data \cite{Adler:2006wg} and our NLO calculation for the $R^{\pi^0}_{\rm dAu}(p_T,\eta)$ for neutral pion production at mid-rapidity for different centrality classes, shown in figure \ref{fig:R_dAu_y0}. When all the uncertainties were properly taken into account, the calculation was found to be consistent with the measurement. The centrality dependence turned out to be rather weak in both data and calculation.
However, the new preliminary data from PHENIX \cite{Sahlmueller:2012ru} from the 2008 run with increased $p_T$ reach showed a surprising centrality dependence at large $p_T$: The slight suppression in most central collisions turns into a clear enhancement in the peripheral collisions. This surprising and not yet fully understood feature is not consistent with the centrality dependence obtained from our nPDFs which is shown in figure 3 of Ref.~\cite{Sahlmueller:2012ru}. It has been proposed that the unexpected behaviour could follow from a correlation of high-$p_T$ particle production at mid-rapidity and particle production at large rapidities that is used for the centrality determination (see e.g. Ref.~\cite{Adare:2013nff}). This in turn implies that the centrality determination based on the event multiplicities is not as well under control in the p/d+$A$ collisions as it is in $A$+$A$ collisions, i.e. that multiplicity alone does not correlate well enough with the collision geometry.

\begin{figure}[htb]
\begin{center}
\includegraphics[width=\textwidth]{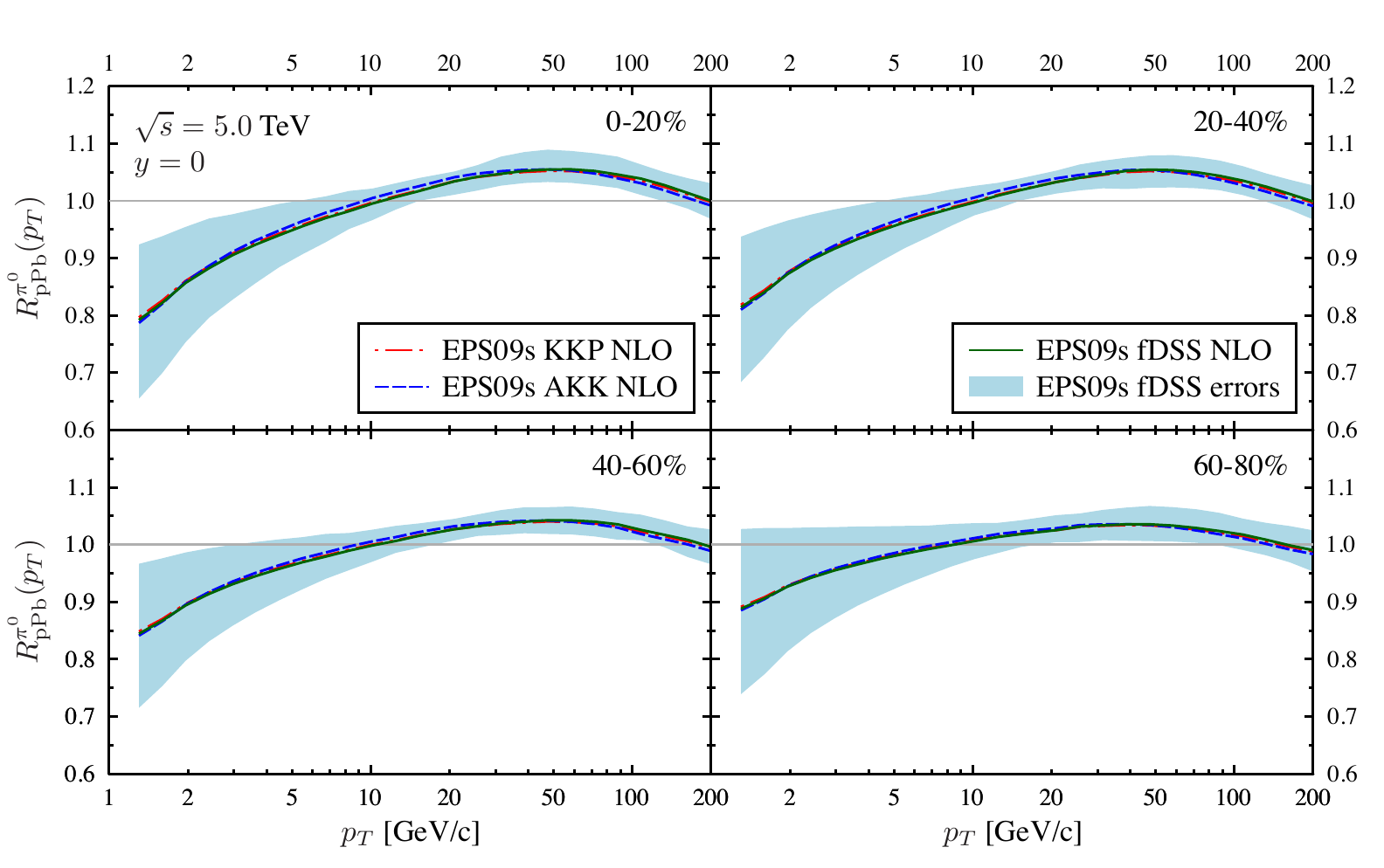}
\caption{The nuclear modification ratio for neutral pion production in p+Pb collisions at $y=0$ and $\sqrt{s_{NN}}=5.0\,\mathrm{TeV}$ in four centrality classes (different panels). The NLO calculations are done with three FFs, KKP (red dot-dashed), AKK05 (blue dashed) and DSS (green solid), the blue band quantifies the uncertainty from the EPS09s nPDFs. Figure from [I].}
\label{fig:R_pPb_y0}
\end{center}
\end{figure}

The other main result in the article [I] (included also in the compilation of p+Pb predictions in Ref.~\cite{Albacete:2013ei}) was the prediction for the nuclear modification factor for neutral pion production at different centrality classes in p+Pb collisions at the LHC, shown in figure \ref{fig:R_pPb_y0}.
Here the centrality dependence at small $p_T$ was found to be slightly stronger than in d+Au collisions at RHIC but still a very precise measurement would be needed to confirm the predicted effect. However, the centrality determination in these p+Pb collisions has proved to be very sensitive to the assumptions in the Glauber model and no centrality dependent data have been published for the considered observable so far. At the moment it is unclear whether the measured multiplicity can be related to the collision geometry at all. Nevertheless, as discussed in section \ref{sec:FF_pA}, our minimum bias prediction for the neutral pions (figure 17 in [I]) seems to agree very well with the preliminary ALICE data \cite{ALICE_RpPb_pi}.

\begin{figure}[htb]
\begin{center}
\includegraphics[width=\textwidth]{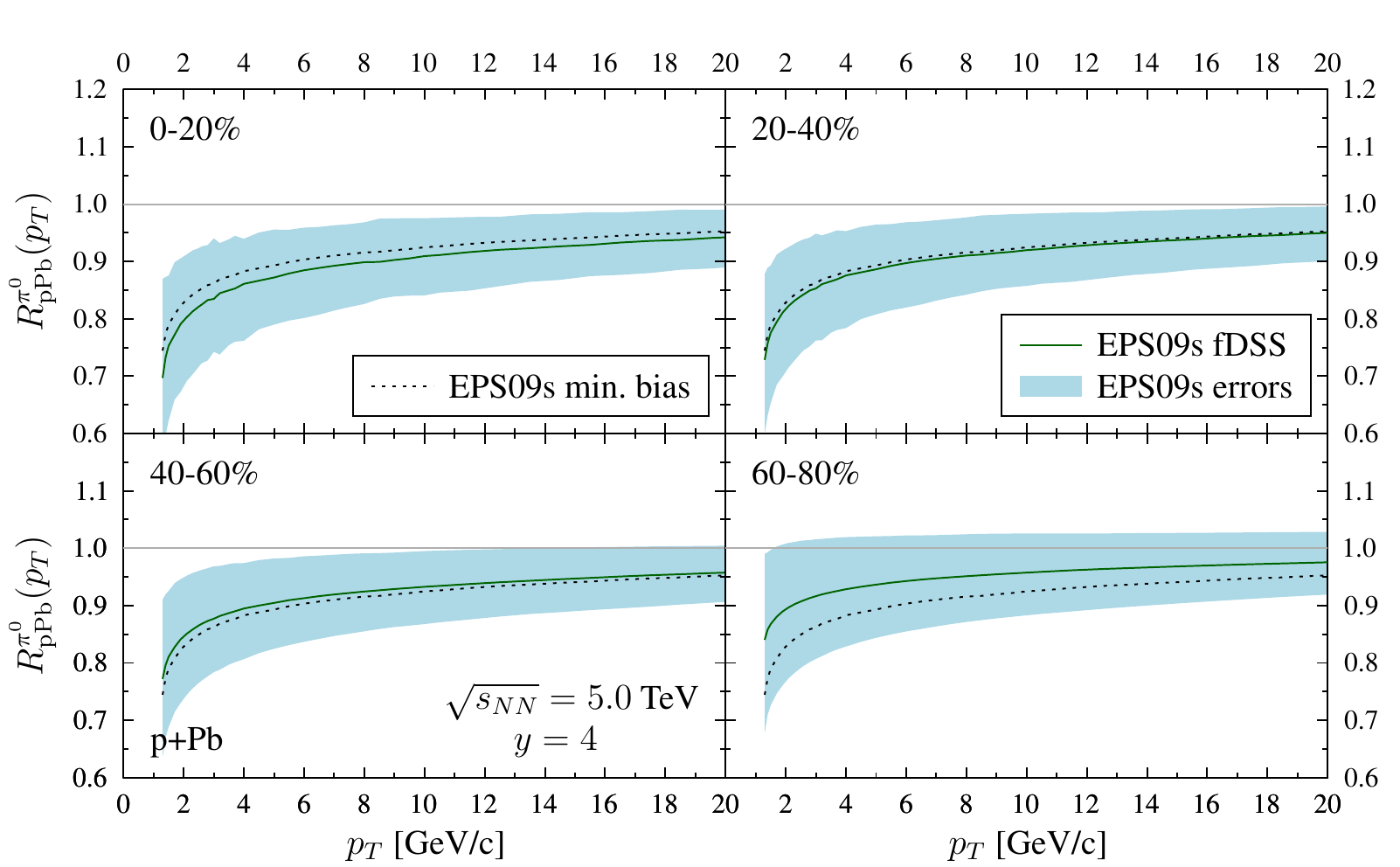}
\caption{The nuclear modification ratio for neutral pion production in p+Pb collisions at $y=4$ and $\sqrt{s_{NN}}=5.0\,\mathrm{TeV}$ in four centrality classes (different panels). The NLO calculations are done with  the DSS FFs (green solid) and the blue band quantifies the uncertainty from the EPS09s nPDFs. Also the minimum bias result is shown (dotted black) in each panel from comparison. Published also in Ref.~\cite{Helenius:2014oea}}
\label{fig:R_pPb_y4}
\end{center}
\end{figure}

After solving the numerical problems in the \texttt{INCNLO} code discussed in section \ref{subsec:incnlo}, we were able to calculate also the centrality dependence of the nuclear modification factors at forward rapidities in p+Pb collisions at the LHC. The result for $\pi^0$ at $y=4$ is shown in figure \ref{fig:R_pPb_y4}.
Here the predicted centrality dependence is stronger than at mid-rapidities as the inclusive pion production at forward rapidities is more sensitive to the small-$x$ region where the nuclear modifications of the PDF are more apparent. However, the effects are still of the order $10\,\%$ and accurate measurements would be required to observe the effect.

\subsection{Direct photon production in nuclear collisions}
\label{subsec:dir_photons_centrality}

In the articles [II] and [III] of this thesis we studied inclusive direct photon production in p/d+$A$ and $A$+$A$ collisions at RHIC and LHC. In the article [II] the goal was to study how the centrality dependence of the nuclear modification ratio turns out for the direct photons and compare this to data where possible.

\begin{figure}[htb]
\begin{center}
\includegraphics[width=\textwidth]{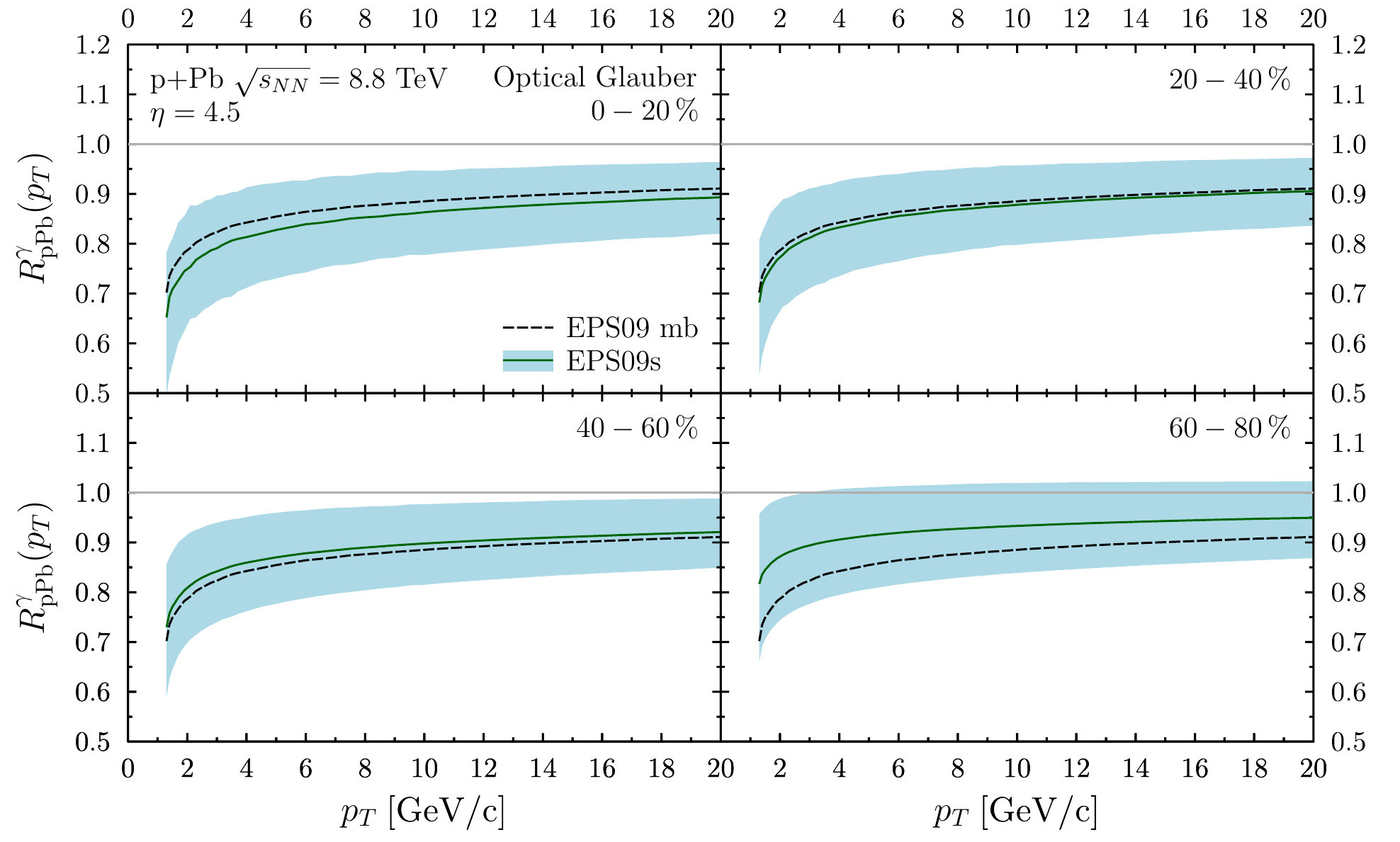}
\caption{The nuclear modification ratio for direct photon production in p+Pb collisions at $y=4.5$ and $\sqrt{s_{NN}}=8.8\,\mathrm{TeV}$ in four centrality classes (different panels) together with the minimum bias result (dashed black). The parton-to-photon FFs are taken from the BFG analysis (set II). Published also in Refs.~\cite{Helenius:2014oea, Helenius:2014cva}}
\label{fig:R_pPb_gamma_y45}
\end{center}
\end{figure}

For the p/d+$A$ collisions the centrality dependence of direct photons at mid-rapidity was observed to be very similar as for pions. The isospin effect has some effects to nuclear modification ratio but does not affect the centrality dependence. As discussed in Chapter \ref{chap:direct_photon} and in the article [V] of this thesis, the direct photons at forward rapidities are more sensitive to the small-$x$ region than the inclusive hadrons at the same rapidity, which results in slightly stronger nuclear effects. Thus one could hope to observe also a stronger centrality dependence here. However, as shown in figure \ref{fig:R_pPb_gamma_y45}, the resulting centrality dependence is only slightly stronger here as the $x$ dependence of EPS09s is rather mild at $x<0.01$. The collision energy $\sqrt{s_{NN}}=8.8\,\mathrm{TeV}$ here corresponds to the nominal LHC energy for p+p collisions ($\sqrt{s}=14\,\mathrm{TeV}$).

So far there are no centrality dependent direct photon data from p/d+$A$ collisions. However, as the photons do not interact directly with the strongly interacting medium which is present in $A$+$A$ collisions at sufficiently high energies, also these collisions can be used to study the centrality dependence of direct photon production. Compared to p/d+$A$ collisions there are two advantages from the nPDF viewpoint: First, as both colliding particles are nuclei, the effects from nuclear modifications in the PDFs should be more pronounced. Second, the centrality determination in $A$+$A$ collisions is in a better control as there exists a clear correlation between the event multiplicity and the collision geometry. However, there are couple of disadvantages too. Even though there are no direct interactions between the photon and the medium, there can be indirect effects, e.g. in the fragmentation component if the momentum of the parent parton is modified before the photon is emitted (see e.g. Refs.~\cite{Jeon:2002dv, Arleo:2006xb}). Also, a large contribution from thermal photons at $p_T<4\,\mathrm{GeV/c}$ has been observed \cite{Wilde:2012wc, Adare:2014fwh}, which prevents the nPDF studies in this region. These are not considered here as the goal is to study how the centrality dependence from spatially dependent nPDFs turns out. The pQCD baseline calculations presented here can, however, be used to quantify the effects from these different medium induced modifications.

\begin{figure}[htb]
\begin{center}
\includegraphics[width=0.9\textwidth]{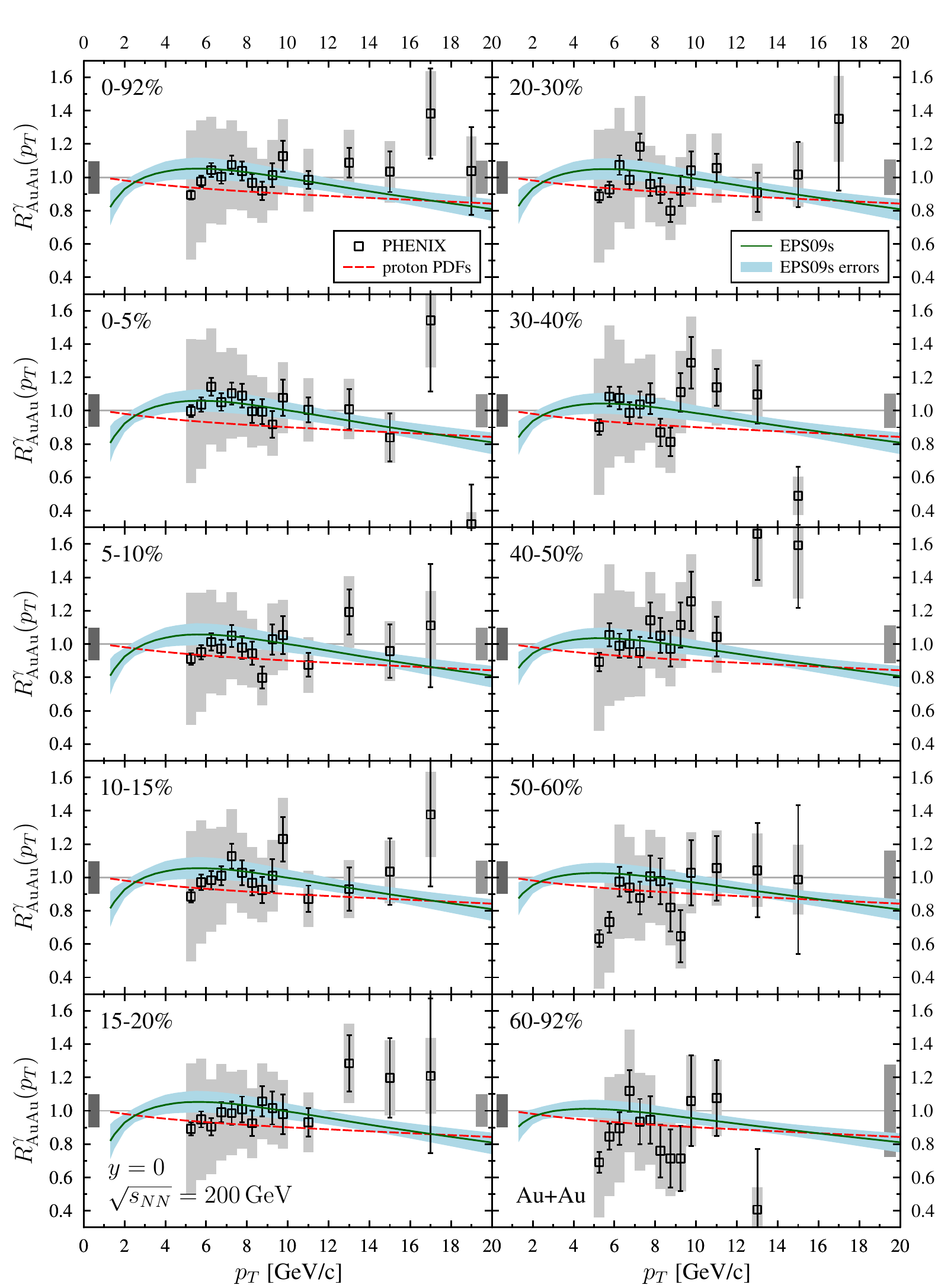}
\caption{The nuclear modification ratio for direct photon production in Au+Au collisions at mid-rapidity with $\sqrt{s_{NN}}=200\,\mathrm{GeV}$ in ten centrality classes (different panels) from our NLO calculation with EPS09s nPDFs (green solid) and their uncertainties (blue band) using the BFGII parton-to-photon FFs. The NLO calculation without the nuclear modifications of the PDFs is shown in each panel to quantify the isospin effect (red dashed). The data is from PHENIX \cite{Afanasiev:2012dg} and the gray boxes in each panel show the normalization uncertainties in the measurement. Figure from [II].}
\label{fig:R_AuAu_gamma_y0}
\end{center}
\end{figure}

Figure \ref{fig:R_AuAu_gamma_y0} shows the centrality dependence of the nuclear modification for direct photon production in Au+Au collisions in ten different centrality classes from the NLO calculation with the EPS09s nPDFs and a PHENIX measurement \cite{Afanasiev:2012dg}. The calculation without the nPDFs is also shown in each panel to quantify the isospin effect, which is centrality independent. The predicted mild centrality dependence is supported by the data: The slight enhancement around $p_T\sim 7\,\mathrm{GeV/c}$ due to the antishadowing in the most central collisions turns into a small suppression due to the isospin effect in peripheral collisions, although the systematic uncertainties in the data are large. As the data are consistent with the NLO pQCD calculations, the indirect, QCD-matter generated, modifications of the direct photon production at $p_T \ge 5\,\mathrm{GeV/c}$ are either small, or the different effects cancel out. Also the CMS data \cite{Chatrchyan:2012vq} for isolated photon nuclear modification ratio was found to be consistent with our (non-isolated) result, although the experimental uncertainties were there even larger, see figure 7 of [II].

\subsection{Thermal photons}

In the article [III] of this thesis the elliptic flow of thermal photons in $A$+$A$ collisions was studied within an event-by-event hydrodynamic framework. The thermal photons are mostly emitted in the early stages of the hydrodynamic evolution and are thus a promising tool to study the fluctuations in the initial nucleon configurations and the properties of the quark-gluon plasma.

\begin{figure}[thb]
\centering
\includegraphics[width=0.8\textwidth]{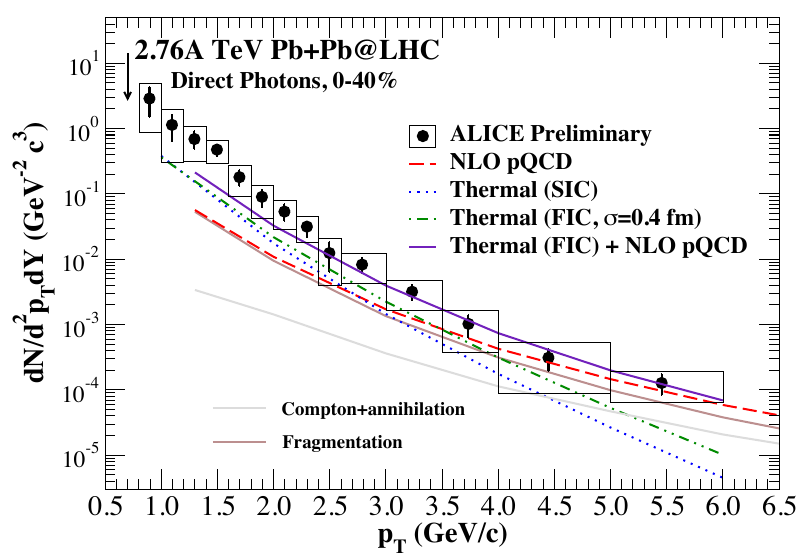}
\caption{The $p_T$ spectrum of direct photons in Pb+Pb collisions at $\sqrt{s_{NN}}=2.76\,\mathrm{GeV}$. The calculations include the thermal component from hydrodynamics with smooth (blue dotted) and fluctuating (green dot-dashed) initial conditions and the NLO pQCD contribution (dashed red) from prompt (gray solid) and fragmentation (brown solid) photons and the combined result (solid magenta). The preliminary data are from ALICE \cite{Wilde:2012wc}. Figure from [III].}
\label{fig:PbPb_spectra}
\end{figure}

The contribution from this thesis work to the thermal photon studies was the calculation of the direct photon spectra from the NLO pQCD using the spatially dependent nPDFs presented in this Chapter. Combining the thermal photon spectra with direct photons from pQCD the measured direct photon spectra are well reproduced at $p_T>2\,\mathrm{GeV/c}$ ($p_T>2.5\,\mathrm{GeV/c}$) for RHIC (LHC) data. Figure \ref{fig:PbPb_spectra} shows the comparison between the combined calculation and the preliminary ALICE data \cite{Wilde:2012wc} (for the corresponding figures for RHIC, see [III]). From the decomposition to the different contributions in the calculation one can notice that the thermal contribution is dominant at $p_T<3.5\,\mathrm{GeV/c}$ but above that the pQCD contribution takes over. It is worth mentioning that the centrality classification in the thermal photon calculation was based on the MC Glauber model while the Optical Glauber model was used for the pQCD part. However, as shown in table \ref{tab:PbPb_centralities}, when considering rather central $A$+$A$ collisions these two approaches give very similar results and they can be safely combined.

\begin{figure}[thb]
\centering
\includegraphics[width=0.8\textwidth]{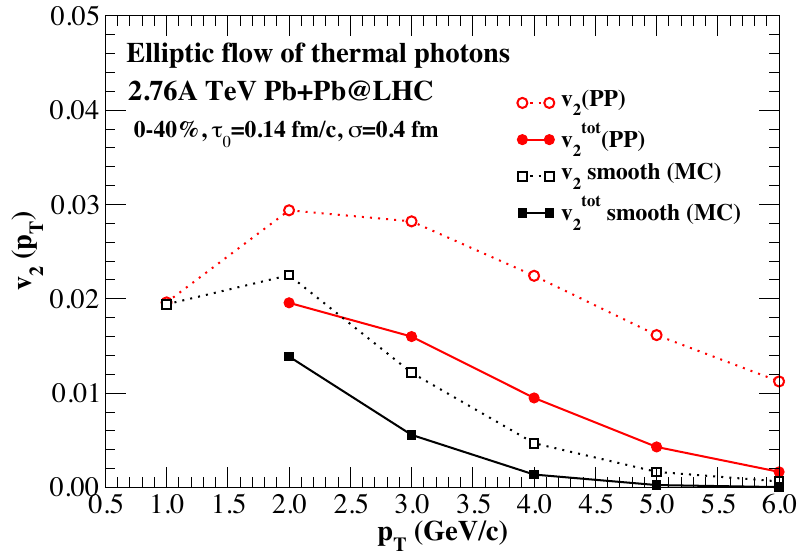}
\caption{The elliptic flow of direct photons from using smooth (black) and fluctuating (red) initial conditions for hydrodynamics. Also the elliptic flow from thermal photons only is shown for both cases (dotted), which demonstrates the big effect of the pQCD photons. Figure from [III].}
\label{fig:v2_gamma}
\end{figure}

The inclusion of the direct photons from pQCD does not affect only the total $p_T$ spectrum itself but also the elliptic flow, $v_2\equiv \langle \cos 2\phi \rangle$ (the second term in the Fourier decomposition) of the direct photons. As the photons from the hard pQCD scatterings are generated isotropically in the transverse momentum space they, unlike the thermal photons, have a zero elliptic flow. The resulting total $v_2$ of direct photons can be therefore calculated from 
\begin{equation}
v_2^{tot} = \frac{v^{th}_2 \mathrm{d}N^{th} + 0}{\mathrm{d}N^{th}+\mathrm{d}N^{pQCD}},
\end{equation}
where $\mathrm{d}N^{th}$ refers to the thermal photon yield, $v^{th}_2$ to its elliptic flow and $\mathrm{d}N^{pQCD}$ to the direct photon yield originating from hard pQCD processes. The effect from the addition of the pQCD component is shown in figure \ref{fig:v2_gamma}. As expected, the total $v_2$ is now suppressed, especially at larger values of $p_T$ where the relative contribution from the pQCD-originating photons is larger. As already the $v_2^{th}$ was below the data, the inclusion of the pQCD component makes the difference between the measured $v_2$ and calculated $v_2$ even larger. However, as can be concluded from the $p_T$ spectra, both contributions clearly should be included and thus the explanation of the discrepancy between the measured and calculated $v_2$ remains unresolved. This is referred to as the ``photon puzzle'' in this field.

To the best of my knowledge, the article [III] was the first study that combined the thermal photon calculations with the full NLO pQCD calculation to obtain the spectra and elliptic flow of direct photons, especially for the LHC case. Earlier studies have used  either LO calculation with a $K$-factor or a parametrized p+p result \cite{vanHees:2011vb}. Certainly, our paper [III] was the first study that applied the spatially dependent nPDFs to consistently calculate the NLO pQCD direct photons in different centrality classes together with the thermal contribution.

\chapter{Conclusions \& Outlook}
\label{chap:outlook}

In the work presented in thesis I have studied the inclusive direct photon and hadron production in high-energy p+p, p/d+$A$ and $A$+$A$ collisions using the NLO pQCD framework and comparing to the recent RHIC and LHC data. The main emphasis has been on nuclear parton distribution functions and especially their spatial dependence.

In the article [I] we published our spatially dependent nPDF sets, \texttt{EPS09s} and \texttt{EKS98s} which can be used to calculate the hard-process cross sections in different centrality classes of nuclear collisions for the first time consistently with the globally analyzed nPDFs. The calculations of the centrality dependent nuclear modification ratios for neutral pion production in d+Au collisions at RHIC were compared to the published PHENIX data and were found to be consistent when all uncertainties were taken into account. Also the predictions for the same observable in p+Pb collisions at the LHC were presented. In general the resulting centrality dependence was found to be rather mild for these observables, especially when considering more central collisions.

In addition to the centrality binned results also the minimum bias $R_{\rm pPb}$ was presented. The preliminary ALICE data for the minimum bias pion $R^{\pi}_{\rm pPb}$ \cite{ALICE_RpPb_pi} seems to be very well consistent with our prediction in figure 17 of [I] and also the published total charged hadron data from ALICE up to $p_T=50\,\mathrm{GeV/c}$ agrees with our predictions nicely. However, the preliminary data by CMS \cite{CMS:2013cka} and ATLAS \cite{ATLAS_RpPb} feature an unexpected excess at $p_T>30\,\mathrm{GeV/c}$ that contradicts the ALICE result and our baseline NLO pQCD prediction. A proper measurement of the p+p baseline at $\sqrt{s_{NN}}=5.0\,\mathrm{TeV}$ can be expected to resolve this discrepancy.

In the article [II] we applied our new spatially dependent nPDFs to study the centrality dependence of inclusive direct photon production in p/d+$A$ and also in $A$+$A$ collisions at mid-rapidity. Again, the centrality dependence was found to be quite weak but consistent with the $A$+$A$ data from PHENIX and CMS. The results were utilized also in article [III] which focused on the $p_T$ spectra and elliptic flow of thermal photons using an event-by-event hydrodynamic simulation. The added NLO pQCD component was found to be necessary to explain the measured direct photon $p_T$ spectra at RHIC and LHC, and to have also a significant impact to direct photon $v_2$.

The current parton-to-hadron fragmentation function sets were systematically studied in the article [IV] in light of the new LHC data for charged hadron production in p+p collisions. We observed that the NLO calculations with the contemporary NLO FF sets tend to overshoot the data due to too hard gluon-to-hadron FFs. This is true especially for the sets that have included charged hadron data from p+p/$\rm \bar{p}$ collisions from RHIC and $\rm Sp\bar{p}S$ that constrain mainly the $p_T<10\,\mathrm{GeV/c}$ region where we find signs of non-perturbative effects as discussed in section \ref{subsec:inde_hadr}. Thus we conclude that a new global analysis including the recent LHC data with a lower cut on the hadron $p_T$ would be required to resolve the issue and recover the validity of the NLO pQCD baseline in high-$p_T$ hadron production.

In the last article of this thesis [V], once having solved the numerical problems in \texttt{INCNLO}, we studied the inclusive direct photon production and the effects from different isolation cuts at forward rapidities in p+Pb collisions at the LHC. The aim was to quantify which $x$-regions are probed at different rapidities and how the isolation cut affects the small-$x$ sensitivity. The direct photons were found to be clearly more sensitive to small-$x$ physics than the inclusive hadrons at the same rapidities, and the isolation cut increased the sensitivity even further. The rapidity dependence at $2 < \eta < 5$ was found to be very weak due to the slow $x$-dependence in the EPS09 nPDFs which follows from the DGLAP evolution. Thus, we conclude that measurements already at $2 < \eta < 3$ region would give important constraints for the so far poorly constrained small-$x$ gluon nPDFs, and that measurements at more forward rapidities would serve as a further test of their DGLAP dynamics. 

As the measured nuclear modification ratios tend to suffer from considerable normalization (systematic) uncertainties, as a new observable we proposed in [V] also a measurement of the yield asymmetry between forward and backward rapidities to study the nPDFs at small $x$. As discussed in section \ref{subsec:dir_photons_centrality}, the direct photon measurements at forward rapidities would be useful to constrain also the spatial dependence of the nPDFs as the centrality dependence turns out to be larger there than at mid-rapidities.

The spatially dependent nuclear PDF sets introduced in the article [I] were based on the smooth Woods-Saxon density distributions. However, as discussed in section \ref{sec:MCGlauber} the fluctuations in the initial nucleon configurations in the colliding nuclei are found necessary to describe the observed triangular flow in heavy-ion collisions. The presence of the dense spots in the nuclei should have an effect also on the nuclear modifications of the PDFs. Thus, a next logical step, as an extension to the work presented here, would be to consider the spatial dependence of the nPDFs in a fluctuating nucleon configuration. 

\begin{figure}[htbp]
\begin{center}
\includegraphics[width=0.7\textwidth]{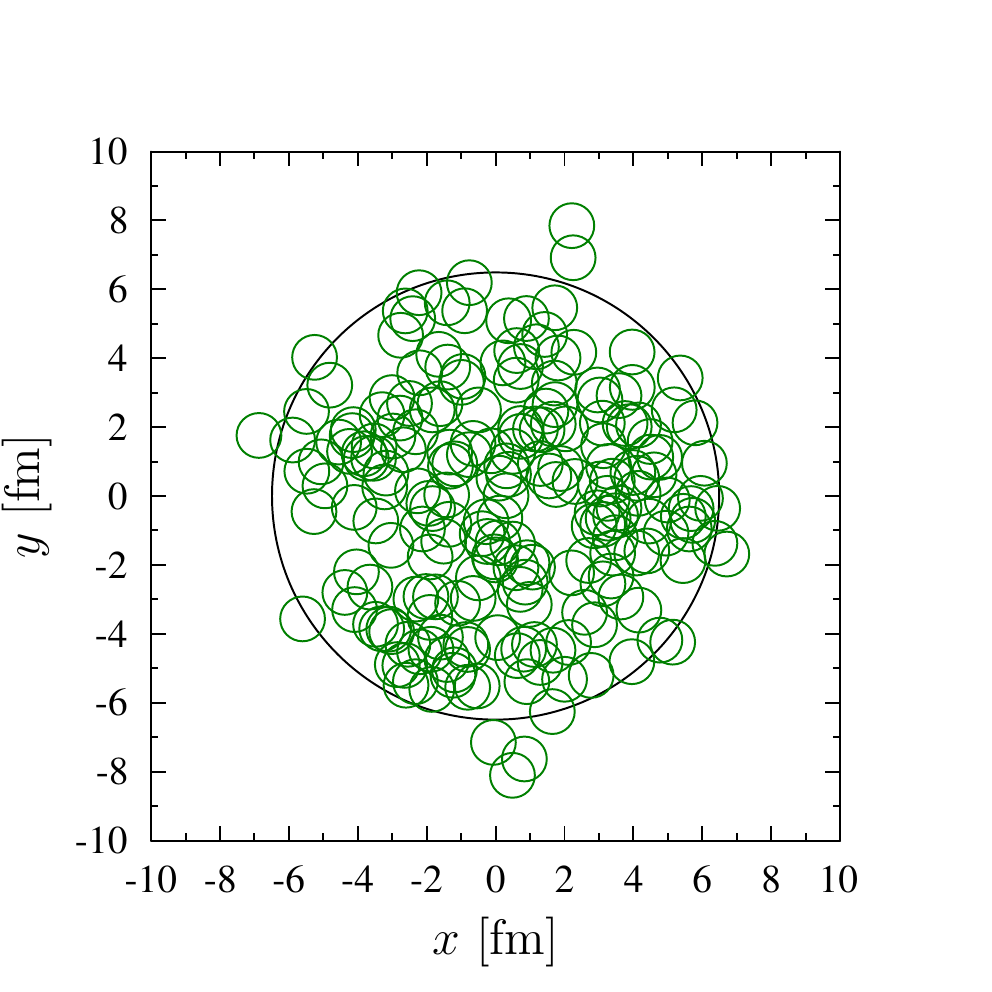}
\caption{A randomly generated nucleon configuration for a Pb-nucleus with $R_A=6.49\,\mathrm{fm}$ and with a nucleon radius $r_n=0.65\,\mathrm{fm}$.}
\label{fig:Pb_configuration}
\end{center}
\end{figure}

To be able to do this one should first figure out a way to relate the generated nucleon configuration to the nuclear thickness function by modeling the thickness of individual nucleons. Then, in principle, one could just replace the smooth thickness function in our power series ansatz by a fluctuating one. However, in our preliminary studies we found out that one can easily generate nucleon configurations that produce very dense spots to the thickness function, an example is shown in figure \ref{fig:Pb_configuration}. As the power series (\ref{eq:r_a}) works well only in the region where we have data constraints ($A\le208$), very large thicknesses may then cause uncontrolled nuclear modifications as the power series of the type (\ref{eq:r_a}) with finite number of terms blows up at sufficiently large values of $T_A(\mathbf{s})$. The problem might be partly solved if the overlapping nucleons are rejected from the generated configuration but if not, one should find an another well behaving dynamically motivated functional form for the thickness function dependence of the nuclear effects. Perhaps one could adopt the exponential form suggested in FGS10~\cite{Frankfurt:2011cs} but then the (EPS09-like) global analysis should also be re-performed with the same input.

Indeed, as the spatially dependent nPDF sets introduced in this thesis were based on the $A$ dependence of earlier global fits, another improvement would be to include the spatial dependence directly into the global analysis. This is bound to increase the number of parameters in the fit but at the same time the $A$ dependence of the parameters would come out naturally. Also one should then include centrality dependent data e.g. the direct photon $R^{\gamma}_{AA}$ measured by PHENIX and the LHC experiments into the analysis to increase the number of data points and hopefully also constraints in the fit. Before the centrality dependent RHIC and LHC p/d+$A$ data can be included in the global analysis, the experimental centrality dependence should be better understood. The feasibility of this kind of an extended global analysis is not trivial and could be tested only by performing such an exercise. This is left as further work.

To summarize, according to our studies and the experimental hard-process RHIC and LHC data from nuclear collisions published so far, collinear factorization with NLO pQCD and universal DGLAP-evolved nPDFs works well. As discussed above, for obtaining well understood, detailed pQCD-baseline from p+p and p/d+A collisions, there are, however, further improvements to be done, both theoretically and experimentally.

\addcontentsline{toc}{chapter}{References}
\bibliographystyle{hunsrt}
\bibliography{thesis}

\end{document}